\documentclass[10pt]{article}
\usepackage{mathptmx}
\usepackage{amssymb,amsmath}
\newtheorem{thm}{Theorem}[section]
\newtheorem{cor}[thm]{Corollary}
\newtheorem{lem}[thm]{Lemma}
\newtheorem{definition}[thm]{Definition}
\newtheorem{prop}[thm]{Proposition}
\def\proof{{\bf Proof. }}

\def\be{\begin{equation}}
\def\ee{\end{equation}}
\def\bea{\begin{eqnarray}}
\def\eea{\end{eqnarray}}
\def\bean{\begin{eqnarray*}}
\def\eean{\end{eqnarray*}}
\def\ea{\end{array}}
\def\ds{\displaystyle}
\def\nm{\noalign{\medskip}}

\def\Z{{\mathcal{Z}}}

\def\F{{\mathcal{F}}}
\def\P{{\mathcal{P}}}
\def\L{{\mathcal{L}}}
\def\H{{\mathcal{H}}}
\def\S{{\mathcal{S}}}
\def\D{{\mathcal{D}}}
\def\N{{\bf N}}
\newcommand{\field}[1]{\mathbb{#1}}
\newcommand{\rz}{\field{R}}
\newcommand{\cz}{\field{C}}
\newcommand{\nz}{\field{N}}
\newcommand{\zz}{\field{Z}}
\def\11{{\rm 1~\hspace{-1.2ex}l} }
\def\d{{\rm{d}}}
\def\p{{\mathbb{P}}}
\def\ra{{\rangle}}
\def\la{{\langle}}

\def\fin{{$\hfill\square$}}
\def\hbarr{{\varepsilon}}
\def\Tr{{\rm{Tr}}}
\def\ccup{\mathop{\cup}}

\begin{document}
\title{ Mean field propagation of Wigner measures and BBGKY hierarchies for general
 bosonic states}
\author{
Z.~Ammari\thanks{zied.ammari@univ-rennes1.fr} \hspace{.3in}
F.~Nier\thanks{francis.nier@univ-rennes1.fr}\\
\\ {\small IRMAR, Universit\'e de Rennes I, }\\ {\small UMR-CNRS 6625,
campus de Beaulieu, 35042 Rennes Cedex, France.
}}
\maketitle
\begin{abstract}
Contrary to the finite dimensional case, Weyl and Wick quantizations
are no more asymptotically equivalent in the infinite dimensional
bosonic second quantization. Moreover neither the Weyl calculus defined
for cylindrical symbols nor the Wick calculus defined for polynomials
are preserved by the action of a nonlinear flow. Nevertheless taking
 advantage carefully of  the information
 brought by these two calculuses in the mean field asymptotics, the
 propagation of Wigner measures for general states can be proved,
 extending to the infinite dimensional case a standard result of
 semiclassical analysis.
\end{abstract}

\section{Introduction}
Our main result is briefly presented in this
introduction.
Accurate definitions will be found in Section~\ref{se.wibbgky}.\\
Let $\H=\Gamma_{s}(\Z)$  be the bosonic Fock space
constructed over the complex separable Hilbert-space $\Z$,
$\Gamma_{s}(\Z)=\oplus_{n=0}^{\infty}\bigvee^{n}\Z$
where  $\bigvee^{n}\Z$  is the symmetric $n$-th
hilbertian tensor power of $\Z$\,.
Consider
the Hamiltonian
$$
H_{\varepsilon}= \d\Gamma(A)+ (\sum_{j=2}^{r}\langle z^{\otimes
  j}\,,\tilde{Q}_{j}z^{\otimes j}\rangle)^{Wick}
$$
defined for the self-adjoint operator $(A,\D(A))$ on $\Z$ and $\tilde{Q_{j}}=\tilde{Q_{j}}^{*}\in
\L(\bigvee^{j} \Z)$. It is the Wick quantized version of
the classical Hamiltonian
$$
h(z,\bar z)= \langle z\,,\, A z\rangle + \sum_{j=2}^{r}\langle z^{\otimes
  j}\,,\tilde{Q}_{j}z^{\otimes j}\rangle\,,\quad z\in \D(A)\subset \Z\,.
$$
When $\Z=L^{2}(\rz^{d})$\,, the operator $H_{\varepsilon}$ is formally
written
$$
H_{\varepsilon}=\int_{\rz^{2d}} A(x,y)a^{*}(x)a(y)~dxdy +
\sum_{j=2}^{r}\int_{\rz^{2dj}} \tilde{Q}_j(x_{1},\ldots, x_{j}\,,\, y_{1},\ldots,
y_{j})
a^{*}(x_{1})\ldots a^{*}(x_{j})a(y_{1})\ldots a(y_{j})~dxdy\,,
$$
with the $\varepsilon$-dependent canonical commutation relations $\left[a(x)\,,\, a^*(y)\right]=\varepsilon\delta(x-y)\,$.
Here $A(.,.)$ and $\tilde Q_j(.,.)$ denote the kernels of the
operators $A$ and $\tilde Q_j$.
The mean field asymptotics is concerned with the limit as
$\varepsilon\to 0$, where $\frac{1}{\varepsilon}=N_{\varepsilon}$
represents a large number of particles and where $\varepsilon$ enters
in the CCR-relations by
$$
\forall f,g\in\Z\,,\quad\left[a(f)\,,\, a^*(g)\right]=\varepsilon\langle f\,,\,g \rangle \,I\,.
$$
The number operator is ${\bf N}=\d\Gamma(I_{\Z})$, with ${\bf
  N}z^{\otimes n}=\varepsilon n z^{\otimes n}$. For a normal state
$\varrho_{\varepsilon}\in \L^{1}(\bigvee^{N_{\varepsilon}} \Z)\subset
\L^{1}(\H)$  with $\Z=L^{2}(\rz^{d})$,
a standard tool considered in the mean field limit is the BBGKY
hierarchy of reduced density matrices:
$$
\gamma^{(p)}_{\varepsilon}(x,y)=\int_{\rz^{2d(N_{\varepsilon}-p)}}\varrho_{\varepsilon}(x,X,y,X)~dX\,,\quad
p\in\nz\,,
$$
and such a definition will be extended to general $\Z$ and
normal states
$\varrho_{\varepsilon}\in \mathcal{L}^{1}(\H)$ fulfilling the condition
 $\Tr\left[\varrho_{\varepsilon}{\bf N}^{k}\right]<+\infty$ for all
 $k\in\nz$\,.\\
For a cylindrical function, $b(z)= b(\wp z)$ for some finite rank
projection $\wp$ and $b$ belonging to the Schwartz class $\mathcal{S}(\wp\Z)$, the Weyl
quantization can be given by
$$
b^{Weyl}=\int_{\wp\Z}\mathcal{F}[b](z) \;W(\sqrt{2}\pi z)~L_{p}(dz)\,,
$$
where
$W(\sqrt{2\pi}z)=e^{i\pi (a(z)+a^{*}(z))}$ and  where $L_{p}$ and $\mathcal{F}$
  are respectively the Lebesgue measure on $\wp\Z$ and the ($\varepsilon$-independent)
 Fourier-transform on $\mathcal{S}(\wp\Z)$.
Associated with a family $(\varrho_{\varepsilon})_{\varepsilon\in
  (0,\bar \varepsilon)}$, Wigner measures can be defined by
$$
\lim_{k\to\infty}\Tr\left[\varrho_{\varepsilon_{k}}b^{Weyl}\right]=\int_{\Z}b(z)~d\mu(z)
$$
after extracting subsequences under the sole uniform estimate
$\Tr\left[\varrho_{\varepsilon}{\bf N}^{\delta}\right]\leq C_{\delta}$
for some $\delta>0$\,.

\bigskip

The problem of the mean field dynamics questions whether the
asymptotic quantities as $\varepsilon\to 0$
associated with
$$
\varrho_{\varepsilon}(t)=
e^{-i\frac{t}{\varepsilon}H_{\varepsilon}}\varrho_{\varepsilon}
e^{-i\frac{t}{\varepsilon}H_{\varepsilon}}\,,\quad t\in\rz
$$
are transported by the flow ${\bf F}_t$ generated by the classical
Hamiltonian
$h(z,\bar z)$ and given, after writing $z_{t}={\bf F}_{t-s}(z_{s})$, by
\begin{equation}
  \label{eq.hartintro}
i\partial_{t}z_{t}= (\partial_{\bar
  z}h)(z_{t},\bar{z}_{t})=Az_{t}+\sum_{j=2}^{r} j\langle
z_{t}^{\otimes j-1}\,,\, \tilde{Q}_{j}\,z_t^{\otimes j}\rangle\,.
\end{equation}
The finite dimensional case enters in the standard framework of
semiclassical analysis and has been studied extensively in the 80's
and 90's by various authors and with various methods
(\cite{Rob}\cite{HMR}\cite{GMMP}\cite{LiPa} \cite{CRR}\cite{Mar}\cite{Fol}
and references therein).\\
It was first considered by Hepp in \cite{Hep} and extended by Ginibre
and Velo in \cite{GiVe1}\cite{GiVe2} by the squeezed coherent states
method well-known as the Hepp method (see also \cite{RoSc}\cite{AmBr}).
More recently the question of
the mean field dynamics has been tackled with the so-called
BBGKY-hierarchy approach inspired by the BBGKY-method of classical
kinetic theory (see \cite{Spo}\cite{BGM}\cite{ErYa}\cite{BEGMY}
\cite{GMP}\cite{ABGT} \cite{AGT}\cite{ESY} and also the related works
\cite{KlMa}\cite{ChPa}).
In \cite{FGS}\cite{FKP}\cite{FKS} a specific use of the structure of
the Wick calculus in the bosonic Fock space was used to make work
truncated Dyson expansions for the mean field dynamics of specific
states.
The aim of our work started in \cite{AmNi1} was to restore the
phase-space geometric nature of the problem
in the spirit of \cite{Ber}\cite{Gro}\cite{KrRa}\cite{Las}
and to extend  as much as possible to the infinite dimensional case,
the methods well understood
for the semiclassical finite dimensional problem. In this first article, we explained
the construction of Wigner
measures, analyzed accurately the gap of information carried by Weyl
observables and Wick observables and use these Wigner (or
semiclassical) measures to reformulate known propagation results.
In \cite{AmNi2}, we reconsidered the truncated Dyson expansion method
of \cite{FGS}\cite{FKP}\cite{FKS} in order to prove the propagation of
Wigner measures for some specific families of states. We are now able
to state the following general result (still
with a regular interaction term
contrary to many other works cited above).
\begin{thm}
\label{th.main}
Let $(\varrho_{\varepsilon})_{\varepsilon\in(0,\bar\varepsilon)}$
be a family of normal states on $\H$
with a single Wigner measure $\mu_{0}$
and such that
\begin{equation}
\label{eq.hypconv}
\forall \alpha\in\nz,
\quad
\lim_{\varepsilon\to 0}
{\rm Tr}[\varrho_{\varepsilon}
\mathbf{N}^\alpha]=\int_{\Z}|z|^{2\alpha}~d\mu_{0}(z) < +\infty\,.
\end{equation}
Then for all $t\in \rz$, the family
$(\varrho_{\varepsilon}(t)=e^{-i\frac{t}{\varepsilon}H_{\varepsilon}}\varrho_{\varepsilon}e^{i\frac{t}{\varepsilon}
H_{\varepsilon}})_{\varepsilon\in
(0,\bar \varepsilon)}$
has a unique Wigner measure $\mu_{t}=({\bf F}_t)_{*}\mu_{0}$, which is
the initial measure $\mu_{0}$ pushed forward by the flow associated
with \eqref{eq.hartintro}.\\
Moreover the convergence
$$
\lim_{\varepsilon\to
  0}\Tr\left[\varrho_{\varepsilon}(t)b^{Wick}\right]=\int_{\Z} b\circ {\bf F}_t(z)~d\mu_{0}(z)
$$
holds for any
$b\in\P_{alg}(\Z)=\oplus_{p,q\in\nz}^{alg}\P_{p,q}(\Z)$\,.\\
Finally, the convergence of the reduced density matrices
$$
\lim_{\varepsilon\to
  0}\gamma^{(p)}_{\varepsilon}(t)=\frac{1}{\int_{\Z}|z|^{2p}~d\mu_{t}(z)}
\int_{\Z}|z^{\otimes p}\rangle\langle z^{\otimes p}|~d\mu_{t}(z)
=: \gamma_{0}^{(p)}(t)\,,
$$
holds in the $\mathcal{L}^{1}(\bigvee^{p}\Z)$-norm for all $p\in\nz$\,.
\end{thm}
\noindent\textbf{Comments:} The existence of Wigner measures
as Borel probability measures requires a uniform estimate
$\Tr\left[\varrho_{\varepsilon}{\bf N}^{\delta}\right]\leq C_{\delta}$
for some $\delta>0$, but such an assumption would be redundant with
the existence of bounded limits stated in \eqref{eq.hypconv}.\\
The uniqueness of the Wigner measure $\mu_{0}$ is not really a strong
assumption since it suffices to  replace the whole family
$(\varrho_{\varepsilon})_{\varepsilon\in (0,\bar \varepsilon)}$ by a
suitable extracted sequence $(\varrho_{\varepsilon_{k}})_{k\in\nz}$,
$\lim_{k\to\infty}\varepsilon_{k}=0$, in order to fulfill this
requirement. Such a reduction argument after extraction will often be
used. \\
The fact that the quantities $\Tr\left[\varrho_{\varepsilon}{\bf
    N}^{\alpha}\right]$ are uniformly bounded w.r.t $\varepsilon\in
(0,\bar \varepsilon)$ is also very natural within the mean field
framework and satisfied by all known physical examples.\\
Actually the strong assumption which is not satisfied in all cases is
that the limit in \eqref{eq.hypconv} equals
$\int_{\Z}|z|^{2\alpha}~d\mu_{0}$.
This condition prevents from the
 phenomenon of ``infinite dimensional defect of
compactness'' identified in \cite{AmNi1} and which was shown to appear
in the physical example of the Bose-Einstein free gas (the non
condensated phase is responsible for a discrepancy between the
left- and right-hand sides of \eqref{eq.hypconv}). The analysis of
this phenomenon is improved in Section~\ref{se.wibbgky}.\\
Finally our proof no more uses truncated Dyson expansions of the quantum
flow and relies only on the good properties of the classical flow,
after exploiting all the a priori information given by the Weyl
and Wick calculus.

\bigskip
\noindent\textbf{Outline:} The Section~\ref{se.wibbgky} introduces the various
objects used for our analysis, Wick and Weyl calculuses, Wigner
measures, reduced density matrices. The conditions presented in
\cite{AmNi2} are reduced to the simple equivalent form
\eqref{eq.hypconv} in Subsection~\ref{se.relywick}.
After this the Subsection~\ref{se.locball} is devoted to
the notion of states localized in a ball.\\
The dynamics is studied in Section~\ref{se.dynmeanfield}. First a
simple condition is proved to ensure, via some equicontinuity
argument, the possibility of a common extraction process
$(\varepsilon_{k})_{k\in\nz}$ for all times $t\in\rz$\,. Then the
propagation of Wigner measures is proved for states localized in a
ball. Then the truncation is removed and all the arguments are
gathered for the proof of Theorem~\ref{th.main} in Subsection \ref{se.thm1}. Finally, additional
simple consequences are listed in Subsection~\ref{se.coro}.\\
Examples are presented in~Section~\ref{se.examples}.
It is recalled that the regular interactions are
physically relevant within the modelling of the rapidly rotating Bose
condensates in the Lowest Landau Level approximation. Details are
given about the propagation of non trivial Wigner measures supported
on a torus, which shows the advantage of this formulation compared to
the BBGKY hierarchy method.
Finally, the propagation
of Wigner measures provides a nice formulation of the Hartree-von
Neumann limit.

\section{Information carried by Wigner measures}
\label{se.wibbgky}
After introducing the symmetric Fock space with
$\varepsilon$-dependent CCR's and recalling some properties of
the Wick quantization, the connection between infinite dimensional
Wigner measures and the BBGKY presentation of the many body problem is
explicitly specified. This section ends with the notion of states
localized in a ball, which will be usefull in the proof of Theorem~\ref{th.main}.

\subsection{Fock space}
Consider  a separable Hilbert space $\mathcal{Z}$ endowed with a
scalar product $\la .,.\ra$ which is anti-linear in the left
argument and linear in the right one and with the associated norm
$|z|=\sqrt{\la z,z\ra}$. Let $\sigma={\rm Im}\la ., .\ra$ and
$S={\rm Re} \la .,.\ra$ respectively denote the canonical symplectic form
and
the real scalar product over $\mathcal{Z}$. The symmetric Fock space on $\mathcal{Z}$
is the Hilbert space
\begin{eqnarray*}
\mathcal{H}=\bigoplus_{n=0}^\infty\bigvee\,^n
\mathcal{Z}=\Gamma_{s}(\Z)\,,
\end{eqnarray*}
where $\bigvee^n \mathcal{Z}$ is
the $n$-fold symmetric tensor product. Almost all the direct sums
and tensor products are completed within the Hilbert framework. This
is omitted in the notation. On the contrary, a specific
$^{\textrm{alg}}$ superscript will be used for the algebraic direct
sums or tensor products.

For any $n\in \nz$, the orthogonal projection of $\bigotimes^n
\Z$ onto the closed subspace $\bigvee^n\Z$ will be denoted
by $\S_{n}$. For any $(\xi_{1},\xi_{2},\ldots,\xi_{n})\in \Z^{n}$,
 the vector $\xi_{1}\vee \xi_{2}\vee\cdots \vee\xi_{n}\in
 \bigvee^{n}\Z$
will be
\begin{equation}
\label{symetrie}
\xi_{1}\vee\xi_{2}\vee\cdots\vee\xi_{n}
=
\S_{n}(\xi_{1}\otimes\xi_{2}\cdots\otimes \xi_{n})
=\frac{1}{n!}\sum_{\pi\in \mathfrak{S}_n}\xi_{\pi(1)}\otimes\xi_{\pi(2)}\cdots\otimes \xi_{\pi(n)}\;,
\end{equation}
where $\mathfrak{S}_n$ is the symmetric group of degree $n$.
The  family of vectors $(\xi_{1}\vee\cdots\vee \xi_{n})_{\xi_{i}\in
\Z}$ is a total family of $\bigvee^{n}\Z$ and thanks to
the polarization identity
\begin{equation}
\label{eq.pola}
\xi_{1}\vee\xi_{2}\vee\cdots\vee\xi_{n}=\ds \frac{1}{2^n n!}
\sum_{\varepsilon_i=\pm 1} \varepsilon_1\cdots \varepsilon_n \;
\big( \sum_{j=1}^n \varepsilon_j
\xi_j)^{\otimes n}\,,
\end{equation}
the same property holds for $\left(\xi^{\otimes n}\right)_{n\in \nz, \xi\in \Z}$\,.

For two operators
$A_{k}:\bigvee^{i_{k}}\Z\to \bigvee^{j_{k}}\Z$, $k=1,2$, the
notation $A_{1}\bigvee A_{2}$ stands for
$$
A_{1}\bigvee A_{2}=\S_{j_{1}+j_{2}}\circ(A_{1}\otimes A_{2})\circ
\S_{i_{1}+i_{2}}\in
{\cal
  L}(\bigvee\,^{i_{1}+i_{2}}\Z, \bigvee\,^{j_{1}+j_{2}}\Z)\,.
$$
Any $z\in \Z$ is identified with the operator $|z\rangle:
\bigvee^{0}\Z=\cz\ni \lambda\mapsto \lambda z\in \Z=\bigvee^{1}\Z$
while $\la z|$ denotes the linear form $\Z\ni\xi\mapsto \la z\,,\,\xi\ra \in \cz$.
The creation and annihilation operators
 $a^*(\xi)$ and $a(\xi)$, parametrized by $\varepsilon>0$,
are then defined by:
\begin{eqnarray*}
a(\xi)_{|\bigvee^n\Z}&=&\sqrt{\varepsilon n} \; \;\la \xi|\otimes
I_{\bigvee^{n-1}\Z}\\
a^*(\xi)_{|\bigvee^n\Z}&=&\sqrt{\varepsilon (n+1)}  \;\;\S_{n+1} \circ
(\;|\xi\rangle\otimes I_{\bigvee^n\Z})=\sqrt{\varepsilon(n+1)} \;\xi\bigvee
I_{\bigvee^{n}\Z}\,
\end{eqnarray*}
and satisfy the canonical commutation relations (CCR):
\begin{eqnarray} \label{ccr}
[a(\xi_1),a(\xi_2)]=[a^*(\xi_1),a^*(\xi_2)]=0, \quad [a(\xi_1),a^*(\xi_2)]=\varepsilon\la \xi_1,\xi_2\ra I. \end{eqnarray}
We  also consider the
canonical quantization of the real
  variables
  $\Phi(\xi)=\frac{1}{\sqrt{2}} (a^*(\xi)+a(\xi))$ and
$\Pi(\xi)=\Phi(i\xi)=\frac{1}{i\sqrt{2}} (a(\xi)-a^*(\xi))$. They are
self-adjoint operators on $\mathcal{H}$ and satisfy the identities:
\begin{eqnarray*} [\Phi(\xi_1),\Phi(\xi_2)]=i \varepsilon \sigma(\xi_1,\xi_2) I,
\hspace{.3in} [\Phi(\xi_1),\Pi(\xi_2)]=i\varepsilon S(\xi_1,\xi_2)
I. \end{eqnarray*}
The
representation of the  Weyl commutation relations in the Fock space
\begin{eqnarray}
\label{eq.Weylcomm}
W(\xi_1) W(\xi_2)&=&e^{-\frac{i\varepsilon}{2} \sigma(\xi_1,\xi_2)} W(\xi_1+\xi_2) \\
\nonumber
 &=& e^{-i \varepsilon \sigma(\xi_1,\xi_2)} W(\xi_2) W(\xi_1),
  \end{eqnarray}
is obtained by setting $W(\xi)=e^{i\Phi(\xi)}$.
The   number operator is also parametrized by $\varepsilon>0$,
\begin{eqnarray*}
 {\bf N}_{|\bigvee^n \Z}=\varepsilon{n} I_{|\bigvee^n \Z}.
\end{eqnarray*}
It is convenient to introduce the subspace
\begin{eqnarray*}
\H_{fin}=\bigoplus_{n\in\nz}^{\textrm{alg}}\bigvee\,^n\Z
\end{eqnarray*}
of $\H$, which is a set of analytic vectors for $\bf N$.\\
For any contraction $S\in {\cal L}(\Z)$,
$\left|S\right|_{\mathcal{L}(\mathcal{H})}\leq 1$,
$\Gamma(S)$ is the
contraction in $\H$  defined by
$$
\Gamma(S)_{|\bigvee^{n}\Z}=S\otimes S\cdots\otimes S\,.
$$
More generally $\Gamma(B)$ can be defined by the same formula as an
operator on $\H_{fin}$ for any $B\in \mathcal{L}(\Z)$.
Meanwhile, for any self-adjoint operator $A:\Z\supset \D(A)\to \Z,$
the operator $\d\Gamma(A)$ is the self-adjoint operator given by \begin{eqnarray*} &&
e^{\frac{it}{\varepsilon}\d\Gamma(A)}=\Gamma(e^{itA})\\
&& \d\Gamma(A)_{|\bigvee^{n,\textrm{alg}}\D(A)}=\varepsilon\left[
\sum_{k=1}^{n}I\otimes\cdots\otimes\underbrace{A}_{k}\otimes
\cdots\otimes I\right]\,. \end{eqnarray*}
 For example $\N=\d\Gamma(I)$\,.

\subsection{Wick operators}
\label{sec.Wick}  The Wick symbolic
calculus on (homogenous) polynomials as introduced in \cite{AmNi1} is
recalled with its basic properties.
\begin{definition}
\label{de.hompol}
For $p,q\in \nz$, $\P_{p,q}(\Z)$ denotes the set of $(p,q)$-homogeneous polynomial
functions on $\Z$ which fulfill :
$$
b(z)=\left\langle z^{\otimes q}\,,\,\tilde{b}z^{\otimes p}\right\rangle \quad \mbox{ with } \quad \tilde b\in
   \L(\bigvee\,^p\Z, \bigvee\,^q\Z)\,.
$$
The subspace of $\P_{p,q}(\Z)$ made of polynomials $b$ such that
$\tilde{b}$ is a compact operator $\tilde{b}\in
\mathcal{L}^{\infty}(\bigvee^p\Z, \bigvee^q\Z)$ (resp. $b\in \mathcal{L}^{r}(\bigvee^p\Z, \bigvee^q\Z)$) is
denoted by $\mathcal{P}^{\infty}_{p,q}(\Z)$
(resp.~$\mathcal{P}^{r}_{p,q}(\Z)$).\\
On those spaces, the natural norms are
$$
|b|_{\mathcal{P}_{p,q}}=|\tilde b|_{ \L(\bigvee^p\Z,
  \bigvee^q\Z)}\quad
\text{ and } \quad
|b|_{\mathcal{P}^{r}_{p,q}}=|\tilde b|_{ \L^{r}(\bigvee^p\Z,
  \bigvee^q\Z)}\,,\quad 1\leq r\,.
$$
The set of non homogeneous polynomials, the algebraic direct sum
$\oplus_{p,q\in\nz}^{alg}\P_{p,q}(\Z)$ (resp. $\oplus_{p,q\in\nz}^{alg}\P_{p,q}^{r}(\Z)$ with
 $1\leq r\leq \infty$),
will be denoted by $\P_{alg}(\Z)$  (resp. $\P_{alg}^{r}(\Z)$)\,.
\end{definition}
Owing to the condition $\tilde{b}\in  \L(\bigvee^p\Z,
\bigvee^q\Z)$ for $b\in \P_{p,q}(\Z)$, this definition implies that any
G\^ateaux differential $\partial_{\overline{z}}^{j}\partial_{z}^{k}b(z)$ at the
point $z\in \Z$ belongs to $  \L(\bigvee\,^{k}\Z,\bigvee\,^{j}\Z)$ with
\begin{equation*}
\langle \varphi, \partial_{\overline{z}}^{j}\partial_{z}^{k}b(z)
\psi\rangle=\frac{p!}{(p-k)!}\frac{q!}{(q-j)!} \langle z^{\otimes q-j}\vee\varphi, \tilde{b} \,z^{\otimes p-k}\vee \psi \rangle\,.
\end{equation*}
In particular, we recover the operator $\tilde b$ from $b(z)$ via the relation
$$
\tilde{b}=\frac{1}{p!}\frac{1}{q!}
\partial_{z}^{p}\partial_{\overline{z}}^{q}b(z)\in
\L(\bigvee\,^p\Z, \bigvee\,^q\Z)\,.
$$
With any "symbol" $b\in \P_{p,q}(\Z)$, a linear operator $b^{Wick}$ called Wick monomial can
be associated according to:
\begin{eqnarray}
\label{def-wick}\nonumber
&&b^{Wick}:\H_{fin}\to\H_{fin},\\
&&b^{Wick}_{|\bigvee^n \Z}=1_{[p,+\infty)}(n)\frac{\sqrt{n!
(n+q-p)!}}{(n-p)!} \;\varepsilon^{\frac{p+q}{2}} \;\left(\tilde{b}\bigvee
I_{\bigvee^{n-p} \Z}\right)\quad \in {\cal
L}(\bigvee\,^{n}\Z,\bigvee\,^{n+q-p}\Z)\,,
\end{eqnarray}
with
$\tilde{b}=(p!)^{-1}(q!)^{-1}\partial_{z}^{p}\partial_{\overline{z}}^{q}b(z)$\,.
The basic symbol-operator correspondence:
\begin{eqnarray*}
&&\begin{array}{ccc}
 \la z,\xi\ra & \longleftrightarrow & a^*(\xi) \\
  \la \xi, z\ra & \longleftrightarrow & a(\xi) \\
\end{array}
\hspace{0.7cm}
\begin{array}{ccc}
\sqrt{2} S(\xi, z)  & \longleftrightarrow & \Phi(\xi)\\
\sqrt{2} \sigma(\xi, z)  & \longleftrightarrow &  \Pi(\xi)\,
\end{array}
\hspace{0.7cm}
\begin{array}{ccc}
\la z, A z\ra   & \longleftrightarrow &  \d\Gamma(A) \\
\left|z\right|^{2}  & \longleftrightarrow & {\bf N}\,,
\end{array}
\end{eqnarray*}
and more generally
$$
\left(\prod_{i=1}^p \la z,\eta_i\ra \times \prod_{j=1}^q \la
\xi_j,z\ra\right)^{Wick}=a^*(\eta_1)\cdots a^*(\eta_p)a(\xi_1)\cdots a(\xi_q).
$$
We have the following properties.
\begin{prop}
\label{wick_prop}
The following identities hold true on $\H_{fin}$ for every $b\in\P_{p,q}(\Z)$:\\
(i) $ \big(b^{Wick}\big)^*=\bar b^{Wick}$.\\
(ii) $\big(C(z) b(z) A(z)\big)^{Wick}=C^{Wick} b^{Wick} A^{Wick},$
if $A\in\P_{\alpha,0}(\Z)$, $
C\in\P_{0,\beta}(\Z).$\\
(iii) $e^{i \frac{t}{\varepsilon} \d\Gamma(A)} b^{Wick} e^{-i
  \frac{t}{\varepsilon} \d\Gamma(A)}=\left(b(e^{-it A}z)\right)^{Wick}$,
if $A$ is a self-adjoint operator on $\Z$.
\end{prop}
A consequence of i) says that $b^{Wick}$ is symmetric when $q=p$ and
$\tilde{b}^{*}=\tilde{b}$. Moreover the definition \eqref{def-wick} gives
\begin{equation}
  \label{eq.poswick1}
\left(q=p\quad\text{and}\quad \tilde{b}\geq 0\right)
\Rightarrow
\left(b^{Wick}\geq 0\;\text{on}\; \H_{fin}\right)\,,
\end{equation}
which is false for general non negative polynomial
symbols\footnote{This property should not be confused with the positivity of
  the finite dimensional \underline{Anti-Wick} quantization which associates a non negative
operator to any non negative symbol.}. For an
increasing net of non negative operators
$(\tilde{b}_{\alpha})_{\alpha}$, $\tilde{b}_{\alpha}\in
\mathcal{L}(\bigvee^{p}\Z)$ (again $q=p$), it also gives
\begin{equation}
\label{eq.supwick1}
\left(\tilde{b}=\sup_{\alpha}~\tilde{b}_{\alpha} \;\text{ in }\; \mathcal{L}(\bigvee\,^{p}\Z)\right)
\Rightarrow
\left(\forall \varphi\in \H_{fin}\,,\quad
\langle \varphi\,,\, b^{Wick}\varphi\rangle
=\sup_{\alpha}\langle \varphi\,,\,b_{\alpha}^{Wick}\varphi\rangle
\right)\,.
\end{equation}
When $\Z=L^{2}(\rz^{d}, dx)$, the general formula for $b^{Wick}$ with
$b\in \P_{p,q}(\Z)$ is simply
\begin{equation*}
  b^{Wick}=\int_{\rz^{d(p,q)}}\tilde{b}(y_{1},\ldots,
  y_{q},x_{1},\ldots,x_{p}) \, a^{*}(y_{1})\ldots
  a^{*}(y_{q}) \, a(x_{1})\ldots a(x_{p})~dx_{1}\cdots dx_{p}\,dy_{1}\cdots dy_{q}\,,
\end{equation*}
where $\tilde{b}(y,x)$ is the Schwartz kernel of $\tilde{b}$ and where
$a(x_{k})=a(\delta_{x_{k}})$ according to the usual convention.
\begin{prop}
\label{wick-estimate}
For $b\in\P_{p,q}(\Z)$, the following number estimate holds
\begin{equation}
  \label{eq.2bis}
\left|\left\langle {\bf N}\right\rangle^{-\frac{q}{2}}b^{Wick}
\left\langle
 {\bf N}\right\rangle^{-\frac{p}{2}}\right|_{\mathcal{L}(\mathcal{H})}\leq
 \left|b\right|_{\P_{p,q}}\,.
\end{equation}
\end{prop}
The relations \eqref{eq.poswick1} and \eqref{eq.supwick1} now become
for $b\in \P_{p,p}(\Z)$ or $b_{\alpha}\in \P_{p,p}(\Z)$
\begin{eqnarray}
\label{eq.poswick}
&\left(q=p\quad\text{and}\quad\tilde{b}\geq 0\right)
\Rightarrow
\left(\langle {\bf N}\rangle^{-p/2}b^{Wick}\langle {\bf N}\rangle^{-p/2}\geq
  0\quad\text{in}\;\L(\H)\right)\,,
&\\
  \label{eq.supwick}
&\left(\tilde{b}=\sup_{\alpha}~\tilde{b}_{\alpha}\;\text{in}\;\mathcal{L}(\bigvee^{p}\Z)\right)
\Rightarrow
\left(\langle {\bf N}\rangle^{-p/2}b^{Wick}\langle {\bf N}\rangle^{-p/2}
=
\sup_{\alpha}~\langle {\bf N}\rangle^{-p/2}b_{\alpha}^{Wick}\langle
 {\bf N}\rangle^{-p/2}
\;\text{in}\; \L(\H)\right)\,.
&
\end{eqnarray}
An important  property of our class of Wick polynomials is
that a composition of
$b_{1}^{Wick}\circ b_{2}^{Wick}$ with
$b_1,b_2\in \P_{alg}(\Z)$
is a Wick polynomial with symbol in $\P_{alg}(\Z)$.
For $b_{1}\in \P_{p_{1},q_{1}}(\Z)$, $b_{2}\in \P_{p_{2},q_{2}}(\Z)$,
$k\in \nz$ and any fixed $z\in \Z$, $\partial_{z}^{k}b_{1}(z)\in
\L(\bigvee^{k}\Z;\cz)$ while $\partial_{\bar z}^{k}b_{2}(z)\in
\bigvee^{k}\Z$.  The $\cz$-bilinear duality product
$\partial_{z}^{k}b_{1}(z).\partial_{\bar z}^{k}b_{2}(z)$ defines a
function of $z\in\Z$ simply denoted by
$\partial_{z}^{k}b_{1}.\partial_{\bar z}^{k}b_{2}$\,.
We also use the following notation for multiple Poisson brackets:
\begin{eqnarray*}
&&\{b_1,b_2\}^{(k)}=\partial^k_z b_1
.\partial^k_{\bar z} b_2 -\; \partial^k_z b_2 .\partial^k_{\bar z}
b_1,\quad k\in\nz\,,\\
&&\{b_1,b_2\}=\{b_1,b_2\}^{(1)}.
\end{eqnarray*}
\begin{prop}
\label{symbcalc}
Let $b_{1}\in \P_{p_{1},q_{1}}(\Z)$ and  $b_{2}\in \P_{p_{2},q_{2}}(\Z)$\,.\\
For any $k\in \left\{0,\ldots, \min\left\{p_{1},q_{2}\right\}\right\}$,
$\partial_{z}^{k}b_{1}.\partial_{\bar z}^{k}b_{2}$ belongs to
$\P_{p_{2}-k,q_{1}-k}(\Z)$ with the estimate
$$
|\partial_{z}^{k}b_{1}.\partial_{\bar z}^{k}b_{2}|_{\P_{p_{2},q_{1}}}\leq
\frac{p_{1}!}{(p_{1}-k)!}\frac{q_{2}!}{(q_{2}-k)!}
|b_{1}|_{\P_{p_{1},q_{1}}}|b_{2}|_{\P_{p_{2},q_{2}}}\,.
$$
The formulas
\begin{eqnarray*}
(i)&&
b_1^{Wick} \circ b_2^{Wick}=\left(\sum_{k=0}^{\min\{p_1,q_2\}}
\;\;\frac{\varepsilon^k}{k!}  \;\;\;\partial^{k}_{z} b_1 .\partial^{k}_{\bar z}
b_2 \right)^{Wick}= \left(e^{\varepsilon \la \partial_z,\partial_{\bar
\omega}\ra}
b_1(z) b_2(\omega)\left|_{z=\omega}\right. \right)^{Wick}\,,\\
(ii)&&
[b_1^{Wick},b_2^{Wick}]=\left(\sum_{k=1}^{\max\{\min\{p_1,q_2\}\,,\,
\min\{p_{2},q_{1}\}\}} \;\;\frac{\varepsilon^k}{k!}  \;\;\{b_1
,b_2\}^{(k)} \right)^{Wick}\,,
\end{eqnarray*}
hold as identities on $\H_{fin}$.
\end{prop}

\subsection{Cylindrical functions and Weyl   quantization}
\label{se.cyl}
Let  $\p$ denote the set of all finite rank orthogonal projections
on $\Z$ and  for a given $p\in\p$ let $L_{p}(dz)$ denote the
Lebesgue measure on the finite dimensional subspace $p\Z$. A
function $f:\Z\to\cz$ is said  cylindrical if there
exists $p\in\p$ and a function $g$ on $p\Z$ such that $ f(z)=g(pz),$
for all $z\in\Z$. In this case we say that $f$ is based on the
subspace $p\Z$. We set $\S_{cyl}(\Z)$ to be the   cylindrical
Schwartz space:
$$
(f\in \S_{cyl}(\Z))\Leftrightarrow
\left(\exists p\in \p,\exists g\in \S(p\Z), \quad f(z)= g(pz)\right)\,.
$$
The Fourier transform of a function $f\in\S_{cyl}(\Z)$ based on the
subspace $p\Z$ is defined as
\begin{eqnarray*}
\F[f](z)=\int_{p\Z} f(\xi) \;\;e^{-2\pi i
\,S(z,\xi)}~L_{p}(d\xi)
\end{eqnarray*}
and its inverse Fourier transform is
\begin{eqnarray*}
f(z)=\int_{p\Z} \F[f](z) \;\;e^{2\pi i
\,S(z,\xi)}~L_{p}(dz)\,.
\end{eqnarray*}
With any
 symbol $b\in\S_{cyl}(\Z)$ based on $p\Z$,
a {\it Weyl observable} can be associated  according to
\begin{eqnarray}
\label{weyl-obs} b^{Weyl}=\int_{p\Z} \F[b](z) \;\;\; W(\sqrt{2}\pi
z)~L_{p}(dz)\,.
\end{eqnarray}
After the tensor decompositions
\begin{eqnarray*}
&&
\H=\Gamma_{s}(\Z)=\Gamma_{s}(p\Z)\otimes
\Gamma_{s}((1-p)\Z)\quad \mbox{ due to }~\Z=p\Z\mathop{\oplus}^{\perp}(1-p)\Z
\\
&&
\forall z\in p\Z\,,\quad W(\sqrt{2}\pi z)= W_{p\Z}(\sqrt{2}\pi
z)\otimes I_{\Gamma_{s}(1-p)\Z}
\end{eqnarray*}
where $W_{p\Z}$ denotes the reduced representation in $\Gamma_{s}(p\Z)$, one sees that
the Weyl quantization of cylindrical observables based on $p\Z$
amounts to the usual finite-dimensional Weyl quantization.
Hence more general classes of symbols can be considered.\\
For $p\in\p$, the symbol classes  defined for $0\leq \nu\leq 1$
on the finite dimensional phase
space $p\Z$\,,
\begin{equation}
\label{eq.classsymbp}
S_{p\Z}^{\nu}= \oplus_{n\in\zz}^{alg}S(\langle z\rangle_{p\Z}^{n},
\frac{dz^{2}}{\langle z\rangle_{p\Z}^{2\nu}})\,,
\end{equation}
 where $\langle
z\rangle_{p}^{2}=1+|z|_{p\Z}^{2}$, are natural Weyl-H{\"o}rmander algebras
associated with the finite dimensional harmonic oscillator
Hamiltonian,
${\bf N}_{p}=(|z|_{p\Z}^{2})^{Wick}=(|z|_{p\Z}^{2})^{Weyl}-\frac{\dim
  p\Z}{2}\varepsilon$.
They contains the
polynomial functions on $p\Z$.
The associated class of Weyl quantized
operators after tensorization  with $I_{\Gamma_{s}((1-p))\Z}$ is denoted by $\textrm{Op}S_{p\Z}^{\nu}$.
For a  cylindrical polynomial $b\in \P_{alg}(\Z)$ based on $p\Z$,
$b(z)=b(pz)$, the asymptotic equivalence of the Weyl and Wick
quantization in finite dimension says for any $\nu\in [0,1]$
\begin{equation}
  \label{eq.compWeWi}
b^{Wick}= b^{Weyl}+\mathcal{O}_{b}(\varepsilon)\quad \text{in}~ \textrm{Op}S^{\nu}_{p\Z}\,.
\end{equation}
Such polynomials have finite rank kernels and make a dense set in
$\P^{\infty}_{alg}(\Z)$ \underline{but not in $\P_{alg}(\Z)$}.

\subsection{Wick observables and BBGKY hierarchy}
\label{se.wickbbgky}
When $\Z=L^{2}(\rz^{d})$, mean field results are often presented or
even analyzed in terms of reduced density matrices or more precisely
in terms of a sequence $(\gamma^{(p)}_{\varepsilon})_{p\in\nz}$ with
$\gamma^{(p)}_{\varepsilon}\in \mathcal{L}^{1}(\bigvee^{p}\Z)$\,. This
follows the general BBGKY approach of the kinetic theory and the
$\gamma^{p}_{\varepsilon}$ correspond in the classical case to the
empirical distributions.\\
The basic example is when $\varrho_{\varepsilon}\in
\mathcal{L}^{1}(\bigvee^{n}\Z)$,
$n=\left[\frac{1}{\varepsilon}\right]$~: For any $p\in\nz$, $p\leq n$,
$\gamma^{(p)}_{\varepsilon}\in \mathcal{L}^{1}(\bigvee^{p}\Z)$ is defined as the partially traced
operator with the kernel
$$
\gamma^{(p)}_{\varepsilon}(x_{1},\ldots,x_{p};y_{1}\ldots y_{p})
:=
\int_{\rz^{d(n-p)}}\varrho_{\varepsilon}(x_{1},\ldots,x_{p},X,y_{1},\ldots
y_{p},X)~L_{\rz^{d(n-p)}}(dX)\,.
$$
With the polarization identity \eqref{eq.pola}, the family
$(|\psi^{\otimes n}\rangle\langle
\psi^{\otimes n}|)_{\psi\in \Z}$ forms a total set of
$\mathcal{L}^{1}(\bigvee^{n}\Z)$. Hence the formal identity
\begin{eqnarray*}
\varepsilon^{p}\frac{n!}{(n-p)!}|\psi|^{2(n-p)}\psi(x_{1})\ldots
\psi(x_{p})\overline{\psi(y_{1})}\ldots\overline{\psi(y_{p})}
&=&
\langle a(y_{1})\ldots a(y_{p})\psi^{\otimes n}\,,\,a(x_{1})\ldots
a(x_{p})\psi^{\otimes n}\rangle
\\
&=&
\Tr\left[a^{*}(y_{1})\ldots a^{*}(y_{p})a(x_{1})\ldots
  a(x_{p})|\psi^{\otimes n}\rangle \langle \psi^{\otimes n}|\right]
\end{eqnarray*}
carries over to $\varrho_{\varepsilon}\in
\mathcal{L}^{1}(\bigvee^{n}\Z)$~:
$$
\forall p\in \left\{1,\ldots, n\right\}\,,\quad \varepsilon^{p}\frac{n!}{(n-p)!} \;
\gamma^{(p)}_{\varepsilon}(x_{1},\ldots,x_{p},y_{1}\ldots, y_{p})=\Tr\left[a^{*}(y_{1})\ldots a^{*}(y_{p})a(x_{1})\ldots
  a(x_{p}) \varrho_{\varepsilon} \right]\,.
$$
The correct meaning of this definition is
$$
\Tr\left[\gamma^{(p)}_{\varepsilon}\tilde{b}\right]
=
\frac{1_{[p,+\infty)}(n)}{\varepsilon^{p}n(n-1)\ldots(n-p+1)}
\Tr\left[\varrho_{\varepsilon}b^{Wick}\right]\,,\quad \forall b\in \P_{p,p}(\Z)\,.
$$
Moreover after noticing that the factor
$\varepsilon^{p}n(n-1)\ldots (n-p+1)$ is nothing but
$\Tr\left[\varrho_{\varepsilon}(|z|^{2p})^{Wick}\right]$ when
$\Tr\left[\varrho_{\varepsilon}\right]=1$  and
$\varrho_{\varepsilon}\in \L^{1}(\bigvee^{n}\Z)$,
it becomes
\begin{equation}
  \label{eq.relwickbbgky}
\Tr\left[\gamma^{(p)}_{\varepsilon}\tilde{b}\right]
=
\frac{\Tr\left[\varrho_{\varepsilon}\right]}{\Tr\left[\varrho_{\varepsilon}(|z|^{2p})^{Wick}\right]}
\Tr\left[\varrho_{\varepsilon}b^{Wick}\right]\,,\quad \forall b\in \P_{p,p}(\Z)\,,
\end{equation}
with the convention that the right-hand side is $0$
when ${\Tr\left[\varrho_{\varepsilon}(|z|^{2p})^{Wick}\right]}=0$\,.
The extension to general $\varrho_{\varepsilon}\in
\mathcal{L}^{1}(\H)$ requires an assumption. Moreover it works for a
general separable Hilbert space $\Z$.
\begin{prop}
\label{pr.defgp} Assume that $\varrho_{\varepsilon}\in
\mathcal{L}^{1}(\H)$ satisfies $\varrho_{\varepsilon}\geq 0$ and
$\N^{k/2}\varrho_{\varepsilon}\N^{k/2}\in \mathcal{L}^{1}(\H)$ for all
$k\in\nz$\,.
Then for any $p\in \nz$, the relation \eqref{eq.relwickbbgky}
defines a unique element $\gamma^{(p)}_{\varepsilon}\geq 0$ of
$\mathcal{L}^{1}(\bigvee^{p}\Z)$\,.
\end{prop}
\proof
Suppose $\Tr\left[\varrho_{\varepsilon}(|z|^{2p})^{Wick}\right]>0$\,.
Writing
$$
\Tr\left[\varrho_{\varepsilon}b^{Wick}\right]=\Tr\left[(1+\N)^{p/2}\varrho_{\varepsilon}(1+\N)^{p/2}
(1+\N)^{-p/2}b^{Wick}(1+\N)^{-p/2} \right]
$$
with our assumptions and the estimates \eqref{eq.2bis}
ensures that $\tilde{b}\to
\Tr\left[\varrho_{\varepsilon}b^{Wick}\right]$
defines  a continuous linear form on $\L(\bigvee^{p}\Z)$\,.
The positivity comes from \eqref{eq.poswick} and the normality of the
associated state after normalization, which says
$\gamma^{(p)}_{\varepsilon}\in \mathcal{L}^{1}(\bigvee^{p}\Z)$, is a
consequence
of \eqref{eq.supwick}\,.
\fin

\bigskip
\noindent We end with this discussion with a natural definition.
\begin{definition}
  \label{de.red} When $\varrho_{\varepsilon}\in
\mathcal{L}^{1}(\Z)$ satisfies $\varrho_{\varepsilon}\geq 0$ and
$\N^{k/2}\varrho_{\varepsilon}\N^{k/2}\in \mathcal{L}^{1}(\H)$ for all
$k\in\nz$, the reduced density matrix
$\gamma^{(p)}_{\varepsilon}$, $p\in\nz$, associated with
$\varrho_{\varepsilon}$ is the element of $\L^{1}(\bigvee^{p}\Z)$
defined by
\begin{equation}
  \label{eq.relwickbbgkydef}
\Tr\left[\gamma^{(p)}_{\varepsilon}\tilde{b}\right]
=
\frac{\Tr\left[\varrho_{\varepsilon}\right]}{\Tr\left[\varrho_{\varepsilon}(|z|^{2p})^{Wick}\right]}
\Tr\left[\varrho_{\varepsilon}b^{Wick}\right]\,,\quad \forall b\in \P_{p,p}(\Z)\,,
\end{equation}
with
$\gamma^{(p)}_{\varepsilon}=0$ in the case
when $\Tr\left[\varrho_{\varepsilon}(|z|^{2p})^{Wick}\right]=0$\,.
\end{definition}

\subsection{Wigner measures}
\label{se.wigmes}
The Wigner measures are defined after the next result proved in
\cite[Theorem 6.2]{AmNi1}.
\begin{thm}
\label{th.wig-measure}
Let $\left(\varrho_{\varepsilon}\right)_{\varepsilon\in (0,\bar\varepsilon)}$ be a family
of normal states on $\mathcal{H}$ parametrized by $\varepsilon$.
Assume   ${\rm Tr}[\varrho_{\varepsilon}\N^\delta]$ $\leq C_{\delta}$ uniformly
w.r.t. $\varepsilon\in (0,\overline{\varepsilon})$
 for some fixed $\delta>0$ and $C_{\delta}\in (0,+\infty)$.
Then for every sequence $(\varepsilon_{n})_{n\in\nz}$ with $\lim_{n\to\infty}\varepsilon_n= 0$
there exists a subsequence $(\varepsilon_{n_k})_{k\in\nz}$ and a Borel probability measure $\mu$ on $\Z$ such that
\begin{eqnarray*}
\lim_{k\to\infty} \Tr[\varrho_{\varepsilon_{n_k}} b^{Weyl}]=
\int_{\Z} b(z) \; d\mu(z)\,,
\end{eqnarray*}
for all $b\in \ccup_{p\in\p}\mathcal{F}^{-1}\left(\mathcal{M}_{b}(p\Z)\right)$.\\
Moreover this probability measure $\mu$ satisfies $\ds \int_{\Z} |z|^{2\delta} \, d\mu(z) <\infty$.
\end{thm}
\begin{definition}
  \label{de.setwig}
The set of Wigner measures associated with a family
$(\varrho_{\varepsilon})_{\varepsilon\in (0,\bar\varepsilon)}$
(resp. a sequence $(\varrho_{\varepsilon_{n}})_{n\in\nz}$)  which
satisfies the assumptions of Theorem~\ref{th.wig-measure} is
denoted by
$$
\mathcal{M}(\varrho_{\varepsilon}, \varepsilon\in
(0,\bar\varepsilon))\,, \quad (\textrm{resp.}\ \mathcal{M}(\varrho_{\varepsilon_{n}}, n\in\nz))\,.
$$
\end{definition}
Wigner measures are in practice identified via their characteristic functions according to the relation
$$
\mathcal{M}(\varrho_{\varepsilon}, \varepsilon\in
(0,\bar\varepsilon))=\{\mu\}\Leftrightarrow \lim_{\hbarr\to 0}\Tr[\varrho_\varepsilon \,W(\sqrt{2}\pi \xi)]=\mathcal{F}(\mu)(\xi)\,.
$$
The expression $\mathcal{M}(\varrho_{\varepsilon}, \varepsilon\in
(0,\bar\varepsilon))=\left\{\mu\right\}$ simply means that the family
$(\varrho_{\varepsilon})_{\varepsilon\in (0,\bar\varepsilon)}$ is "pure" in the sense
$$
\lim_{\varepsilon\to
  0}\Tr\left[\varrho_{\varepsilon} b^{Weyl}\right]=\int_\Z b(z)~d\mu\,,
$$
for all cylindrical symbol $b$ without extracting a
subsequence. Actually the general case can be reduced to this after
reducing the range of parameter to $\varepsilon\in \left\{\varepsilon_{n_{k}},
  k\in\nz\right\}$\,.\\
A simple a priori estimate argument allows to extend the convergence
to symbols which have a polynomial growth and to test to Wick
quantized symbols with compact kernels belonging to $\P_{alg}^{\infty}(\Z)$  (see \cite[Corollary 6.14]{AmNi1}).
\begin{prop}
 \label{pr.polycomp}
Let $\left(\varrho_{\varepsilon}\right)_{\varepsilon\in (0,\bar\varepsilon)}$ be a family
of normal states on $\L(\mathcal{H})$ parametrized by $\varepsilon$
such that ${\rm Tr}[\varrho_{\varepsilon}\N^{\alpha} ]\leq C_{\alpha}$ holds
uniformly with respect to $\varepsilon\in (0,\bar \varepsilon)$ for
all $\alpha\in\nz$ and such that $\mathcal{M}(\varrho_{\varepsilon},
\varepsilon\in (0,\bar \varepsilon))=\left\{\mu\right\}$.
Then the convergence
\begin{equation}
  \label{eq.convpolWW}
\lim_{\varepsilon\to 0}
\Tr[\varrho_{\varepsilon} b^{\textrm{quantized}}]=
\int_{\Z} b(z) \; d\mu(z)\,,
\end{equation}
holds for the Weyl quantization of any $b\in S_{p\Z}^{\nu}$ with
$p\in\p$ and
$\nu\in[0,1]$, and for the Wick quantization of any $b\in \P_{alg}^{\infty}(\Z)$.
\end{prop}
Wigner measures are completely identified by testing with
Weyl-quantized observable and possibly by restricting to some
countable subset  $\ccup_{n\in\nz}D_{n,p_{n}}$ where $D_{n,p_{n}}$
is a countable dense subset of
$\mathcal{F}^{-1}(\mathcal{M}_{b}(p_{n}\Z))$, and $(p_{n})_{n\in\nz}$
is a sequence of $\p$ such that $\sup_{n\in\nz}p_{n}=I_{\Z}$\,
(see \cite{AmNi1}). One may question whether testing on all the
$b^{Wick}$
with $b\in \P^{\infty}_{alg}(\Z)$ also identifies the Wigner measures. When $\Z$ is finite dimensional, this
amounts to the well-known Hamb\"{u}rger moment problem of identifying
a probability measure $\nu$ on $\rz$ from its moments $a_{n}=
\int_{\rz}x^{n}d\nu(x)$, $n\in\nz$, for which uniqueness fails without
growth conditions on the sequence $(a_{n})_{n\in\nz}$
(\cite{RS} \cite{Akh}), which can be translated in our case to growth
conditions of $(\sup_{\varepsilon\in (0,\bar\varepsilon)}
\Tr\left[\varrho_{\varepsilon}\N^{\alpha}\right])_{\alpha\in\nz}$.
 We shall circumvent this difficulty, by identifying the Wigner
 measures in two steps by approximating the states
 $(\varrho_{\varepsilon})_{\varepsilon\in (0,\bar\varepsilon)}$ by
 states $(\varrho_{\varepsilon}^{app})_{\varepsilon\in
   (0,\bar\varepsilon)}$ for which the growth condition is satisfied.
 We shall reconsider the moment problem later, but the comparison
 argument is given below.
 \begin{prop}
\label{pr.compwig}
   Let $(\varrho_{\varepsilon}^{j})_{\varepsilon\in
     (0,\bar\varepsilon)}$, $j=1,2$, be two families (or sequences) of normal states
   on $\H$ such that
   $\Tr\left[\varrho_{\varepsilon}^{j}\,\N^{\delta}\right]\leq
   C_{\delta}$ uniformly w.r.t. $\varepsilon\in (0,\bar \varepsilon)$
   for some $\delta>0$ and $C_{\delta}\in (0,+\infty)$. Assume further
   $\mathcal{M}(\varrho_{\varepsilon}^{j}, \varepsilon\in (0,\bar
   \varepsilon))=\left\{\mu_{j}\right\}$ for $j=1,2$. Then
$$
\int|\mu_{1}-\mu_{2}|\leq \liminf_{\varepsilon\to 0}
|\varrho_{\varepsilon}^{1}-\varrho_{\varepsilon}^{2}|_{\mathcal{L}^{1}(\H)}\,.
$$
 \end{prop}
\proof
For a symbol $b\in \mathcal{S}_{cyl}(\Z)$, the finite dimensional Weyl semiclassical
calculus says
$|b^{Weyl}|_{\L(\H)}\leq
\|b\|_{\infty}+\mathcal{O}_{b}(\varepsilon)$ with
$\|b\|_{\infty}=\|b\|_{L^{\infty}(p\Z)}$\,.
This implies for a given $b\in \mathcal{S}_{cyl}(\Z)$,
$$
|\int_{\Z}b(z) d(\mu_{1}-\mu_{2})(z)|=\lim_{\varepsilon\to
  0}\left|\Tr[(\varrho_{\varepsilon}^{1}-\varrho_{\varepsilon}^{2})b^{Weyl}]\right|\leq
\|b\|_{\infty}
\liminf_{\varepsilon\to 0}|\varrho_{\varepsilon}^{1}-\varrho_{\varepsilon}^{2}|_{\L^{1}(\H)}\,.
$$
The measure $\mu_{1}-\mu_{2}$ is absolutely continuous with respect to
the Borel probability measure $\frac{\mu_{1}+\mu_{2}}{2}$. Hence there
exists a Borel function $\lambda$ on $\Z$ such that $\mu_{1}-\mu_{2}=
\lambda(z)\frac{\mu_{1}+\mu_{2}}{2}$ with the additional property
$|\lambda(z)|\leq 2$ $\frac{\mu_{1}+\mu_{2}}{2}$-almost everywhere.
But for any Borel probability measure $\nu$ on $\Z$, it was checked in
\cite{AmNi1}
that $\mathcal{S}_{cyl}(\Z)$ is dense in $L^{p}(\Z,\nu)$ for
$p\in[1,\infty)$ on the basis of a general measurable version of
Stone-Weierstrass theorem (see for instance \cite{Cou}). Hence there exists
a sequence $(\beta_{n})_{n\in\nz}$ in $\mathcal{S}_{cyl}(\Z)$ such that
$$
\lim_{n\to\infty}\|\beta_{n}-\frac{|\lambda|}{\lambda}1_{\{\lambda\neq 0\}}\|_{L^{1}(\Z,\frac{\mu_{1}+\mu_{2}}{2})}=0
$$
and after extraction
$\lim_{k\to\infty}\beta_{n_{k}}(z)=\frac{|\lambda|}{\lambda}(z)1_{\{\lambda\neq 0\}}(z)$,
$\frac{\mu_{1}+\mu_{2}}{2}$-almost everywhere.
By setting $b_{k}=2\frac{\beta_{n_{k}}}{1+|\beta_{n_{k}}|^{2}}$, we
get a sequence $(b_{k})_{k\in\nz}$ such that
\begin{eqnarray*}
  &&\forall k\in \nz,\quad b_{k}\in
  \mathcal{S}_{cyl}\quad\text{and}\quad \|b_{k}\|_{\infty}\leq 1\,,\\
&&
\lim_{k\to\infty}b_{k}(z)=\frac{|\lambda|(z)}{\lambda(z)}1_{\left\{\lambda\neq
  0\right\}}(z)\quad \frac{\mu_{1}+\mu_{2}}{2}~\text{a.e.}
\end{eqnarray*}
We conclude with
$$
\int |\mu_{1}-\mu_{2}|=\int_{\Z}|\lambda(z)|~d\frac{\mu_{1}+\mu_{2}}{2}(z)=
\left|\lim_{k\to\infty}\int_{\Z}b_{k}(z)~d(\mu^{1}-\mu^{2})(z)\right|
\leq 1\times
\liminf_{\varepsilon\to 0}|\varrho_{\varepsilon}^{1}-\varrho_{\varepsilon}^{2}|_{\L^{1}(\H)}\,.
$$
\fin

\noindent
When the two sets $\mathcal{M}(\varrho_{\varepsilon}^{j},
\varepsilon\in (0,\bar\varepsilon))$ have more than one element, the
extraction of subsequences, $(\varepsilon_{n})_{n\in\nz}$, can be made simultaneously and the result
has to be modified into:
\begin{equation}
  \label{eq.limsup}
  \inf_{(\mu_{1},\mu_{2})\in \mathcal{M}(\varrho^{1}_{\varepsilon},
  \varepsilon\in (0,\bar \varepsilon))\times \mathcal{M}(\varrho^{2}_{\varepsilon},
  \varepsilon\in (0,\bar \varepsilon))}
 \int|\mu_{1}-\mu_{2}|\leq \limsup_{\varepsilon\to 0}|\varrho_{\varepsilon}^{1}-\varrho^{2}_{\varepsilon}|_{\L^{1}(\H)}\,.
\end{equation}

\subsection{Wigner measures and the BBGKY hierarchy}
\label{se.WigBBGKY}
The compactness condition
$b\in\P_{alg}^{\infty}(\Z)$  for the
Wick quantization in Proposition~\ref{pr.polycomp} is not a technical restriction and the
convergence is no more true for  a general $b\in \P_{alg}(\Z)$.
It was identified in \cite{AmNi1} as a
 ``dimensional defect of compactness''  and illustrated with examples,
one of them being related with the
Bose-Einstein condensation of the free Bose gas.\\
This terminology comes from the idea that this defect of compactness
does not come from the infinity in the phase space like in the finite
dimensional case (see \cite{Tar}\cite{Ger}) but from the non
compactness in the norm topology of balls in infinite dimension.
Actually this was made more accurate in \cite{AmNi2}: under the assumptions
$\mathcal{M}(\varrho_{\varepsilon}, \varepsilon\in
(0,\bar\varepsilon))=\left\{\mu\right\}$ and
$\Tr\left[\varrho_{\varepsilon}\N^{k}\right]\leq \lambda^{k}$,  we
proved  $(T)\Rightarrow (P)$ with
\begin{eqnarray*}
(P) && \forall b\in \P_{alg}(\Z),\quad
\lim_{\varepsilon\to
  0}\Tr\left[\varrho_{\varepsilon}b^{Wick}\right]=\int_\Z b(z)~d\mu(z) \,;
\\
(T) && \forall \eta>0,\exists P_{\eta}\in \p,\quad
\Tr\left[(1-\Gamma(P_{\eta}))\varrho_{\varepsilon}\right] < \eta\,,
\end{eqnarray*}
where $(T)$ appears as a quantum Prokhorov condition (or \underline{t}ightness
condition in the strong topology).\\
The condition $(P)$  which will be simplified in the next subsection,
 actually contains, for all $\alpha\in \nz$, the uniform bound w.r.t. $\varepsilon$
 of
$\Tr\left[\varrho_{\varepsilon}\N^{\alpha}\right]$ since
$\N^{\alpha}=[(|z|^{2})^{Wick}]^{\alpha}$.
It implies actually a strong relationship between the
Wigner measure formulation and the convergence of reduced density
matrices.
\begin{prop}
\label{pr.wignergp0}
Assume that $(\varrho_{\varepsilon})_{\varepsilon\in
  (0,\bar\varepsilon)}$ is a family of $\L^{1}(\H)$ with
$\varrho_{\varepsilon}\geq 0$, $\Tr[\varrho_{\varepsilon}]=1$, $\mathcal{M}(\varrho_{\varepsilon},
\varepsilon\in (0,\bar\varepsilon))=\left\{\mu\right\}$ with the
 condition $(P)$ and assume $\mu\neq \delta_{0}$. Define for $p\in \nz$
 \begin{equation}
   \label{eq.defga0}
   \gamma^{(p)}_{0}:=
\frac{1}{\int_{\Z}|z|^{2p}~d\mu(z)}\int_{\Z}|z^{\otimes
  p}\rangle\langle z^{\otimes p}|~d\mu(z)
 \end{equation}
Then for all $p\in\nz$, the reduced density matrix
 $\gamma^{(p)}_{\varepsilon}$ converges to $\gamma^{(p)}_{0}$
in the $\mathcal{L}^{1}$-norm.
\end{prop}
\proof
For $p=0$, the result is nothing but $1=\int\mu=\lim_{\varepsilon\to
  0}\Tr[\varrho_{\varepsilon}]=1$.\\
For $p\in\nz^{*}$, the condition $(P)$ with $\mu\neq \delta_{0}$ says first
$$
\lim_{\varepsilon\to
  0}\Tr\left[\varrho_{\varepsilon}(|z|^{2p})^{Wick}\right]=\int_{\Z}
|z|^{2p}~d\mu(z)> 0\,.
$$
Hence, the reduced density matrix
$\gamma^{(p)}_{\varepsilon}$ is well defined according to
Definition~\ref{de.red}
for  $\varepsilon< \bar\varepsilon_{p}$ small enough (with
$\Tr[\varrho_{\varepsilon}]=1$).
The condition $(P)$ gives the general convergence:
$$
\lim_{\varepsilon\to 0}
\Tr\left[\gamma^{(p)}_{\varepsilon}\tilde{b}\right]=
\lim_{\varepsilon\to
  0}\frac{\Tr\left[\varrho_{\varepsilon}b^{Wick}\right]}{\Tr\left[\varrho_{\varepsilon}(|z|^{2p})^{Wick}\right]}
=
\frac{\int_{\Z}b(z)~d\mu(z)}{\int_{\Z}|z|^{2p}~d\mu(z)}
=\Tr\left[\gamma_{0}^{(p)}\tilde{b}\right]
\,,
$$
for all $b\in \P_{p,p}(\Z)$, where the last equality is a
$\mu$-integration of the equality of continuous functions
$$
b(z)=\langle z^{\otimes p}\,,\,
\tilde{b}z^{\otimes p}\rangle=\Tr\left[|z^{\otimes p}\rangle\langle
  z^{\otimes p}|\tilde{b}\right]\,.
$$
This proves the weak
convergence of $\gamma^{(p)}_{\varepsilon}$ to $\gamma_{0}^{(p)}$ in
$\mathcal{L}^{1}(\bigvee^{p}\Z)$.
But since $\gamma^{(p)_{\varepsilon}}$ and $\gamma^{(p)}_{0}$ are non
negative with
$\Tr[\gamma^{(p)}_{\varepsilon}]=1=\Tr\left[\gamma^{(p)}_{0}\right]$,
this implies the norm convergence according to
\cite{Sim}\cite{Ak}\cite{DA}).\footnote{
In a more general framework, it is  said that
$\L^{1}(\bigvee^{p}\Z)$ has a uniform Kadec-Klee property (see
\cite{Len} and references therein).}
\fin
\subsection{A simple criterion for the reliability of  Wick
  observables}
\label{se.relywick}
The proof of Proposition~\ref{pr.wignergp0} can be adapted in order to make an equivalent
condition to $(P)$ with a weaker and easier to handle formulation:
\begin{equation*}
  (PI)\quad \forall \alpha\in\nz,\quad
  \lim_{\varepsilon\to 0}\Tr\left[\varrho_{\varepsilon}{\bf N}^{\alpha}\right]
=\int_\Z |z|^{2\alpha}~d\mu(z) < +\infty.
\end{equation*}
\begin{prop}
  \label{pr.eqPPI}
For a family $(\varrho_{\varepsilon})_{\varepsilon\in
  (0,\bar\varepsilon)}$ in  $\L^{1}(\H)$ such that
$\varrho_{\varepsilon}\geq 0$, $\Tr[\varrho_{\varepsilon}]=1$, $\mathcal{M}(\varrho_{\varepsilon},
\varepsilon\in (0,\bar\varepsilon))=\left\{\mu\right\}$, the condition
$(P)$ and $(PI)$ are equivalent:
$$
\left(\forall \alpha\in\nz,~
  \lim_{\varepsilon\to 0}\Tr\left[\varrho_{\varepsilon}{\bf N}^{\alpha}\right]
=\int_\Z |z|^{2\alpha}~d\mu(z)\right)
\Leftrightarrow
\left(
\forall b\in \P_{alg}(\Z),~
\lim_{\varepsilon\to
  0}\Tr\left[\varrho_{\varepsilon}b^{Wick}\right]=\int_\Z b~d\mu\right)
$$
\end{prop}
\proof The condition $(PI)$ is a particular case of $(P)$. Let us prove
$(PI)\Rightarrow (P)$\,.\\
We start with two remarks:
\begin{itemize}
\item
For $k\in \nz^{*}$,
$(|z|^{2k})^{Wick}={\bf N}({\bf N}-\varepsilon)\ldots({\bf
  N}-(k-1)\varepsilon)$.
Hence the condition $(PI)$
is equivalent to
$$
\forall \alpha\in\nz,\quad
\lim_{\varepsilon\to 0}\Tr\left[\varrho_{\varepsilon}(|z|^{2\alpha})^{Wick}\right]=\int_\Z |z|^{2\alpha}~d\mu(z)\,.
$$
\item For $p=0$ (resp $q=0$) the operators in $\L(\cz,\bigvee^{q}\Z)$ (resp.
  in $\L(\bigvee^{p}\Z,\cz)$) are compact and
  $\P_{0,q}(\Z)=\P_{0,q}^{\infty}(\Z)$
  (resp. $\P_{p,0}(\Z)=\P_{p,0}^{\infty}(\Z)$). Hence the convergence
$\lim_{\varepsilon\to 0}\Tr[\varrho_{\varepsilon}b^{Wick}]=\int
b~d\mu$, is consequence of Proposition~\ref{pr.polycomp} when $p=0$ or $q=0$.
\end{itemize}
According to Proposition~\ref{pr.wignergp0}, there are two cases.\\
\noindent\textbf{If $\mu=\delta_{0}$:} Then for $b\in \P_{p,p}(\Z)$,
$p\in\nz^{*}$,
such that $\tilde{b}\geq 0$, the inequality $0\leq \tilde{b}\leq
|b|_{\P_{p,p}}I_{\bigvee^{p}\Z}$ and the positivity \eqref{eq.poswick} says
$$
0\leq \lim_{\varepsilon\to
  0}\Tr\left[\varrho_{\varepsilon}b^{Wick}\right]\leq
\lim_{\varepsilon\to
  0}
|b|_{p,p}\Tr\left[\varrho_{\varepsilon}(|z|^{2p})^{Wick}\right]=\int_{\Z}|z|^{2p}\delta_{0}(z)=0\,.
$$
For a general $b\in \P_{p,p}(\Z)$, $p\in\nz^{*}$, the decomposition
$\tilde{b}=\tilde{b}_{R,+}-\tilde{b}_{R,-}+i\tilde{b}_{I,+}-i\tilde{b}_{I,-}$
with all the $\tilde{b}_{\bullet}\geq 0$ now gives
$$
\forall p\in \nz^{*}, \forall b\in \P_{p,p}(\Z),\quad
\lim_{\varepsilon\to
  0}\Tr\left[\varrho_{\varepsilon}b^{Wick}\right]=0\,.
$$
For $p\neq q$, $p, q\in \nz^{*}$,  write
$$
\left|\Tr\left[\varrho_{\varepsilon}b^{Wick}\right]\right|
=\left|\Tr\left[\varrho_{\varepsilon}^{1/2}(\varrho_{\varepsilon}^{1/2}b^{Wick})\right]\right|
\leq \Tr[\varrho_{\varepsilon}]^{1/2}\Tr\left[\varrho_{\varepsilon}b^{Wick}b^{Wick,*}\right]^{1/2}\,.
$$
Proposition~\ref{symbcalc} says that
$b^{Wick}b^{Wick,*}=\sum_{\ell=0}^{p}\frac{\varepsilon^{\ell}}{\ell
  !}\partial_{z}^{\ell}b.\partial_{\bar z}^{\ell}\bar{b}$ belongs to
$\oplus_{k=0}^{p+q}\P_{k,k}(\Z)$ with an $\mathcal{O}(\varepsilon)$
term in $\P_{0,0}(\Z)$.
We have proved
$$
\forall p,q\in \nz^{*},\, \forall b\in
\P_{p,q}(\Z),\quad
\lim_{\varepsilon\to
  0}\Tr\left[\varrho_{\varepsilon}b^{Wick}\right]=0=\int_{\Z} b(z)\delta_{0}(z)\,,
$$
while the cases $(0,q)$ and $(p,0)$ are already known.

\bigskip
\noindent\textbf{If $\mu\neq \delta_{0}$:} Then we know by
 Proposition~\ref{pr.wignergp0} that $\lim_{\varepsilon\to
   0}\|\gamma_{\varepsilon}^{(p)}-\gamma_{0}^{(p)}\|_{\L^{1}}=0$,
 which implies
$$
\forall b\in \P_{p,p}(\Z)\,,\quad
\lim_{\varepsilon\to 0}\Tr\left[\varrho_{\varepsilon}b^{Wick}\right]=
\lim_{\varepsilon\to
  0}\Tr\left[\gamma^{(p)}_{\varepsilon}\tilde{b}\right]
= \Tr\left[\gamma_{0}^{(p)}\tilde{b}\right]=\int_{\Z}b(z)~d\mu(z)\,.
$$
Let us consider the general case $b\in \P_{p,q}(\Z)$. The above
convergence is still true when
the kernel $\tilde{b}$ is compact by Proposition~\ref{pr.polycomp}.
Consider now a general $b\in
\P_{p,q}(\Z)$. Since $\int_{\Z}|z^{\otimes p}\rangle\langle z^{\otimes
q}|~d\mu(z)$ is nuclear (or trace-class in $\bigvee^{q}\Z\oplus \bigvee^{p}\Z$), for any
$n\in\nz$ there exists a compact operator $\tilde{b}_{n}\in
\L^{\infty}(\bigvee^{p}\Z,\bigvee^{q}\Z)$ such that
$|b_{n}|_{\P_{p,q}}=|\tilde{b}_{n}|_{\L(\bigvee^{p}\Z,\bigvee^{q}\Z)}=|\tilde{b}|_{\L(\bigvee^{p}\Z,\bigvee^{q}\Z)}=|b|_{\P_{p,q}}$ and
$$
\left|\int_{\Z}(b(z)-b_{n}(z))~ d\mu(z)\right|
=
\left|
\Tr\left[\int_{\Z}|z^{\otimes p}\rangle\langle z^{\otimes
q}|~d\mu(z) [\tilde{b}-\tilde{b}_{n}]\right]
\right|\leq \frac{1}{n+1}\,.
$$
The Lebesgue convergence theorem with
\begin{eqnarray*}
  &&\forall n\in\nz,\quad |b(z)-b_{n}(z)|^{r}\leq
  (2|b|_{\P_{p,q}})^{r}|z|^{r(p+q)}\,,\quad \int_{\Z}|z|^{r(p+q)}~d\mu(z)<\infty\,,\\
&&
\forall z\in\Z\,,\quad
\lim_{n\to\infty}b_{n}(z)=\lim_{n\to\infty}\langle z^{\otimes q}\,,
\tilde{b}_{n}z^{\otimes p}\rangle=b(z)\,,
\end{eqnarray*}
yields
$$
\lim_{n\to\infty}\int_{\Z}|b(z)-b_{n}(z)|^{r}~d\mu(z)=0\,.
$$
Set $\eta_{r}(n)= \int_{\Z}|b(z)-b_{n}(z)|^{r}~d\mu(z)$ and use again the
Cauchy-Schwarz inequality
$$
\left|\Tr\left[\varrho_{\varepsilon}(b^{Wick}-b_{n}^{Wick})\right]\right|
\leq
\Tr\left[\varrho_{\varepsilon}(b^{Wick}-b_{n}^{Wick})(b^{Wick,*}-b_{n}^{Wick,*})\right]^{1/2}\,.
$$
Owing to the result valid when $p=q$ we deduce
$$
\limsup_{\varepsilon\to  0}
\left|\Tr\left[\varrho_{\varepsilon}(b^{Wick}-b_{n}^{Wick})\right]\right|
\leq
\left[\int_{\Z}
\left|b(z)-b_{n}(z)\right|^{2}
~d\mu(z)\right]^{1/2}= \eta_{2}(n)^{1/2}\,.
$$
Since for $n\in\nz$ fixed,
$\lim_{\varepsilon\to
  0}\Tr\left[\varrho_{\varepsilon}b_{n}^{Wick}\right]=\int_{\Z}b_{n}(z)~d\mu(z)$,
we deduce
$$
\forall n\in\nz,\quad
\limsup_{\varepsilon\to
  0}\left|\Tr\left[\varrho_{\varepsilon}b^{Wick}\right]-\int_{\Z}b(z)~d\mu(z)\right|
\leq \frac{1}{n+1}+\eta_{2}(n)^{1/2}\,,
$$
while the right-hand side goes to $0$ as $n\to\infty$\,.
\fin
\subsection{States localized in a ball}
\label{se.locball}
The condition, $\Tr\left[\varrho_{\varepsilon}\N^{\alpha}\right]\leq
\lambda^{\alpha}$ for all $\alpha\in\nz$, used in \cite{AmNi2} is
actually equivalent to $$\varrho_{\varepsilon}=1_{[0,\lambda]}({\bf
  N})\varrho_{\varepsilon}1_{[0,\lambda]}({\bf N})$$ (locate the
spectral measure of $\varrho_{\varepsilon}$ for the self-adjoint
operator ${\bf N}$). Such
an assumption remains an important step in the present analysis, and
${\N}=(|z|^{2})^{Wick}$ suggests that such a state is localized in
ball of the phase-space.
\begin{definition}
\label{de.locb}
  A family $(\varrho_{\varepsilon})_{\varepsilon\in
    (0,\bar\varepsilon)}$ (or a sequence
  $(\varrho_{\varepsilon_{n}})_{n\in\nz}$) of normal states on $\H$,
 is said to be localized in the  ball of radius $R>0$, if
$\varrho_{\varepsilon}=1_{[0,R^{2}]}(\N)\varrho_{\varepsilon} 1_{[0,R^{2}]}(\N)$ for all
$\varepsilon\in (0,\bar\varepsilon)$\,.
\end{definition}
The meaning of the geometric intuition contained in
the terminology
``localized in a ball of radius $R$'', can be made more accurate.
\begin{lem}
  \label{le.locball}
For a family$(\varrho_{\varepsilon})_{\varepsilon\in
  (0,\bar\varepsilon)}$  (or a sequence $(\varrho_{\varepsilon_{n}})_{n\in\nz}$) of normal
states on $\H$ localized in a ball of radius $R>0$, all its
Wigner measures are supported in the ball $\left\{|z|\leq R\right\}$\,.
\end{lem}
\proof
A family $(\varrho_{\varepsilon})_{\varepsilon\in
  (0,\bar\varepsilon)}$ localized in a ball of radius $R$ satisfies
$\Tr\left[\varrho_{\varepsilon}\N^{\delta}\right]\leq R^{2\delta}$ for
 all $\delta>0$. Therefore the set of Wigner measures
$\mathcal{M}(\varrho_{\varepsilon}, \varepsilon\in
(0,\bar\varepsilon))$ is well defined and the convergence after
extraction can be
tested with Weyl-quantized cylindrical functions in the symbol class
$S_{p}^{\nu}$ introduced in \eqref{eq.classsymbp} for any $p\in \mathbb{P}$.
 Let $\mu\in
\mathcal{M}(\varrho_{\varepsilon}, \varepsilon\in
(0,\bar\varepsilon))$ be associated with the sequence
$(\varepsilon_{n})_{n\in\nz}$.
 For any finite rank projection $p\in\mathbb{P}$,
the Wick quantized operator $(|pz|^2)^{Wick}$ is
${\bf N}_{p}\otimes I_{\Gamma_s((1-p)\Z)}$ where ${\bf N}_{p}$ is the
 number  operator on  $\Gamma_s(p\Z)$ and equals
 $(|z|_{p\Z}^{2}-C_{p}\varepsilon)^{Weyl}$ in the finite dimensional
 framework of $p\Z$.
For any cut-off function $\chi\in\mathcal{C}^{\infty}_{0}(\rz)$
such that $\chi\equiv 1$ on $\left[0,R^2\right]$, the finite dimensional
Weyl semiclassical calculus tells us
$(1-\chi)({\bf N}_{p})=(1-\chi)(|z|_{p\Z}^{2})^{Weyl}+
\mathcal{O}_{p}(\varepsilon)$ in $\L(\Gamma_s(p\Z))$.
Further the commutative decomposition ${\bf N}= {\bf N}_{p}\otimes
I_{\Gamma_s((1-p)\Z)}+ I_{\Gamma_s(p\Z)}\otimes{\bf N}_{(1-p)}\geq {\bf N}_{p}\otimes
I_{\Gamma_s((1-p)\Z)}$ and choosing $\chi$ decreasing on $[0,+\infty)$
implies
$$
(1-\chi)(|pz|^{2})^{Weyl}+ \mathcal{O}_{p}(\varepsilon)
\leq (1-\chi)({\bf N}_{p}\otimes
I_{\Gamma((1-p)\Z)})
\leq (1-\chi)({\bf N}) \,.
$$
We deduce
$$
0\leq \int_{\Z}(1-\chi(|pz)|^{2})~d\mu(z)
=\lim_{n\to\infty}\Tr\left[\varrho_{\varepsilon}(1-\chi(|pz|^{2}))^{Weyl}\right]
\leq \lim_{n\to \infty}~\Tr\left[\varrho_{\varepsilon_{n}}1_{[0,R^2]}({\bf N})(1-\chi({\bf N}))\right]=0\,.
$$
Hence the measure $\mu$ vanishes outside a cylinder $\left\{|pz|\geq
  R\right\}$. This yields the result. \fin

With such localized states we can solve the moment problem.
\begin{prop}
\label{pr.moment}
Let $(\varrho_{\varepsilon})_{\varepsilon\in (0,\bar \varepsilon)}$
be a family (or a sequence $(\varrho_{\varepsilon_{n}})_{n\in\nz}$)
of normal states on $\H$, localized in the ball of radius $R>0$\,.
If there exists a Borel measure $\mu$ on $\Z$ such that
$$
\forall b\in \P^{\infty}_{alg}(\Z)\,,\quad \lim_{\varepsilon\to
  0}\Tr\left[\varrho_{\varepsilon} b^{Wick}\right]=\int_{\Z}b(z)~d\mu(z)\,,
$$
then
$$
\mathcal{M}(\varrho_{\varepsilon}, \varepsilon\in (0,\bar\varepsilon))=\left\{\mu\right\}\,.
$$
\end{prop}
\proof
Although this is shown in \cite[Proposition 6.15]{AmNi1}, we provide here a different proof.\\
Let $p\in\p$ and consider the direct image by $p$ of the measure
$\mu$:
$$
\forall E\in \mathcal{B}(p\Z)\,,\quad \mu_{p}(E)=\int_{\Z}1_{p^{-1}(E)}(z)~d\mu(z)\,,
$$
where $\mathcal{B}(p\Z)$ denotes the Borel $\sigma$-set on $p\Z$\,.\\
For any $b\in\P^{\infty}_{alg}(\Z)$, such that $b(pz)=b(z)$ we have
$$
\lim_{\varepsilon\to
  0}\Tr\left[\varrho_{\varepsilon} b^{Wick}\right]=\int_{p\Z}b(z)~d\mu_{p}(z)\,.
$$
This holds in particular when $b(z)=|pz|^{2k}$ with $b^{Wick}={\bf
  N}_{p}^{k} + \mathcal{O}(\varepsilon)\leq
{\bf N}^{k}+\mathcal{O}(\varepsilon)$ with
$$
\int_{p\Z}|z|^{2k}~d\mu_{p}(z)\leq \lim_{\varepsilon\to
  0}\Tr\left[\varrho_{\varepsilon}{\bf N}_{p}^{k}\right]
\leq
\lim_{\varepsilon\to
  0}\Tr\left[\varrho_{\varepsilon}{\bf N}^{k}\right]\leq R^{2k}\,.
$$
Hence all the moments $\int_{p\Z}|z|^{2k}~d\mu_{p}(z)$ are bounded by
$R^{2k}$ and the finite dimensional moment problem applies (see
\cite{RS}\cite{Akh}):
$\mu_{p}$ is completely determined by the set of values
$\left\{\int_{p\Z}b~d\mu_{p}\,,\, b~\text{polynomial}\right\}$\,.
Let $\mu'$ be a Wigner measure of the family
$(\varrho_{\varepsilon})_{\varepsilon\in (0, \bar \varepsilon)}$. It
is supported in the ball $\left\{z\in\Z,\, |z|\leq R\right\}$ so that
its direct image by $p$, $\mu'_{p}$ is supported in the ball
$\left\{z\in p\Z,\, |z|\leq R\right\}$. Moreover there exists a
sequence $(\varepsilon_{n})_{n\in\nz}$, such that
$$
\forall b\in S_{p\Z}^{\nu}\,,\quad \lim_{n\to \infty}
\Tr\left[\varrho_{\varepsilon_{n}}b^{Weyl}\right]= \int_{p\Z}b(z)~d\mu_{p}'(z)\,,
$$
where the $b^{Weyl}$ can be replaced by $b^{Wick}$ for any polynomial $b$
such that $b(z)= b(pz)$ according to the finite dimensional comparison
of the Weyl and Wick calculus in \eqref{eq.compWeWi}.
We deduce $\mu_{p}=\mu'_{p}$. Since this holds for all the
$p\in\p$, this ends the proof.
\fin

\bigskip
Let $\chi$ be a continuous cut-off function supported in $[0,1]$,  with $0\leq \chi\leq
1$  and such that
$\chi\equiv 1$ in $[0,\frac{1}{2}]$.
Within the assumptions of Theorem~\ref{th.wig-measure} and especially
$\Tr[\varrho_{\varepsilon}{\bf N}^{\delta}]\leq C_{\delta}$, the difference between
the state $\varrho_{\varepsilon}$ and the localized state
$\varrho^{\chi,R}_{\varepsilon}=\frac{1}{\Tr\left[\varrho_{\varepsilon}
    \chi^{2}(\frac{{\bf N}}{R^{2}})\right]}\chi(\frac{{\bf
    N}}{R^{2}})\varrho_{\varepsilon}\chi(\frac{{\bf N}}{R^{2}})$ can be made
arbitrarily small according to
\begin{equation}
  \label{eq.cutoffst}
\forall \varepsilon\in (0,\bar\varepsilon)\,,
\quad|\varrho_{\varepsilon}-\varrho_{\varepsilon}^{\chi,R}|_{\L^{1}(\H)}\leq
\frac{C_{\delta}}{(R/2)^{2\delta}-C_{\delta}}\,,
\end{equation}
where the right-hand side vanishes as $R\to\infty$.
Then the comparison result in Proposition~\ref{pr.compwig} or its
variant \eqref{eq.limsup}
says that the Wigner measures $(\varrho_{\varepsilon})_{\varepsilon\in
(0,\bar \varepsilon)}$ can be identified by its
approximation by states localized in balls:
\begin{equation}
  \label{eq.comXR}
  \inf_{(\mu,\mu')\in \mathcal{M}(\varrho_{\varepsilon},
  \varepsilon\in (0,\bar \varepsilon))\times \mathcal{M}(\varrho^{\chi,R}_{\varepsilon},
  \varepsilon\in (0,\bar \varepsilon))}
 \int|\mu-\mu'|\leq \frac{C_{\delta}}{(R/2)^{2\delta}-C_{\delta}}\,.
\end{equation}
Then the question arises whether the family
$(\varrho_{\varepsilon}^{\chi,R})_{\varepsilon\in
  (0,\bar\varepsilon)}$, or an extracted subsequence, satisfies the condition
$(PI)$ (or equivalently $(P)$) if the family $(\varrho_{\varepsilon})_{\varepsilon\in
  (0,\bar\varepsilon)}$ does.
\begin{prop}
\label{pr.locPI}
Assume that the family $(\varrho_{\varepsilon})_{\varepsilon\in
  (0,\bar \varepsilon)}$ of normal states on $\H$ satisfies
$\mathcal{M}(\varrho_{\varepsilon}, \varepsilon\in (0,
\bar\varepsilon))=\left\{\mu\right\}$
and the condition $(PI)$\,. Let the function $f\in \mathcal{C}^{0}([0,+\infty),\rz)$
 be polynomially bounded such that the quantity $\Tr[\varrho_{\varepsilon}f^{2}({\bf
   N})]$ is uniformly bounded from below for $\varepsilon\in (0,\bar \varepsilon)$\,.
Then the family
$(\varrho_{\varepsilon}^{f})_{\varepsilon\in (0, \bar
   \varepsilon)}$ given by
 $\varrho_{\varepsilon}^{f}=
\frac{1}{\Tr\left[\varrho_{\varepsilon}
     f^{2}({\bf N})\right]}f({\bf N})
\varrho_{\varepsilon}f({\bf N})$
  has a unique Wigner measure
$\mathcal{M}(\varrho_{\varepsilon}^{f}, \varepsilon\in
(0,\bar\varepsilon))=
\left\{\frac{f^{2}(|z|^{2})\mu}{\int f^{2}(|z|^{2})d\mu}\right\}$ and satisfies the condition
$(PI)$\,.
\end{prop}
We will need the next lemma
\begin{lem}
  \label{le.exptN}
Let the family $(\varrho_{\varepsilon})_{\varepsilon\in (0,\bar \varepsilon)}$
(or a sequence $(\varrho_{\varepsilon_{n}})_{n\in\nz}$) of normal
states be localized in the ball of radius $R$ and assume the condition
$(PI)$ with $\mathcal{M}\left(\varrho_{\varepsilon}, \varepsilon\in
  (0,\bar\varepsilon)\right)=\left\{\mu\right\}$. Then the equality
\begin{equation}
  \label{eq.exptN}
\lim_{\varepsilon\to
  0}\Tr\left[e^{\alpha_{1}\N}\varrho_{\varepsilon}e^{\alpha_{2}\N}b^{Wick}\right]
=\int_{\Z}e^{(\alpha_{1}+\alpha_{2})|z|^{2}}b(z)~d\mu(z)
\end{equation}
holds for all $\alpha_{1},\alpha_{2}\in\cz$ and all $b\in \P_{alg}(\Z)$\,.
\end{lem}
\proof
The right-hand side of \eqref{eq.exptN} is the sum of the
double series
$$
\sum_{k_{1},k_{2}\in\nz}\frac{(\alpha_{1})^{k_{1}}(\alpha_{2})^{k_{2}}}{k_{1}!
k_{2}!}\int_{\Z}|z|^{2k_{1}+2k_{2}}b(z)~d\mu(z)\,,
$$
for $\mu$ is a Borel probability measure supported in $\left\{|z|\leq
  R\right\}$ and $b$ is a polynomial function.\\
Due to $\varrho_{\varepsilon}=\varrho_{\varepsilon}1_{[0,R^{2}]}({\bf N})$,
the sum
$$
S_{K_{2},k}=\sum_{k_{2}=0}^{K_{2}}\varrho_{\varepsilon}
\frac{(\alpha_{2}{\bf N})^{k_{2}}}{k_{2}!}(1+{\bf N})^{k}\,,\quad K_{2}, k\in \nz
$$
and the remainder term
$$
R_{K_{2},k}=\varrho_{\varepsilon}e^{\alpha_{2}{\bf N}}(1+{\bf N})^{k}-S_{K_{2},k}
=\int_{0}^{1}\frac{(1-t)^{K_{2}}}{K_{2}!}
\varrho_{\varepsilon}(\alpha_{2}{\bf N})^{K_{2}+1}e^{\alpha_{2}t{\bf N}}(1+{\bf N})^{k}~dt\,
$$
satisfy
\begin{eqnarray*}
&& 1_{[0,R^{2}]}({\bf N})S_{K_{2}, k}= S_{K_{2}, k}\quad
\text{with}\quad |S_{K_{2,k}}|_{\L^{1}(\H)}\leq
e^{|\alpha_{2}|R^{2}}(1+R^{2})^{k}\,,\\
\text{and}
&& 1_{[0,R^{2}]}({\bf N})R_{K_{2}, k}=R_{K_{2},k} \quad
\text{with}
\quad
|R_{K_{2},k}|_{\L^{1}(\H)}\leq e^{|\alpha_{2}|R^{2}}(1+R^{2})^{k}\frac{(|\alpha_{2}|R^{2})^{K_{2}+1}}{(K_{2}+1)!}\,.
\end{eqnarray*}
Repeating the same estimate on the left hand side with
$S_{K_{2},k}$ and $R_{K_{2},k}$ instead of $\varrho_{\varepsilon}$
implies that the $\L^{1}(\H)$ norm of
\begin{equation*}
  (1+{\bf N})^{k}\left[e^{\alpha_{1}{\bf
      N}}\varrho_{\varepsilon}e^{\alpha_{2}{\bf N}}
-
\sum_{k_{1}=0}^{K_{1}}\sum_{k_{2}=0}^{K_{2}}\frac{(\alpha_{1}{\bf
    N})^{k_{1}}}{k_{1}!}\varrho_{\varepsilon} \frac{(\alpha_{2}{\bf
    N})^{k_{2}}}{k_{2}!}
\right]
(1+{\bf N})^{k}
\end{equation*}
is bounded by
$$
e^{|\alpha_{2}|R^{2}+|\alpha_{1}|R^{2}}(1+R^{2})^{2k}
\left[\frac{(|\alpha_{1}|R^{2})^{K_{1}+1}}{(K_{1}+1)!}
+\frac{(|\alpha_{2}|R^{2})^{K_{2}+1}}{(K_{2}+1)!}
+\frac{(|\alpha_{1}|R^{2})^{K_{1}+1}(|\alpha_{2}|R^{2})^{K_{2}+1}}{(K_{1}+1)!(K_{2}+1)!}
\right]\,,
$$
which vanishes as $\min(K_{1},K_{2})\to \infty$\,. We conclude with a
$\delta/3$-argument after noticing that $(1+{\bf N})^{-k} b^{Wick}$
$(1+{\bf N})^{-k}$
is bounded for $k\geq k_{b}$ and that the convergence
as $\varepsilon\to 0$
holds  for $b\in \P_{alg}(\Z)$ fixed and for the finite sums
$\sum_{k_{1}=0}^{K_{1}}\sum_{k_{2}=0}^{K_{2}}$ owing to the
condition $(PI)$.
\fin

\bigskip
\noindent\textbf{Proof of Proposition~\ref{pr.locPI}:}
Let $C_{f}>1$ be a constant such that
$\Tr\left[\varrho_{\varepsilon}f^{2}({\bf N})\right]\geq
\frac{1}{C_{f}}$ and $\sup_{s\in [0,+\infty)} f(s)(1+s)^{-\nu}\leq
C_{f}$\,.
The inequalities
$\Tr[\varrho^{f}_{\varepsilon}{\bf N}^{\alpha}]\leq C_{f}^{2}
\Tr\left[\varrho_{\varepsilon}{\bf N}^{\alpha}(1+{\bf
    N})^{2\nu}\right]$, $\alpha\in\nz$,
ensure that the family $(\varrho_{\varepsilon}^{f})_{\varepsilon\in
(0,\bar \varepsilon)}$ admits Wigner measures without any way to
identify them for the moment. So take a sequence
$(\varepsilon_{n})_{n\in\nz}$, such that $\lim_{n\to
  \infty}\varepsilon_{n}=0$ and
$\mathcal{M}(\varrho_{\varepsilon_{n}}^{f}, n\in\nz)=
\left\{\mu^{f}\right\}$\,. We first
prove that the sequence
$(\varrho_{\varepsilon_{n}}^{f})_{n\in\nz}$ satisfies the
condition $(PI)$, then check that
$\mu^{f}=\frac{f^{2}(|z|^{2})\mu}{\int f^{2}(|z|^{2})d\mu}$ in the
cases when $(\varrho_{\varepsilon})_{\varepsilon\in
  (0,\bar\varepsilon)}$
is localized in a ball and then when $f$ is compactly supported,
and finally conclude with approximation arguments.\\
\noindent\textbf{1) The condition $(PI)$ for the sequence:}
The uniform control of $\Tr\left[\varrho_{\varepsilon_{n}}^{f}{\bf
    N}^{\alpha}\right]\leq C_{\alpha}$, $\alpha\in \nz$, implies
$\int_{\Z}|z|^{2\alpha}~d\mu^{f}(z)<+\infty$ and the
Proposition~\ref{pr.polycomp} says that the convergence
$$
 \lim_{n\to
  \infty}\Tr\left[\varrho_{\varepsilon_{n}}^{f}b^{Wick}\right]=
\int_{\Z} b(z)~d\mu^{f}(z)
$$
holds for any $b\in \P_{alg}^{\infty}(\Z)$ with a compact kernel. In
particular for $b(z)=|pz|^{2k}$ with $p\in\p$ and $k\in\nz$,
\begin{equation}
  \label{eq.convpz}
\lim_{n\to \infty}
\Tr\left[\varrho_{\varepsilon_{n}}^{f}\left((|pz|^{2})^{Wick}\right)^{k}\right]
=
\lim_{n\to\infty}
\Tr\left[\varrho_{\varepsilon_{n}}^{f}(|pz|^{2k})^{Wick}\right]
=
\int_{\Z}|pz|^{2k}~d\mu^{f}(z)\,,
\end{equation}
while we assumed
\begin{equation}
  \label{eq.recondp}
\forall b\in \P_{alg}(\Z)\,,\quad
\lim_{n\to\infty}
\Tr\left[\varrho_{\varepsilon_{n}}b^{Wick}\right]
=
\int_{\Z}b(z)~d\mu(z)\,.
\end{equation}
Fix $\alpha\in\nz^{*}$ and take $\delta>0$. By Lebesgue's convergence,
there exists $p\in\p$ such that
\begin{eqnarray*}
&&\int_{\Z}\left||z|^{2\alpha}-|pz|^{2\alpha}\right|~d\mu^{f}(z)\leq \delta\\
\text{and}&&
\int_{\Z}\left||z|^{2\alpha}-|pz|^{2\alpha}\right|(1+|z|^{2})^{2\nu}~d\mu(z)\leq \delta\,.
\end{eqnarray*}
Remember that $(|pz|^{2})^{Wick}={\bf N}_{p}\otimes I_{\Gamma_{s}((1-p)\Z)}={\bf N}_{p}$
with
${\bf N}_{p}^{\alpha}\leq {\bf N}^{\alpha}$ where both sides
commute with $f({\N})$ and we get:
\begin{eqnarray*}
0\leq \Tr\left[\varrho_{\varepsilon_{n}}^{f}({\bf N}^{\alpha}-
{\bf N}_{p}^{\alpha})\right]
&\leq & C_{f}\Tr\left[f({\bf N})
({\bf N}^{\alpha}-{\bf
  N}_{p}^{\alpha})^{1/2}\varrho_{\varepsilon_{n}}({\bf
  N}^{\alpha}-{\bf N}_{p}^{\alpha})^{1/2}
f({\bf N})
\right]
\\
&\leq&  C_{f}|f({\bf N})(1+{\bf N})^{-\nu}|_{\L(\H)}^{2} \\&&\times
\,\Tr\left[(1+{\bf N})^{\nu}
({\bf N}^{\alpha}-{\bf
  N}_{p}^{\alpha})^{1/2}\varrho_{\varepsilon_{n}}({\bf
  N}^{\alpha}-{\bf N}_{p}^{\alpha})^{1/2}(1+{\bf N})^{\nu}
\right]
\\
&\leq&
C_{f}^{3}
\Tr\left[\varrho_{\varepsilon_{n}}({\bf N}^{\alpha}-{\bf N}_{p}^{\alpha})(1+{\bf N})^{2\nu}\right]\,.
\end{eqnarray*}
But we know  by \eqref{eq.recondp} that the right-hand side converges as $n\to \infty$ to
$$
C_{f}^{3}\int_{\Z}(|z|^{2\alpha}-|pz|^{2\alpha})(1+|z|^{2})^{2\nu}~d\mu(z)\leq C_{f}^{3}\delta\,,
$$
while \eqref{eq.convpz} with $(|pz|^{2})^{Wick}={\bf N}_{p}$ gives
$$
\lim_{n\to\infty}\Tr\left[\varrho_{\varepsilon_{n}}^{f}{\bf
    N}_{p}^{\alpha}\right]=
\int_{\Z}|pz|^{2\alpha}~d\mu^{f}(z)\,.
$$
Hence there exist $n_{\delta}\in\nz$ such that
$$
\forall n\geq n_{\delta}\,,\quad
\left|\Tr\left[\varrho_{\varepsilon_{n}}^{f}{\bf N}^{\alpha}\right]-
\int_{\Z}|pz|^{2\alpha}~d\mu^{f}(z)\right|\leq (C_{f}^{3}+1)\delta\,.
$$
From $\int_{\Z}||z|^{2\alpha}-|pz|^{2\alpha}|~d\mu^{f}(z)\leq \delta$, we deduce
$$
\limsup_{n\to\infty}|\Tr\left[\varrho_{\varepsilon_{n}}^{f}{\bf
    N}^{\alpha}\right]
-\int_{\Z}|z|^{2\alpha}~d\mu(z)|\leq (C_{f}^{3}+2)\delta\,.
$$
Letting $\delta\to 0$ ends the proof of this part.

\noindent\textbf{2) Identification of $\mu^{f}$ when
  $(\varrho_{\varepsilon})_{\varepsilon\in (0,\bar \varepsilon)}$ is
  localized in a ball:}
Assume that $(\varrho_{\varepsilon})_{\varepsilon\in
  (0,\bar\varepsilon)}$ is localized in a ball of radius $R>0$.
The Lemma~\ref{le.exptN} tells us
$$
\forall t_{1}, t_{2}\in \rz,\;\forall b\in \P_{alg}(\Z)\,,\quad
\lim_{n\to \infty}\Tr\left[e^{it_{2}{\bf N}}\varrho_{\varepsilon_{n}}e^{it_{1}{\bf
      N}}b^{Wick}\right]
=\int_{\Z} e^{i(t_{1}+t_{2})|z|^{2}}b(z)~\mu(z)\,,
$$
while the uniform boundedness of  $(1+{\bf N})^{-k_{b}}b^{Wick}(1+{\bf
  N})^{-k_{b}}$ entail
$$
\left|\Tr\left[e^{it_{2}{\bf N}}\varrho_{\varepsilon_{n}}e^{it_{1}{\bf
      N}}b^{Wick}\right]\right|
\leq C_{b}\Tr\left[\varrho_{\varepsilon_{n}}(1+{\bf
    N})^{2k_{b}}\right]\leq C_{b}(1+R^{2})^{2k_{b}}\,.
$$
Hence for $f\in \mathcal{F}^{-1}(L^{1}(\rz))$, we get
$$
\lim_{n\to\infty}\frac{\Tr\left[f({\bf N})\varrho_{\varepsilon_{n}}f({\bf
  N})b^{Wick}\right]}{\Tr\left[f({\bf N})\varrho_{\varepsilon_{n}}f({\bf
  N})\right]}
=
\frac{\int_{\Z} f(|z|^{2})^2 \, b(z)~d\mu(z)}{\int_{\Z} f(|z|^{2})^2~d\mu(z)}\,.
$$
We have proved:
$$
\forall b\in \P_{alg}(\Z)\,, \quad
\lim_{n\to\infty}
\Tr\left[\varrho_{\varepsilon_{n}}^{f}b^{Wick}\right]
=
\frac{\int_{\Z} f(|z|^{2})^2 \,b(z)~d\mu(z)}{\int_{\Z}f(|z|^{2})^2~d\mu(z)}\,.
$$
The part 1) and
$1_{[0,R^{2}]}({\bf N})\varrho_{\varepsilon_{n}}^{f}1_{[0,R^{2}]}({\bf
  N})=
\varrho_{\varepsilon_{n}}^f$
 ensure that
$(\varrho_{\varepsilon_{n}}^f)_{n\in\nz}$ satisfies
the sufficient conditions for solving the moment problem
(Proposition~\ref{pr.moment}) and
$\mu^{f}=\frac{f(|z|^{2})^2 \,\mu}{\int_{\Z}f(|z|^{2})^2~d\mu}$ in this
  case.\\
\noindent\textbf{3) Identification of $\mu^{f}$ when $f$ is compactly
  supported:}
Assume that $f\in \mathcal{C}_{c}^{0}([0,+\infty))$ is supported in
$[0, R_{0}]$. Consider for $\chi\in \mathcal{C}^{0}_{c}([0,+\infty))$,
$0\leq \chi\leq 1$, $\chi\equiv 1$ on $[0,1/2]$ and for
 $R>0$, the truncated states
$$
\varrho_{\varepsilon_{n}}^{R}=
\frac{1}{\Tr\left[\varrho_{\varepsilon_{n}}\chi^{2}(\frac{\bf N}{R^2})\right]}
\chi(\frac{\bf N}{R^2})\varrho_{\varepsilon_{n}}\chi(\frac{\bf
  N}{R^2})\,,\quad n\in\nz\,.
$$
For $R\geq 2R_{0}$, we have
$$
\forall
n\in\nz^{*}\,,\quad\varrho^{f}_{\varepsilon_{n}}=
\frac{1}{\Tr\left[\varrho_{\varepsilon_{n}}^{R}
f^{2}({\bf N})\right]}f(\N)\varrho_{\varepsilon_{n}}^{R}f(\N)\,.
$$
By extracting a subsequence we can assume
$\mathcal{M}\left(\varrho_{\varepsilon_{n_{k}}}^{R},
  k\in\nz\right)=\left\{\mu^{R}\right\}$, and Part~1) applied to
$(\varrho_{\varepsilon_{n_{k}}}^{R})_{k\in\nz}$ ensures that
 the pair $(\varrho_{\varepsilon_{n_{k}}}^{f}, \varrho_{\varepsilon_{n_{k}}}^{R})$
fulfills all the
assumptions of Part~2) if $f\in \mathcal{C}^{0}_{c}([0,+\infty))\cap \mathcal{F}^{-1}L^{1}(\rz)$. Thus the measure $\mu^{f}$ equals
$\frac{|f(|z|^{2})|^{2}\mu^{R}}{\int_{\Z}|f(|z|^{2})|^{2}~d\mu^{R}}$\,.
From the comparison \eqref{eq.comXR} we know
$\int|\mu^{R}-\mu|=\mathcal{O}(R^{-1})$ and since $f$ is a bounded
function
$$
\int
\left|\mu^{f}-\frac{|f(|z|^{2})^{2}\mu}{\int
    |f(|z|^{2})|^{2}~d\mu}\right|
\leq \frac{C}{R}.
$$
Taking the limit as $R\to 0$ gives the result when $f\in
\mathcal{C}^{0}_{c}([0,+\infty))\cap \mathcal{F}^{-1}L^{1}(\rz)$\,.
Removing the condition $f\in \mathcal{F}^{-1}L^{1}(\rz)$ is obtained
by a comparison argument between $\varrho_{\varepsilon_{n}}^{f}$ and
  $\varrho_{\varepsilon_{n}}^{f_{\ell}}$ with
$f_{\ell}\in
  \mathcal{C}^{0}_{c}\cap \mathcal{F}^{-1}L^{1}(\rz)$ and
$\sup_{s\in[0,+\infty]}|f(s)-f_{\ell}(s)|\leq \frac{1}{\ell+1}$\,, for
$\ell\in\nz$\,.\\
\noindent\textbf{4) Final approximation argument and uniqueness of
  $\mu^{f}$:}
Consider now the complete problem with the extracted sequence
$(\varrho_{\varepsilon_{n}}^{f})_{n\in\nz}$. We again use the cut-off
$\chi(\frac{\bf N}{R^2})$ but now to compare
$\varrho^{f}_{\varepsilon_{n}}$
with $\varrho_{\varepsilon_{n}}^{f\chi(R^{-2}.)}$\,.
After extracting a subsequence, we can assume $\mathcal{M}\left(
\varrho_{\varepsilon_{n_{k}}}^{f\chi(R^{-2}.)}, k\in\nz
\right)=\left\{\mu^{f\chi(R^{-2}.)}\right\}$\,.
The pair
$(\varrho_{\varepsilon_{n}}^{f\chi(R^{-2}.)},
\varrho_{\varepsilon_{n}})$
fulfills the assumptions of Part~3) and
$$
\mu^{f\chi(R^{-2}.)}=\frac{f^{2}(|z|^{2})\chi^{2}(R^{-2}|z|^{2})\mu}{\int
f^{2}(|z|^{2})\chi^{2}(R^{-2}|z|^{2})~d\mu}\,.
$$
But from the inequalities $f(s)(1-\chi(R^{-2}s))(1+s)^{-\nu-1}\leq
CR^{-2}$ and $\Tr\left[\varrho_{\varepsilon}(1+\N)^{2\nu+2}\right]\leq
\tilde{C}_{\nu}$ we deduce the uniform estimate:
$$
 \forall k\in\nz\,,\quad
\left|\varrho_{\varepsilon_{n_{k}}}^{f}-\varrho_{\varepsilon_{n_{k}}}^{f\chi(R^{-2}.)}\right|_{\L^{1}(\H)}
\leq \frac{C_{f}'}{R^{2}}\,.
$$
Again the comparison argument \eqref{eq.comXR} gives
$$
\int\left|\mu^{f}-\frac{f^{2}(|z|^{2})\chi^{2}(R^{-2}|z|^{2})\mu}{\int
f^{2}(|z|^{2})\chi^{2}(R^{-2}|z|^{2})~d\mu}\right|
\leq \frac{C_{f}'}{R^{2}}\,,
$$
and we take the limit as $R\to \infty$. We have proved $\mu^{f}=\frac{f^{2}(|z|^{2})\mu}{\int
f^{2}(|z|^{2})~d\mu}$ for any sequence extracted
from
$(\varrho_{\varepsilon}^{f})_{\varepsilon\in (0,\bar \varepsilon)}$
with a single Wigner measure. This proves
$\mathcal{M}(\varrho_{\varepsilon}, \varepsilon\in (0,\bar \varepsilon))=\left\{\frac{f^{2}(|z|^{2})\mu}{\int
f^{2}(|z|^{2})~d\mu}\right\}$ while the condition $(PI)$ was checked
in Part~1).
\fin

\section{Dynamical mean field limit}
\label{se.dynmeanfield}
Let $Q$ be a real-valued polynomial in $\P_{alg}(\Z)$ given by
$$
Q=\sum_{j=2}^r Q_j, \quad \mbox{ with } \quad Q_j\in\P_{j,j}(\Z)\,.
$$
We consider the many-body quantum Hamiltonian for a system of bosons
\begin{eqnarray}
\label{hamiltonian}
H_\varepsilon=\d\Gamma(A)+ Q^{Wick}\,,
\end{eqnarray}
with $A$ a given self-adjoint operator on $\Z$. Here $Q^{Wick}$ is the operator
$\sum_{j=2}^r Q_j^{Wick}$ with $Q_j^{Wick}$ given by (\ref{def-wick}).
Clearly, $H_\varepsilon$ acts as a self-adjoint operator on the symmetric  Fock space $\H$. When
$\Z=L^2(\rz^d)$, the Schr\"odinger Hamiltonian  $A=-\Delta+V(x)$ and
the semi-relativistic Hamiltonian $A=\sqrt{-\Delta+m^2}+V(x)$  are among the typical examples (e.g.~\cite{ElSc}).

\subsection{Existence of Wigner measure for all times.}
\label{se.existwignt}
The first step to prove Theorem \ref{th.main} is to show the existence
of Wigner measures for all times. This is
accomplished in the Proposition \ref{prop-1} by following the same lines as in the proof of Theorem \ref{th.wig-measure}.
For this task two useful lemmas are stated below with the first one being proved in \cite[Proposition 2.10]{AmNi1}.
\begin{lem}
  \label{pr.transla}
  For any $b\in \P_{alg}(\Z)$ we have:\\
(i) $b^{Wick}$ is a closable operator with the domain of its closure containing
$$
\H_0={\rm vect}\{W(\varphi)\psi, \psi\in\H_{fin},\varphi\in\Z\}\,.
$$
(ii) For any $\varphi\in \Z$ the identity
$$
W(\xi)^{*}\,b^{Wick}\,W(\xi)=(b(z+\frac{i\varepsilon}{\sqrt{2}}\xi))^{Wick}
$$
holds on $\H_0$ with $b(\cdot+\frac{i\varepsilon}{\sqrt{2}}\xi)\in \P_{alg}(\Z)$\,.
\end{lem}

\begin{lem}
\label{number-weyl}
 For any  $k\in\nz$ there exists a $\varepsilon$-independent constant $C_k>0$ such that
\begin{equation}
\label{estnumweyl}
W(\xi)^*\langle\N\rangle^k \,W(\xi) \leq C_k \la \bar\varepsilon\ra^{k} \la\xi\ra^k \langle\N\rangle^k\,,
\end{equation}
for any $\xi\in \Z$ and uniformly in $\varepsilon\in (0,\bar\varepsilon)$.
\end{lem}
\proof
Since $\N$ is a self-adjoint operator, the functional calculus provides the inequality
$$
\langle\N\rangle^k\leq (1+\N)^k \,.
$$
Therefore, it is enough to prove (\ref{estnumweyl}) with $\langle\N\rangle$ in the l.h.s replaced by $(1+\N)$.
The Wick calculus in Proposition \ref{symbcalc} tell us that $(1+\N)^k$ is a Wick operator with symbol
$b_k(z)$ in $\oplus_{j=0}^k\P_{j,j}(\Z)$, i.e.:
$$
b_k(z)=\sum_{j=0}^k \la z^{\otimes j}, \tilde b_k^{(j)} z^{\otimes j}\ra\, \quad \mbox{ with }\quad  b_k^{(j)}\in\P_{j,j}(\Z).
$$
Now, applying Lemma \ref{pr.transla}  yields
\begin{eqnarray*}
W(\xi)^{*}\,(1+\N)^k\,W(\xi)=W(\xi)^{*}\,b_k^{Wick}\,W(\xi)=(b_k(z+\frac{i\varepsilon}{\sqrt{2}}\xi))^{Wick}\,.
\end{eqnarray*}
A Taylor expansion of the symbol gives us
$$
b_k(z+\frac{i\varepsilon}{\sqrt{2}}\xi)=\sum_{j=0}^k \frac{(i\varepsilon)^j}{j!\sqrt{2^j}}\; D^{(j)}b_k(z)[\xi]\,,
$$
with $D^{(j)}$ is the $j$th derivatives and $D^{(j)}b_k(z)[\xi]\in \oplus_{m,n=0}^{k-j}\P_{m,n}(\Z)$.
So, by the number estimate (\ref{wick-estimate}) we can derive the following bound
$$
\left|\langle\N\rangle^{-k/2} \; \left(D^{(j)}b_k(z)[\xi]\right)^{Wick} \;
\langle\N\rangle^{-k/2}\right|\leq \tilde C_k \la \xi\ra^j \,
$$
with $\tilde C_k$ only depending on $k\in\nz$. Hence, we obtain
\begin{eqnarray*}
\left|\langle\N\rangle^{-k/2} \sum_{j=0}^k \frac{(i\varepsilon)^j}{j!\sqrt{2^j}}\; \left(D^{(j)}b_k(z)[\xi]\right)^{Wick} \;
\langle\N\rangle^{-k/2}\right|\leq C_k \la \bar\varepsilon\ra^{k} \la\xi\ra^{k}\,,
\end{eqnarray*}
with $C_k$ only depending on $k\in\nz$.  Thus, we conclude that $W(\xi)^{*}\,(1+\N)^k\,W(\xi)$ as a positive
quadratic form is bounded by $C_k \la \bar\varepsilon\ra^{k} \la\xi\ra^{k} \la \N\ra^k$.
\fin
\begin{prop}
\label{prop-1}
Let  $\left(\varrho_{\varepsilon}\right)_{\varepsilon\in(0,\bar
  \varepsilon)}$ be a family of normal states on $\H$ satisfying  the
uniform estimate ${\rm Tr}[\varrho_{\varepsilon}\mathbf{N}^r]$ $\leq
C_{r}$ for some $r>0$\,.\\
Then for any sequence $(\varepsilon_{n})_{n\in\nz}$ in $(0,\bar\varepsilon)$ such that $\lim_{n\to\infty} \varepsilon_n=0$
there exists a subsequence $(\varepsilon_{n_k})_{k\in\nz}$ and a family of Borel probability measures $(\mu_t)_{t\in\rz}$ satisfying
\begin{eqnarray*}
\mathcal{M}( e^{-i\frac{t}{\varepsilon_n}H_{\varepsilon_n}} \varrho_{\varepsilon_{n}}
e^{i\frac{t}{\varepsilon_{n}}H_{\varepsilon_n}}, n\in\nz)=\{\mu_t\}\,,
\end{eqnarray*}
for any $ t\in\rz$. Moreover, we have
\begin{eqnarray*}
\int_\Z  |z|^{2r}\,d\mu_t(z)\leq C_{r}\,.
\end{eqnarray*}
\end{prop}
\proof
We set
\begin{eqnarray*}
\varrho_{\varepsilon}(t)= e^{-i\frac{t}{\varepsilon}H_{\varepsilon}} \,\varrho_{\varepsilon} \, e^{i\frac{t}{\varepsilon}H_{\varepsilon}}\quad \mbox{ and } \quad
\tilde\varrho_{\varepsilon}(t)= e^{i\frac{t}{\varepsilon}\d\Gamma(A)} e^{-i\frac{t}{\varepsilon}H_{\varepsilon}}\, \varrho_{\varepsilon}\,
 e^{i\frac{t}{\varepsilon}H_{\varepsilon}} e^{-i\frac{t}{\varepsilon}\d\Gamma(A)}\;.
\end{eqnarray*}
{\bf (i)}
Consider for $\varepsilon>0$ the function
$$
G_{\varepsilon}(t,\xi)=\Tr
\left[\tilde\varrho_{\varepsilon}(t) \,W(\sqrt{2}\pi\xi)\right]\,.
$$
Write for any $(s,\xi), (t,\eta)\in\rz\times\Z$
\begin{eqnarray*}
|G_\varepsilon(t,\eta)-G_\varepsilon(s,\xi)|\leq  \left|{\rm Tr}\left[\left(\tilde\varrho_\varepsilon(t)-\tilde \varrho_\varepsilon(s)\right)
W(\sqrt{2}\pi\eta)\right]\right|+
\left|{\rm Tr}\left[\tilde\varrho_\varepsilon(s) \left(W(\sqrt{2}\pi\eta)- W(\sqrt{2}\pi\xi)\right)\right] \right|\,.
\end{eqnarray*}
By differentiation, we get
\begin{eqnarray}
\label{equi-eq1}
\left|{\rm Tr}\left[[\tilde\varrho_\varepsilon(t)-\tilde \varrho_\varepsilon(s)]
W(\sqrt{2}\pi\eta)\right]\right|\leq \frac{1}{\varepsilon} \left|\int_s^t {\rm Tr}\left[ \tilde\varrho_\varepsilon(t')
[Q_{t'}^{Wick},W(\sqrt{2}\pi\eta)]\right] \,dt'\right|
\end{eqnarray}
with $Q_{t'}(z)=Q(e^{-i t' A} z)$, while the second term is estimated by
\begin{eqnarray}
\label{equi-eq2}
\left|{\rm Tr}\left[\tilde\varrho_\varepsilon(s) \left( W(\sqrt{2}\pi\eta)- W(\sqrt{2}\pi\xi)\right)\right] \right|
\leq (1+C_{r}) \, \left|[W(\sqrt{2}\pi\eta)- W(\sqrt{2}\pi\xi) ]
 ({\bf N}+1)^{-1}\right|_{\L(\H)} \;.
\end{eqnarray}
Now, we claim that there exists a constant $c>0$  such that the r.h.s of (\ref{equi-eq1}) is bounded by
\begin{eqnarray}
\label{bound-1}
c |t-s| \, (\sum_{j=2}^r ||\tilde Q_j||)  \, \sum_{i=1}^{2r} \varepsilon^{i-1} |\eta|^i  \,.
\end{eqnarray}
This can be proved by first writing
\begin{eqnarray}
\label{eq.wigt}
&&{\rm Tr}\left[ \tilde\varrho_\varepsilon(t')
[Q_{t'}^{Wick},W(\sqrt{2}\pi\eta)]\right]=\\  \nonumber
&& {\rm Tr}\left[\la \N\ra^{r} \tilde\varrho_{\varepsilon}(t') \la \N\ra^{r}  \left(\la \N\ra^{-r} W(\sqrt{2}\pi\eta)
\la \N\ra^{r} \right) \la \N\ra^{-r} [W(\sqrt{2}\pi\eta)^* Q_{t'}^{Wick} W(\sqrt{2}\pi\eta)-Q_{t'}^{Wick}]\,
\la \N\ra^{-r}\right]
\end{eqnarray}
and second estimating the r.h.s of  (\ref{eq.wigt}) using Lemma \ref{number-weyl} and  Lemma \ref{pr.transla} (ii)
so that
$$
\left|\int_s^t {\rm Tr}\left[ \tilde\varrho_\varepsilon(t')
[Q_{t'}^{Wick},W(\sqrt{2}\pi\eta)]\right] \,dt'\right|\leq
c |t-s| \sup_{t'\in[s,t]} \left|
\la \N\ra^{-r} [Q_{t'}(.+\frac{i\varepsilon}{\sqrt{2}}\eta)^{Wick}-Q_{t'}^{Wick}]
\la \N\ra^{-r}\right|_{\L(\H)}.
$$
Thus, the bound (\ref{bound-1}) follows from the
number estimate in Proposition \ref{wick-estimate}.\\
We recall the inequality proved in  \cite[Lemma 3.1]{AmNi1},
\begin{eqnarray*}
\left|[W(\sqrt{2}\pi\eta)-W(\sqrt{2}\pi \xi)] (\N+1)^{-1/2} \right| \leq
\tilde C\; \,\left|\eta-\xi\right| \;
[\min(\varepsilon |\eta|, \varepsilon |\xi|)+\max(1,\varepsilon)]\,.
\end{eqnarray*}
This leads to the following bound on  the r.h.s of (\ref{equi-eq2})
$$
\tilde C \, \la\bar\varepsilon\ra \,|\eta-\xi| \; \left(1+\sqrt{|\eta|^2+|\xi|^2}\right)\,.
$$
Thus, we conclude that $\forall (s,\xi),(t,\eta)\in \rz\times\Z,$
\begin{equation}
 \label{eq.unifcont}
\left|G_{\varepsilon}(t,\eta)-G_{\varepsilon}(s,\xi)\right|
\leq \tilde c \left( |t-s| (|\eta|+1)^{2r}+\, |\eta-\xi|\, \sqrt{|\eta|^2+|\xi|^2}\right),
\end{equation}
uniformly w.r.t. $\varepsilon\in (0,\overline{\varepsilon})$. Recall also that we have
the uniform estimate $\left|G_{\varepsilon}(s,\xi)\right|\leq
1$.\\
\noindent
Now, we apply an Ascoli type argument:
\begin{itemize}
\item Since $\rz\times\Z$ is separable, it admits a countable dense set
 $\mathcal{N}=\left\{(t_\ell,\xi_{\ell}),\; \ell\in \nz \right\}$. For any
 $\ell\in\nz$ the set $\{G_{\varepsilon}(t_\ell,\xi_{\ell})\}_{\varepsilon\in(0,\bar\varepsilon)}$ remains
 in $\left\{\sigma\in \cz, \left|\sigma\right|\leq 1\right\}$. Hence
for any sequence $(\varepsilon_n)_{n\in\nz}$ such that $\varepsilon_n\to 0$ there exists  by a diagonal extraction process
a subsequence, still denoted by $(\varepsilon_{n})_{n\in \nz}$, such that for all $\ell\in \nz$,
$G_{\varepsilon_{n}}(t_\ell,\xi_{\ell})$ converges in $\left\{\sigma\in
 \cz, \left|\sigma\right|\leq 1\right\}$ as $n\to \infty$. Set
$$
G(t_\ell,\xi_{\ell}):=\lim_{n\to\infty}G_{\varepsilon_{n}}(t_\ell,\xi_{\ell})
$$
for all $\ell\in \nz$.
\item The uniform estimate \eqref{eq.unifcont} implies that the limit
 $G$ is uniformly continuous on any set
 $$
 \mathcal{N}\cap
 \left\{(t,z)\in\rz\times\Z:\left|t\right|+\left|z\right|\leq R\right\}.
 $$
 Hence it admits a continuous
 extension still denoted $G$ in $(\rz\times\Z,\left|~\right|_{\rz\times\Z})$. An ``epsilon$/3$''-argument
 shows that for any $(t,\xi)\in\rz\times\Z$,
 $\lim_{n\to\infty}G_{\varepsilon_{n}}(t,\xi)$ exists and equals
 $G(t,\xi)$.
\end{itemize}
Finally for any $t\in\rz$, $G(t,.)$ is a norm continuous normalized function of positive type
since
\begin{eqnarray*}
 &&
G(t,0)=\lim_{n\to\infty}\Tr\left[\tilde\varrho_{\varepsilon}(t) \right]=1
\\
&&
\sum_{i,j=1}^{N}\lambda_{i}\overline{\lambda_{j}} \, G(t,\xi_{i}-\xi_{j})=\lim_{n\to\infty}
\sum_{i,j=1}^{N}\lambda_{i}\overline{\lambda_{j}}\; \Tr
\left[
\tilde\varrho_{\varepsilon_n}(t) W(\sqrt{2}\pi (\xi_i-\xi_j))\right] e^{i\varepsilon_n\pi^2 \sigma(\xi_i,\xi_j)} \geq 0\,.
\end{eqnarray*}
The positivity in the last statement follows by Weyl commutation relations (\ref{eq.Weylcomm}).
Therefore, according to the Bochner theorem (e.g. \cite[Corollary 1.4.2]{BSZ}) for any $t\in\rz$,  $G(t,.)$ is a characteristic
function of a  weak distribution or equivalently a cylindrical measure  $\tilde\mu_t$ on $\Z$ (see \cite{Sko} and also
\cite[Section 6]{AmNi1} for specific information).    \\
{\bf (ii)} The fact that $\tilde\mu_t$ are Borel probability measures satisfying
\begin{eqnarray}
\label{ctrmas}
\tilde\mu_t(|z|^{2r})\leq C_{r}<\infty\,,
\end{eqnarray}
follows directly by \cite[Theorem 2.5 Chap.VI]{Par} or by part {\bf (iv)} in the proof of \cite[Theorem 6.2]{AmNi1}.\\
{\bf (iii)}
Using (\ref{weyl-obs}) we see that for
any $b\in\S_{cyl}(\Z)$ based on a finite dimensional subspace $p\Z$ with $p\in\mathbb{P}$
\begin{eqnarray*}
\lim_{n\to\infty } {\rm Tr}\left[\tilde\varrho_{\varepsilon_n}(t) \, b^{Weyl}\right]&=&\lim_{n\to\infty }
\int_{p\Z} G_{\varepsilon_n}(t, \xi) \; \F[b](\xi) \, L_p(d\xi)\\
&=& \int_{p\Z} G(t, \xi) \; \F[b](\xi) \, L_p(d\xi)=\int_\Z b(z) \,d\tilde\mu_t(z)\,.
\end{eqnarray*}
Therefore, according to Definition \ref{de.setwig} of Wigner measures we conclude that
$$
\forall t\in\rz, \quad \mathcal{M}(\tilde\varrho_{\varepsilon_n}(t), n\in\nz)=\{\tilde\mu_t\}\,.
$$
{\bf (iv)} Finally the family of measures $\mu_t$ which satisfy the claimed statement in the proposition are
the push-forward measures
\begin{eqnarray*}
\mu_t=(e^{-itA})_*\tilde\mu_t\,.
\end{eqnarray*}
Furthermore, an analogue of (\ref{ctrmas}) can be easily checked for the measures $\mu_t$.
\fin

\subsection{Polynomial approximations of the classical flow.}
\label{se.polapprox}
With the classical hamiltonian
$$
h(z)= \langle z\,,\, A z\rangle + Q(z)= \langle z\,,\, A z\rangle +
\sum_{j=2}^{r}\langle z^{\otimes j}\,,\, \tilde{Q_{j}}z^{\otimes j}\rangle, \quad z\in\D(A)\,,
$$ the related nonlinear field equation is
\begin{eqnarray*}
\label{hartree}
\left\{
  \begin{array}[c]{l}
    i\partial_{t}z_{t}=Az_t+\partial_{\overline
      z}Q(z_{t})\\
   z_{t=0}=z_{0}\,.
  \end{array}
\right.
\end{eqnarray*}
Actually this Cauchy problem is better studied when reformulated as an integral equation
\begin{eqnarray}
\label{hartree.int}
 z_t=e^{-i t A} z-i\int_0^t e^{-i (t-s) A} \,\; \partial_{\bar z}Q(z_s) \, ds, \;\mbox{ for } \;z\in\Z\,,
\end{eqnarray}
which admits a classical $\mathcal{C}^0$-flow ${\bf F}_{t}:\rz\times\Z\to \Z$:
1) since the $\tilde{Q}_{j}$ are bounded a fixed point argument gives the
local in time existence and uniqueness; 2) then the conservation
$|z_{t}|=|z_{0}|$ ensures the global in time result.
As a classical $\mathcal{C}^{0}$-flow, $\mathbf{F}$ is a $\mathcal{C}^0$-map satisfying $\mathbf{F}_{t+s}(z)=
\mathbf{F}_{t}\circ\mathbf{F}_{s}(z)$ and $\mathbf{F}_t(z)$ solves
(\ref{hartree.int}) for any $z\in\Z$\,. \\
Moreover, if $z_t$ solves $(\ref{hartree.int})$,  and $Q_{t}(z)=Q(e^{-it A} z)$,
then  $w_t=e^{i tA} z_t$ solves the differential equation
$$
\frac{d}{dt}\, w_t=-i \partial_{\bar z} Q_t(w_t)\,.
$$
Therefore for any $b\in\P_{p,q}(\Z)$, the following identity holds
\begin{eqnarray*}
\frac{d}{dt}\, b(w_t)&=&\partial_{\bar z} b(w_t)[-i \partial_{\bar z}Q_t(w_t)]+
\partial_{z} b(w_t)[-i \partial_{\bar z}Q_t(w_t)]\\&=&i \{Q_t,b\}(w_t).
\end{eqnarray*}
Hence, we obtain the Duhamel formula
\begin{eqnarray}
\label{class-integ-form}
b(z_t)=b_t(z)+i\int_{0}^{t}\left\{Q_{t_1},b_{t}\right\}(e^{it_1 A} z_{t_1})~dt_{1}\,.
\end{eqnarray}
A simple iteration in (\ref{class-integ-form}), using 
$$
\{Q_{t_1},b_t\}(w_{t_1})=\{Q_{t_1},b_t\}(w_{0})+i\int_0^{t_1} 
\{Q_{t_2},\{Q_{t_1},b_t\}\}(w_{t_2}) dt_2\,,
$$
yields
\begin{eqnarray*}
b(z_{t})
=b_t(z)+  i \int_0^t  \; \{Q_{t_1},b_t\}(z)\;dt_1+ i^2 \int_0^t
dt_1\int_0^{t_{1}} dt_2 \; \{Q_{t_2},\{Q_{t_1},b_t\}\}(e^{i t_2 A} z_{t_2})\,.
\end{eqnarray*}
Therefore, by induction and after setting $\mathbf{F}_{t}(z)=z_t$, we obtain for any $K>1$:
\begin{eqnarray*}
b\circ{\bf F}_t(z)
&=&b_t(z)+\sum_{k=1}^{K-1}  i^k  \;\;
 \int_0^t dt_1\cdots\int_0^{t_{k-1}} dt_k \; \;\{Q_{t_{k}},\{\ldots,\{Q_{t_{1}},b_{t}\}\ldots \}\}(z) \label{idf.1}\\
&+&  i^{K} \int_0^t
dt_1\cdots\int_0^{t_{K-1}} dt_K \; \;\{Q_{t_{K}},\{\ldots,\{Q_{t_{1}},b_{t}\}\ldots \}\}(e^{i t_K A} z_{t_K}) \,.\label{idf.2}
\end{eqnarray*}
With the polynomial $Q$ we associate the norm
\begin{equation}
  \label{eq.normQ}
\|Q\|=\max_{j\in\left\{2,\ldots, r\right\}}|Q_{j}|_{\P_{j,j}}
=
\max_{j\in\left\{2,\ldots, r\right\}}|\tilde{Q}_{j}|_{\L(\bigvee^{j}\Z,\bigvee^{j}\Z)}
\end{equation}
and we note that $\|Q_{t}\|=\|Q\|$ for all $t\in\rz$\,.
Notice that the flow ${\bf F}_{t}$ preserves the norm
$$
\forall z\in \Z\,,\quad |{\bf F}_{t}(z)|=|z|\,,
$$
and is gauge invariant
$$
\forall z\in \Z,\, \forall \theta\in \rz,\quad
{\bf F}_{t}(e^{i\theta}z)=e^{i\theta}{\bf F}_{t}(z)\,.
$$
\underline{But} for a given polynomial $b(z)$,
the map $z\mapsto b(z_t)$ does not remain a polynomial.
Starting from a polynomial $b(z)\in {\cal P}_{p,q}(\Z)$, we study
polynomial approximations of $b(z_t)$. \\
Consider the expression
\begin{eqnarray}
  \label{eq.approxpol}
b^{K}(t,z)
&=& b_t(z)+\sum_{k=1}^{K-1}  i^k  \;\;
 \int_0^t dt_1\cdots\int_0^{t_{k-1}} dt_k \; \;\{Q_{t_{k}},\{\ldots,\{Q_{t_{1}},b_{t}\}\ldots \}\}(z)
=\sum_{k=0}^{K-1}b_{k}(t,z)\,\\
R^K(t,z)&=& i^{K} \int_0^t
dt_1\cdots\int_0^{t_{K-1}} dt_K \; \;\{Q_{t_{K}},\{\ldots,\{Q_{t_{1}},b_{t}\}\ldots \}\}(e^{i t_K A} z_{t_K}) \,.
\end{eqnarray}
The two approximation results that we will use are given in the two
next propositions.
\begin{prop}
\label{pr.polapprox1}
  For $b\in \P_{p,q}(\Z)$, the polynomial $b^{K}(t,z)=\sum_{k=0}^{K-1}b_{k}(t,z)$
  defined in \eqref{eq.approxpol} belongs to\\
  $\oplus_{j=1}^{K(r-1)}\P_{j+p,j+q}(\Z)$ with the estimates
\begin{eqnarray}
  \label{eq.estimpol}
  &&|b_{k}(t,z)|\leq 2^{\frac{p+q}{2(r-1)}} (p+q) \,(4 r^3)^k \; \|Q\|^{k} \; |b|_{\P_{p,q}} |t|^{k}\langle
  z\rangle^{2k(r-1)+p+q}\,.
\end{eqnarray}
Moreover, we have for $R^{K}(t,z)$ the estimates
\begin{eqnarray}
  \label{eq.estimrestepol}
  &&|R^{K}(t,z)|\leq
2^{\frac{p+q}{2(r-1)}} (p+q) \,(4 r^3)^K \; \|Q\|^{K} \; |b|_{\P_{p,q}} |t|^{K}\langle
  z\rangle^{2K(r-1)+p+q}\,.
\end{eqnarray}
\end{prop}
\proof
With $b\in \P_{p,q}(\Z)$ and $Q_{t}=\sum_{j=2}^{r}Q_{j,t}$, the
polynomial
$$
b_{k}(t)=(i)^{k} \int_0^t dt_1\cdots\int_0^{t_{k-1}} dt_k \; \{Q_{t_{k}},\{\ldots,\{Q_{t_{1}},b_{t}\}\ldots \}\}(z)
$$
is the sum of $(r-1)^{k}\leq r^{k}$ monomials
\begin{eqnarray*}
&&b_{k}(t)=\sum_{\alpha\in \left\{2,\ldots,
    r\right\}^{k}}b_{k,\alpha}(t)\\
\text{with}
&&
b_{k,\alpha}(t)
=
(i)^{k} \int_0^t dt_1\cdots\int_0^{t_{k-1}} dt_k  \{Q_{\alpha_{k},t_{k}},\{\ldots,\{Q_{\alpha_{1},t_{1}},b_{t}\}\ldots
\}\} \quad\in \P_{|\alpha|-k+p,|\alpha|-k+q}(\Z)\,.
\end{eqnarray*}
A consequence of Proposition~\ref{symbcalc} says for $c\in \P_{p',q'}(\Z)$\,,
$$
|\left\{Q_{\alpha_1,t_1}, c\right\}(z)| \leq r (p'+q') \;|Q_{\alpha_1}|_{\P_{\alpha_1,\alpha_1}} \;|c|_{\P_{p',q'}}\;\la z\ra^{p'+q'+2(\alpha_1-1)}.
$$
We deduce
\begin{eqnarray*}
|b_{k,\alpha}(t,z)|&\leq& \int_0^t dt_1\cdots\int_0^{t_{k-1}} dt_k   \;r^k (p+q)\cdots (p+q+2 k (r-1)) \;
\|Q\|^{k} \; |b|_{\P_{p,q}}\; \la z\ra^{p+q+2|\alpha|-2k} \\
&\leq& (p+q) r^{k} (2 (r-1))^{k-1} |t|^k
\frac{\Gamma(a+k+2)}{\Gamma(k+1) \Gamma(a+1)} \frac{1}{a+k+1}
\|Q\|^{k} \; |b|_{\P_{p,q}}\;\la z\ra^{p+q+2 k(r-1)}
\end{eqnarray*}
with $a=\frac{p+q}{2 (r-1)}$ and $\Gamma$ denotes the Gamma function.
Now, we notice the relation with the Beta function
$$
B(k+1,a+1)=\frac{\Gamma(k+1) \Gamma(a+1)}{\Gamma(a+k+2)}=\int_0^1 t^{k}(1-t)^{a}\,dt \geq  \frac{1}{2^{a+k+1} (a+k+1)}\,,
$$
which yields \eqref{eq.estimpol}. \\
The remainder
$$
R^{K}(t,z)=i\left\{Q_{t},b^{K}\right\} =
i^{K} \int_0^t
dt_1\cdots\int_0^{t_{K-1}} dt_K \; \;\{Q_{t_{K}},\{\ldots,\{Q_{t_{1}},b_{t}\}\ldots \}\}(e^{i t_K A} z_{t_K})
$$
is analyzed like the term $b_{k}(t)$\,.
\fin

\begin{prop}
  \label{pr.estimflot}
Let $\mu$ be a positive Borel measure on $\Z$ supported in the ball
$\left\{|z|\leq R\right\}$, $R>0$, then for any polynomial $b\in
\P_{p,q}(\Z)$,
$$
\int_{\Z}|R^K(t,z)|~d\mu(z)
\leq
\la R\ra^{p+q} \;2^{\frac{p+q}{2(r-1)}} (p+q) |b|_{\P_{p,q}} \, \left[4 r^3 \|Q\| \,\la R\ra^{2(r-1)}|t|\right]^{K}\,.
$$
\end{prop}
\proof
It easily follows from (\ref{eq.estimrestepol}).
\fin

\subsection{Transport  for a state localized in a ball}
\label{se.translocball}
The previous approximation result allows to prove partly
Theorem~\ref{th.main} for states localized in a ball, introduced
according to Definition~\ref{de.locb} and studied in
Subsection~\ref{se.locball}.
\begin{prop}
\label{pr.ball}
Let $(\varrho_{\varepsilon_n})_{n\in\nz}$  be a sequence of normal states on
$\H$ localized in a ball with radius $R>0$ and such that
\begin{eqnarray*}
&&
\forall t\in [-T,T]\,,\quad
\mathcal{M}(e^{-i\frac{t}{\varepsilon_{n}}H_{\varepsilon_{n}}}\varrho_{\varepsilon_{n}}e^{i\frac{t}{\varepsilon_{n}}H_{\varepsilon_{n}}}\,,\,n\in\nz)=\left\{\mu_{t}\right\}\,,\\
\text{and}&&\forall \alpha\in\nz\,,\quad\lim_{k\to\infty} \Tr[\varrho_{\varepsilon_{n_k}}\,\mathbf{N}^\alpha
]=\int_{\Z}|z|^{2\alpha}~d\mu_{0}(z)\,. \\
\end{eqnarray*}
Then for all $t\in[-T,T]$, the probability measure $\mu_{t}$ is the
push-forward by the flow ${\bf F}_t$ of the measure $\mu_0$, i.e., $\mu_{t}=({\bf F}_t)_*\mu_0\,$. Moreover the identity
$$
\lim_{n\to\infty}\Tr\left[e^{-i\frac{t}{\varepsilon_{n}}H_{\varepsilon_{n}}}\varrho_{\varepsilon_{n}}e^{i\frac{t}{\varepsilon_{n}}H_{\varepsilon_{n}}}
b^{\textrm{quantized}}\right]=\int_{\Z}b(z)~d\mu_{t}(z)=\int_{\Z}b({\bf F}_{t}(z))~d\mu_{0}(z)\,,
$$
holds for Weyl quantized cylindrical functions $b\in\bigcup_{p\in\mathbb{P}}\F^{-1}(\mathcal{M}_b(p\Z))$ and general Wick quantized polynomials
$b\in \P_{alg}(\Z)$.
\end{prop}
\proof
We set
$$
\tilde\varrho_{\varepsilon_{n}}(t):=e^{i\frac{t}{\varepsilon_{n}} \d\Gamma(A)}
e^{-i\frac{t}{\varepsilon_{n}} H_{\varepsilon_{n}}}
\varrho_\varepsilon  e^{i\frac{t}{\varepsilon_{n}} H_{\varepsilon_{n}}} e^{-i\frac{t}{\varepsilon_{n}}\d\Gamma(A)}\,.
$$
It is worth noticing that for all $t\in\rz$, the sequence
$(\tilde{\varrho}_{\varepsilon_{n}}(t))_{n\in\nz}$ is localized in the
ball with radius $R$.\\
For a \underline{fixed} $b\in \P_{p,q}(\Z)$,
differentiating with respect to $t$
the quantity ${\rm Tr}[\tilde\varrho_\varepsilon(t) \, b^{Wick}]$, we obtain
\begin{equation}
\label{eq.0.intder}
{\rm Tr}[\tilde\varrho_{\varepsilon_{n}}(t) \, b^{Wick}]= {\rm Tr}[\tilde\varrho_{\varepsilon_{n}}(0) \, b^{Wick}]+\frac{i}{\varepsilon_{n}}\int_0^t \,
{\rm Tr}\left[\tilde\varrho_{\varepsilon_{n}}(s) \,[Q_s^{Wick},
  b^{Wick}]\right] \; ds
\end{equation}
and replacing $b$ by $b_t$ we end up with
\begin{eqnarray}
  \label{eq.intder}
{\rm Tr}[\varrho_{\varepsilon_{n}}(t) \, b^{Wick}]
&=& {\rm Tr}[\varrho_{\varepsilon_{n}}(0) \, b_t^{Wick}]+i\int_0^t \,
{\rm Tr}\left[\tilde\varrho_{\varepsilon_{n}}(s) \,\{Q_s,
  b_t\}^{Wick}\right]~ds\\
\nonumber
&&+
i\sum_{j=2}^r\frac{\varepsilon_{n}^{j-1}}{j!} \int_0^t \,
{\rm Tr}\left[\tilde\varrho_{\varepsilon_{n}}(s) \,\left(\{Q_s, b_t\}^{(j)}\right)^{Wick}\right]~ds\,.
\end{eqnarray}
Consider now the case when $b\in \P_{p,q}^{\infty}(\Z)$ with a compact
kernel, $\tilde{b}\in \L^{\infty}(\bigvee^{p}\Z; \bigvee^{q}\Z)$\,.
Then we know that the left-hand side converges to
$\int_{\Z}b(z)~ d{\mu}_{t}(z)$. The number estimate of
Proposition~\ref{wick-estimate}
 with $\Tr\left[{\bf N}^{\alpha}\varrho_{\varepsilon_{n}}\right]\leq
 R^{2\alpha}$ implies that the last term of the right-hand side
 converges to $0$ as $n\to\infty$. Finally the first term of the
 right-hand side converges to $\int_{\Z}b(z)~d\mu_{0}(z)$, even when
 $\tilde{b}$ is not compact.\\ We conclude that the limit of the
 second term of the r.h.s exists with
$$
\int_{\Z} b(z)~d{\mu}_{t}(z)=\int_{\Z}b_t(z)~d\mu_{0}(z)
+\lim_{n\to\infty}i\int_0^t \,
{\rm Tr}\left[\tilde\varrho_{\varepsilon_{n}}(s) \,\{Q_s, b_t\}^{Wick}\right]\;ds\,,
$$
and this initiates our induction process.\\
Given $K>1$, take the approximation
$b^{K}(t)=\sum_{k=0}^{K-1}b_{k}(t)$ to $b({\bf F}_{t}(z))$ given in
\eqref{eq.approxpol}, and assume
\begin{eqnarray}
  \label{eq.inductionbK}
\int_{\Z} b(z)~d\mu_{t}(z)&=&\int_{\Z}b^{K}(t,z)~d\mu_{0}(z)
\\&&+\lim_{n\to\infty} i^{K} \int_0^t
dt_1\cdots\int_0^{t_{K-1}} dt_K \; {\rm Tr}\left[\tilde\varrho_{\varepsilon_n}(t_K)
\left(\{Q_{t_K}, \cdots \{Q_{t_1},
b_t\}\cdots\}\right)^{Wick} \,\right]\,.
\end{eqnarray}
A simple differentiation with respect to $t_K$ gives for $\Theta \in
\P_{alg}(\Z)$,
\begin{eqnarray*}
{\rm Tr}\left[\tilde\varrho_{\varepsilon_n}(t_K)
\Theta^{Wick} \,\right]={\rm Tr}\left[\tilde\varrho_{\varepsilon_n}(0) \Theta^{Wick} \,\right]+ i \int_0^{t_{K}}dt_{K+1} {\rm Tr}\left[\tilde\varrho_{\varepsilon_n}(t_{K+1})
\left(\{Q_{t_{K+1}}, \Theta\}\right)^{Wick} \,\right]\\+
i\sum_{j=2}^r\frac{\varepsilon_{n}^{j-1}}{j!} \int_0^{t_K} \,
{\rm Tr}\left[\tilde\varrho_{\varepsilon_{n}}(t_{K+1}) \,\left(\{Q_{t_{K+1}}, \Theta\}^{(j)}\right)^{Wick}\right]~dt_{K+1}\,.
\end{eqnarray*}
Hence, choosing $\Theta= \{Q_{t_K}, \cdots \{Q_{t_1},
b_t\}\cdots\}$ yields
\begin{eqnarray*}
\int_{\Z} b(z)~d\mu_{t}(z)&=&\int_{\Z}b^{K}(t,z)~d\mu_{0}(z)\\&+&\lim_{n\to\infty}\left\{ i^{K} \int_0^t
dt_1\cdots\int_0^{t_{K-1}} dt_K \; {\rm Tr}\left[\tilde\varrho_{\varepsilon_n}(0)
\left(\{Q_{t_K}, \cdots \{Q_{t_1},
b_t\}\cdots\}\right)^{Wick} \,\right]\right.
\\&+& \left.i^{K+1}\sum_{j=2}^r\frac{\varepsilon_{n}^{j-1}}{j!} \int_0^t
dt_1\cdots\int_0^{t_{K}}dt_{K+1} {\rm Tr}\left[\tilde\varrho_{\varepsilon_n}(t_{K+1})
\left(\{Q_{t_{K+1}}, \cdots \{Q_{t_1},
b_t\}\cdots\}\right)^{Wick} \,\right]\right.\,\\
&+&\left.i^{K+1} \int_0^t
dt_1\cdots\int_0^{t_{K}} dt_{K+1} \; {\rm Tr}\left[\tilde\varrho_{\varepsilon_n}(t_{K+1})
\left(\{Q_{t_{K+1}}, \cdots \{Q_{t_1},
b_t\}\cdots\}\right)^{Wick} \,\right]\right\}\\
&=:& {\rm I}+\lim_{n\to\infty} ({\rm II}+{\rm III}+{\rm IV}).
\end{eqnarray*}
For any $K$, when $n\to\infty$, the second term (II) converges to $\int_{\Z}\Theta(z)~d\mu_{0}(z)$ because
the initial states ${\tilde\varrho}_{\varepsilon_{n}}(0)=\varrho_{\varepsilon_{n}}$ satisfies
$\lim_{\varepsilon\to
  0}\Tr\left[\varrho_{\varepsilon_{n}}c^{Wick}\right]=\int_{\Z}c(z)~d\mu_{0}(z)$
according to Proposition~\ref{pr.eqPPI}. Moreover, the third term (III) vanishes, when $n\to\infty$,
thanks to the number estimate in Proposition~\ref{wick-estimate} and the fact that $\Tr\left[\varrho_{\varepsilon_n} {\bf N}^{\alpha}\right]\leq R^{2\alpha}$\,.
Therefore, we have
\begin{eqnarray*}
  \int_{\Z} b(z)~d{\mu}_{t}(z)&=&\int_{\Z}b^{K+1}(t,z)~d\mu_{0}(z)
\\&&+\lim_{n\to\infty} i^{K+1} \int_0^t
dt_1\cdots\int_0^{t_{K}} dt_{K+1} \; {\rm Tr}\left[\tilde\varrho_{\varepsilon_n}(t_{K+1})
\left(\{Q_{t_{K+1}}, \cdots \{Q_{t_1},
b_t\}\cdots\}\right)^{Wick} \,\right]\,.
\end{eqnarray*}
By Proposition~\ref{pr.estimflot} and the fact that $\mu_{0}$ is
supported in $\left\{|z|\leq R\right\}$, we deduce
\begin{eqnarray}
\label{approxmu}
\left|\int_{\Z}b(z)~ d{\mu}_{t}(z)-\int_{\Z}b({\bf F}_t(z))~d\mu_{0}\right|
\leq
\la R\ra^{p+q} 2^{\frac{p+q}{2(r-1)}} (p+q) |b|_{\P_{p,q}} \, \left[4 r^3 \|Q\| \,\la R\ra^{2(r-1)}|t|
\right]^{K}\\ \nonumber
\hspace{.2in}
+\left|\lim_{n\to\infty}\int_0^t
dt_1\cdots\int_0^{t_{K-1}} dt_K \; {\rm Tr}\left[\tilde\varrho_{\varepsilon_n}(t_K)
\left(\{Q_{t_K}, \cdots \{Q_{t_1},
b_t\}\cdots\}\right)^{Wick} \,\right]\right|\,.
\end{eqnarray}
The number estimate of Proposition \ref{wick-estimate}
with the inequality \eqref{eq.estimpol} of
Proposition~\ref{pr.polapprox1} implies
\begin{eqnarray*}
&&\left|\langle
{\bf N}\rangle^{-\frac{q+K(r-1)}{2}} \;
\left(\{Q_{t_K}, \cdots \{Q_{t_1},
b_t\}\cdots\}\right)^{Wick}
\;\langle
{\bf N}\rangle^{-\frac{p+K(r-1)}{2}}\right|_{\L(\H)}\leq
2^{\frac{p+q}{2(r-1)}} (p+q) \,(4 r^3)^K \; \|Q\|^{K} \; |b|_{\P_{p,q}} \,.
\end{eqnarray*}
This provides for the last term in the r.h.s of (\ref{approxmu}) the upper bound
$$
\la R\ra^{\frac{p+q}{2}+K(r-1)} \;2^{\frac{p+q}{2(r-1)}} (p+q) \,(4 r^3)^K \; \|Q\|^{K} \; |b|_{\P_{p,q}} \,|t|^K\,.
$$
For small times, $|t|\leq T_{\delta}=\frac{\delta}{(4 r^3)\|Q\| \la R\ra^{r-1}}$ with
$\delta<1$, taking the limit as $K\to \infty$ now gives
$$
\forall b\in \P_{p,q}^{\infty}(\Z)\,,\quad
\int_{\Z}b(z)~d{\mu}_{t}(z)=\int_{\Z}b({\bf F}_t(z))~d\mu_{0}(z)\,.
$$
But according to Proposition~\ref{pr.ball}, the measure
${\mu}_{t}$ is a Borel probability measure supported  in the ball
$\left\{|z|\leq R\right\}$ which is weakly compact. Meanwhile
cylindrical polynomials which are
contained in $\P_{alg}^{\infty}(\Z)$, because they
are associated with finite rank kernels, make a dense set in the
$\mathcal{C}^{0}(B(0,R)_{weak},\cz)$ and therefore in
$L^{1}(\Z,d\mu)$\,. Thus, we have proved
$$
\forall
t\in[-T_{\delta},T_{\delta}]\,,\quad
\mu_{t}=({\bf F}_{t})_{*}\mu_{0}\,.
$$
Finally, since $|{\bf F}_{t}(z)|=|z|$ and $[H_{\varepsilon}, {\bf N}]=0$,
the pair $\left((\varrho_{\varepsilon_{n}}(t))_{n\in\nz},\mu_{t}\right)$
satisfies
 the same assumptions as $\left((\varrho_{\varepsilon_{n}})_{n\in\nz},\mu_{0}\right)$.
Since the time $T_{\delta}$ depends only on $Q$ and $R$ the result extends to all $t\in\rz$\,.
\fin

\subsection{Proof of the main result}
\label{se.thm1}
Gathering all the information of Section \ref{se.wibbgky} and \ref{se.dynmeanfield}, we are now in position to prove Theorem \ref{th.main}.\\
{\bf Proof of Theorem \ref{th.main}}:\\
Let $(\varrho_\varepsilon)_{\varepsilon\in(0,\bar\varepsilon)}$ be a family of normal states satisfying hypothesis of Theorem \ref{th.main} and
let $\chi\in\mathcal{C}^0([0,\infty),\rz)$ be a continuous cutoff  function such that $0\leq\chi\leq 1$, $\chi(x)=1$ if $x\leq 1/2$ and
$\chi(x)=0$ if $x\geq 1$. For $R>0$, consider the family of normal states
\begin{eqnarray*}
\varrho_\varepsilon^R=\frac{\chi(\N/R^2) \varrho_\varepsilon \chi(\N/R^2)}{\Tr[\chi(\N/R^2) \varrho_\varepsilon \chi(\N/R^2)]}\,,
\end{eqnarray*}
localized in the ball of radius $R$. By Proposition \ref{pr.locPI}, we know that
\begin{eqnarray*}
&(i)& \mathcal{M}(\varrho_\varepsilon^R,\varepsilon\in(0,\bar\varepsilon))=\left\{\frac{\chi^2(|z|^2/R^2)}{\int_\Z \chi^2(|z|^2/R^2) d\mu_0} \,\mu_0
\right\}=:\{\mu_0^R\} \\
&(ii)& \forall \alpha\in\nz, \quad \lim_{\varepsilon\to 0}\Tr[\varrho_\varepsilon^R \,\N^\alpha]=\int_\Z |z|^{2\alpha} \,d\mu_0^R(z)\,.
\end{eqnarray*}
Next, we use the notations
\begin{eqnarray*}
\varrho_\varepsilon(t)= e^{-i\frac{t}{\varepsilon}H_{\varepsilon}} \;\varrho_{\varepsilon} \;e^{i\frac{t}{\varepsilon}H_{\varepsilon}}
\quad \mbox{ and } \quad \varrho_{\varepsilon}^R(t)=e^{-i\frac{t}{\varepsilon}H_{\varepsilon}} \;\varrho^R_{\varepsilon} \; e^{i\frac{t}{\varepsilon}H_{\varepsilon}} \,.
\end{eqnarray*}
For any sequence $(\varepsilon_n)_{n\in\nz}$ there exists  by Proposition \ref{prop-1} a subsequence $(\varepsilon_{n_k})_{k\in\nz}$ and a family
of Borel probability measures  $(\mu_t^R)_{t\in\rz}$  such that
\begin{eqnarray*}
&(i)'& \mathcal{M}(\varrho_{\varepsilon_{n_k}}^R(t),k\in\nz)=\{\mu_t^R\} \\
&(ii)'& \forall \alpha\in\nz, \quad \lim_{k\to\infty}\Tr[\varrho_{\varepsilon_{n_k}}^R \,\N^\alpha]=
\int_\Z |z|^{2\alpha} \,d\mu_0^R(z)\,.
\end{eqnarray*}
Applying now Proposition \ref{pr.ball} with $(i)'-(ii)'$, we obtain that
\begin{eqnarray}
\label{subseq}
\mathcal{M}(\varrho_{\varepsilon_{n_k}}^R(t),k\in\nz)=\{({\bf F}_t)_*\mu_0^R\}\,,
\end{eqnarray}
for any time $t\in\rz$. Since for any  sequence $(\varepsilon_n)_{n\in\nz}$ we can extract a subsequence $(\varepsilon_{n_k})_{k\in\nz}$ such that
  (\ref{subseq}) holds we conclude that
\begin{eqnarray}
\label{cutwig}
\mathcal{M}(\varrho_{\varepsilon}^R(t),\varepsilon\in(0,\bar\varepsilon))=\{({\bf F}_t)_*\mu_0^R\}\,,
\end{eqnarray}
for any $R>0$ and $t\in\rz$. Again applying Proposition \ref{prop-1} for $(\varrho_\varepsilon)_{\varepsilon\in(0,\bar\varepsilon)}$, there
exists for any sequence $(\varepsilon_n)_{n\in\nz}$ a subsequence $(\varepsilon_{n_k})_{k\in\nz}$ and a family of Borel probability
measures $(\mu_t)_{t\in\rz}$ such that
\begin{eqnarray*}
\mathcal{M}(\varrho_{\varepsilon_{n_k}}(t), k\in\nz)=\{\mu_t\}\,.
\end{eqnarray*}
The identification of the measures $(\mu_t)_{t\in\rz}$ follows by a $\delta/3$ argument. For any
$b\in\S_{cyl}(\Z)$ based in $p\Z$, $p\in\mathbb{P}$, we write
\begin{eqnarray}\label{ap-t1}
\left|\Tr[\varrho_{\varepsilon_{n_k}}(t) b^{Weyl}]-\int_\Z b(z) d({\bf F}_t)_*\mu_0\right|
&\leq& \left|\Tr[\varrho_{\varepsilon_{n_k}}(t) b^{Weyl}]- \Tr[\varrho_{\varepsilon_{n_k}}^R(t) b^{Weyl}]\right| \\ \label{ap-t2}
&&+ \left|\Tr[\varrho^R_{\varepsilon_{n_k}}(t) b^{Weyl}]-\int_\Z b(z) d\mu^R_t\right| \\\label{ap-t3}
&&+\left|\int_\Z b({\bf F}_t(z)) d\mu_0^R-\int_\Z b({\bf F}_t(z)) d\mu_0\right|\,.
\end{eqnarray}
Each term (\ref{ap-t1})-(\ref{ap-t3}) can be made arbitrarily small by choosing $R$ and $k$ large enough and
respectively using the bound (\ref{eq.cutoffst}), the relation (\ref{cutwig}) and the dominated convergence theorem.
So, we conclude that $\mu_t=({\bf F}_t)_*\mu_0$ and hence we have proved
$$
\mathcal{M}(\varrho_\varepsilon,\varepsilon\in(0,\bar\varepsilon))=\{({\bf F}_t)_*\mu_0\}\,.
$$
Finally, the use of Proposition \ref{pr.eqPPI} with $\varrho_{\varepsilon}(t)$ yields
$$
\lim_{\varepsilon\to
  0}\Tr\left[\varrho_{\varepsilon}(t)b^{Wick}\right]=\int_{\Z} b\circ {\bf F}_t(z)~d\mu_{0}(z)\,,
$$
since $\lim_{\varepsilon\to\ 0} \Tr[\varrho_{\varepsilon}(t) \N^{\alpha}]=\lim_{\varepsilon\to\ 0} \Tr[\varrho_{\varepsilon} \N^\alpha]=
\int_\Z |z|^{2\alpha} d\mu_0=\int_\Z |z|^{2\alpha} d\mu_t$, for all $\alpha\in\nz$.
The reformulation of this result in terms of BBGKY hierarchy of reduced matrices is a consequence of Proposition
\ref{pr.wignergp0}. \fin

\subsection{Additional results}
\label{se.coro}
Although it was not written in Theorem \ref{th.main}, remember that
the existence of Wigner measures contains a result for Weyl observables.
\begin{cor}
\label{wigner-measure-id}
Let $(\varrho_{\varepsilon})_{\varepsilon\in(0,\bar\varepsilon)}$  be a family of normal states on $\H$ satisfying the hypothesis
of Theorem \ref{th.main}. \\
 The limit
\begin{eqnarray*}
\lim_{\varepsilon \to 0} {\rm Tr}[e^{-i\frac{t}{\varepsilon}
 H_{\varepsilon}}\;\varrho_{\varepsilon} \;e^{i\frac{t}{\varepsilon}
 H_{\varepsilon}} \; b^{Weyl} ]= \int_\Z b\circ\mathbf{F}_t(z) \; d\mu_0\,
\end{eqnarray*}
holds for any $b$ in  the cylindrical Schwartz space $\S_{cyl}(\Z)$\,,
 any $t\in\rz$ and any $b\in S^{\nu}_{p\Z}$, $\nu\in [0,1]$, $p\in\p$\,.
\end{cor}
The next result, shows that the class of observables can be extended
to functions of Wick-quantized symbols.
\begin{cor}
Let $(\varrho_{\varepsilon})_{\varepsilon\in(0,\bar\varepsilon)}$  be a family of normal states on $\H$ satisfying the hypothesis
of Theorem \ref{th.main}. Then
\begin{description}
\item[i)]
The limit
\begin{equation}
\label{eq.funcwick}
\lim_{\varepsilon \to 0} {\rm Tr}[e^{-i\frac{t}{\varepsilon}
 H_{\varepsilon}}\;\varrho_{\varepsilon} \;e^{i\frac{t}{\varepsilon}
 H_{\varepsilon}} \; f(b^{Wick})]= \int_\Z f(b\circ{\bf F}_t(z)) \; d\mu_0\,
\end{equation}
holds for any $f\in\mathcal{F}^{-1}(\mathcal{M}_b(\rz))$ and any
$b\in\P_{p,p}(\Z)$ such that $\tilde b^*=\tilde b$.
\item[ii)]
 If additionally $(\varrho_{\varepsilon})_{\varepsilon\in(0,\bar\varepsilon)}$ is a family of localized states
on a ball of radius $R>0$, then the limit \eqref{eq.funcwick}
 holds for any entire function $f(x)=\sum_{k=0}^\infty a_k x^k$ over $\cz$ and any
$b\in\P_{p,p}(\Z)$ such that $\tilde b^*=\tilde b$.
\end{description}
\end{cor}
\proof
\noindent\textbf{i)}
Let $\chi\in\mathcal{C}^0([0,\infty),\rz)$ be a continuous cutoff  function such that $0\leq\chi\leq 1$, $\chi(x)=1$ if $x\leq 1/2$ and
$\chi(x)=0$ if $x\geq 1$. Consider the family $(\varrho_{\varepsilon}(t)=e^{-i\frac{t}{\varepsilon}H_{\varepsilon}}\varrho_{\varepsilon}e^{i\frac{t}{\varepsilon}H^{_{\varepsilon}}})_{\varepsilon\in
(0,\bar \varepsilon)}$ with
$$
\varrho_\varepsilon^R=\frac{\chi(\N/R^2) \varrho_\varepsilon \chi(\N/R^2)}{\Tr[\chi(\N/R^2) \varrho_\varepsilon \chi(\N/R^2)]}\,,\quad R>0.
$$
Let $b\in\P_{p,p}(\Z)$ such that $\tilde b^*=\tilde b$, then $b^{Wick}$ extends to a
self-adjoint operator on $\H$ satisfying $[\N,b^{Wick}]=0$. We claim that
\begin{eqnarray}
\label{func-eq1}
\forall\theta\in\rz, \quad \Tr[\varrho^R_\varepsilon(t) e^{i \theta b^{Wick}}]
= \sum_{k=0}^\infty \frac{i^k}{k!} \theta^k \Tr[\varrho^R_\varepsilon(t) (b^{Wick})^k]\,.
\end{eqnarray}
Thanks to the estimate
\begin{eqnarray}
\label{func-eq2}
\left|\Tr[\varrho^R_\varepsilon(t) (b^{Wick})^k]\right|&=&\left|\Tr[\la \N\ra^{pk/2}\varrho^R_\varepsilon(t)\la \N\ra^{pk/2}  (\la \N\ra^{-p/2}b^{Wick}
\la \N\ra^{-p/2})^k]\right| \nonumber\\
&\leq& \la R\ra^{pk} \, |b|_{\P_{p,p}}^k\,,
\end{eqnarray}
the l.h.s of (\ref{func-eq1}) is an absolutely convergent series uniformly in $\varepsilon\in(0,\bar\varepsilon)$.  Moreover, on can easily
show the strong limit
$$
s-\lim_{N\to \infty} \sum_{k=0}^N \frac{i^k}{k!} \theta^k (b^{Wick})^k 1_{[0,R^2]}(\N)=e^{i\theta b^{Wick}}\,1_{[0,R^2]}(\N)\,.
$$
Therefore, we see that
$$
\sum_{k=0}^\infty \frac{i^k}{k!} \theta^k \Tr[\varrho^R_\varepsilon(t) (b^{Wick})^k ]=
\sum_{k=0}^\infty \frac{i^k}{k!} \theta^k \Tr[\varrho^R_\varepsilon(t) (b^{Wick})^k \,1_{[0,R^2]}(\N)]=
\Tr[\varrho^R_\varepsilon(t) e^{i \theta b^{Wick}}]\,.
$$
This proves (\ref{func-eq1}) and  again by the uniform estimate (\ref{func-eq2}) with respect to
$\varepsilon\in(0,\bar\varepsilon)$, we obtain
$$
\lim_{\varepsilon\to 0}\Tr[\varrho^R_\varepsilon(t) e^{i \theta b^{Wick}}]=\sum_{k=0}^\infty \frac{i^k}{k!} \theta^k \int_\Z
b({\bf F}_t(z))^k  d\mu_0=\int_\Z e^{-i\theta b({\bf F}_t(z))} d\mu_0\,.
$$
Now, a similar $\delta/3$ argument as in the proof of Theorem \ref{th.main}
\begin{eqnarray*}
\left|\Tr[\varrho_{\varepsilon_{n_k}}(t) e^{i \theta b^{Wick}}]-\int_\Z e^{i \theta b({\bf F}_t(z))} d\mu_0\right|
&\leq& \left|\varrho_{\varepsilon_{n_k}}-\varrho_{\varepsilon_{n_k}}^R \right|_{\L^1(\H)} \\
&&+ \left|\Tr[\varrho^R_{\varepsilon_{n_k}}(t) e^{i \theta b^{Wick}}]-\int_\Z e^{i \theta b({\bf F}_t(z))} d\mu^R_0\right| \\
&&+\left|\int_\Z e^{i \theta b({\bf F}_t(z))} d\mu_0^R-\int_\Z e^{i \theta b({\bf F}_t(z))} d\mu_0\right|\,,
\end{eqnarray*}
using the bound (\ref{eq.cutoffst}), the relation (\ref{cutwig}) and the dominated convergence theorem,  yields the limit
$$
\lim_{\varepsilon\to 0}\Tr[\varrho_{\varepsilon_{n_k}}(t) e^{i \theta b^{Wick}}]=\int_\Z e^{i \theta b({\bf F}_t(z))} d\mu_0\,.
$$
By integrating with respect to $\F(f)\in\mathcal{M}_b(\rz)$, we end the
proof.\\
\noindent\textbf{ii)} The proof is similar to (i). Indeed, one shows
\begin{eqnarray}
\label{func-eq3}
\Tr[\varrho_\varepsilon(t) f(b^{Wick})]
= \sum_{k=0}^\infty a_k \Tr[\varrho_\varepsilon(t) (b^{Wick})^k]\,,
\end{eqnarray}
with a l.h.s absolutely convergent series uniformly in $\varepsilon\in(0,\bar\varepsilon)$.
Letting $\varepsilon\to 0$ in (\ref{func-eq3}) yields the result.
\fin

\section{Examples}
\label{se.examples}
We review a series of examples. Firstly, the propagation of coherent
states and Hermite states is recalled.  Secondly, bounded interactions occur
naturally within the modelling of rapidly rotating Bose-Einstein
condensates, owing to some hypercontractivity property.
Thirdly, the tensor decomposition of the Fock space allows to specify some
Wigner measures for which the propagation cannot be translated in terms
of the reduced density matrices without writing all the BBGKY
hierarchy.
Finally, the result of Theorem~\ref{th.main} provides a new way to
consider the Hartree-von Neumann limit in the mean field regime.
\subsection{Coherent and Hermite states}
 The coherent states on the Fock space, $\Gamma_{s}(\Z)$ are  given by
$E(\xi)=W(\frac{\sqrt{2}}{i\varepsilon}
\xi)\Omega=e^{\frac{a^{*}(\xi)-a(\xi)}{\varepsilon}}\Omega$, where
$\Omega$ is the vacuum vector of $\Gamma_{s}(\Z)$, $\xi\in\Z$ and
$[a(f),a^{*}(g)]=\varepsilon\langle f,g\rangle\, I$\,. The Hepp method
(\cite{Hep}\cite{GiVe1}\cite{GiVe2})
consists in studying the propagation of squeezed coherent states
a slightly larger class which includes covariance deformations.
The normal state made with $E(\xi)$ is
$$
\varrho_\varepsilon(\xi)
:=
\big|W(\frac{\sqrt{2}}{i\varepsilon} \xi)\Omega\big\rangle\big\langle W(\frac{\sqrt{2}}{i\varepsilon} \xi)\Omega\big|\,.
$$
We proved in \cite{AmNi1} that
 $\mathcal{M}(\varrho_\varepsilon(\xi),\varepsilon\in(0,\bar\varepsilon))=\{\delta_\xi\}$ and
a simple computation shows that the property $(PI)$ is satisfied:
$$
\lim_{\varepsilon\to 0}\Tr[\varrho_\varepsilon \N^k]=|\xi|^{2k}=\delta_\xi(|z|^{2k})\,.
$$
A second example is given by Hermite states, also well studied within
the propagation of chaos technique or other works
(e.g., \cite{MaSh}\cite{BGM}\cite{ESY}). They are given by
\begin{equation}
\label{chaos}
\varrho_N(\varphi):=|\varphi^{\otimes N}\rangle\langle\varphi^{\otimes N}|\,,
\end{equation}
with $\varphi\in\Z$, $|\varphi|_\Z=1$ and discrete values for
$\varepsilon=\frac{1}{N}$.
 We know from \cite{AmNi1} that $\mathcal{M}(\varrho_N(\varphi),N\in\nz)=
\{\frac{1}{2\pi}\int_0^{2\pi}
\,\delta_{e^{i\theta}\varphi}\;d\theta\}$ where the rotation
invariance is the phase-space translation of the
gauge invariance of the Hermite states $\varphi\to e^{i\theta}\varphi$\,.
One easily checks the property $(PI)$:
$$
\lim_{N\to\infty}\Tr[\varrho_N(\varphi)\, \N^k]=1=\frac{1}{2\pi}\int_0^{2\pi} \,|z|^{2k}\delta_{e^{i\theta}\varphi}(z) \,d\theta\,.
$$
It is convenient to introduce a notation for this Wigner measure.
\begin{definition}
\label{de.deltaS}
  For $\varphi\in \Z$, the symbol $\delta_{\varphi}^{S^{1}}$ denotes
  the Borel probability measure
$$
\delta_{\varphi}^{S^{1}}=\frac{1}{2\pi}\int_{0}^{2\pi}\delta_{e^{i\theta}\varphi}~d\theta\,.
$$
\end{definition}
Theorem~\ref{th.main} applies and the Wigner measures associated with
$$
(e^{-i\frac{t}{\varepsilon}H_{\varepsilon}}\varrho_{\varepsilon}(\xi)
e^{i\frac{t}{\varepsilon}H_{\varepsilon}})_{\varepsilon\in(0,\bar\varepsilon)}
\quad \mbox{ and }  \quad(e^{-i\frac{t}{\varepsilon}H_{\varepsilon}}\varrho_{N}(\varphi)
e^{i\frac{t}{\varepsilon}H_{\varepsilon}})_{\varepsilon=1/N,\, N\in\nz^{*}}
$$
 are respectively  $\delta_{\xi_{t}}$ and
 $\delta^{S^{1}}_{\varphi_{t}}$, where $\xi_{t}$
or
 $\varphi_{t}$ evolves according to the classical flow.\\
For example when
$$H_{\varepsilon}=\d\Gamma(-\Delta)+\frac{1}{2}\int_{\rz^{2d}}V(x-y)a^{*}(x)a^*(y)a(x)a(y)~dxdy$$
with $\Z=L^{2}(\rz^{d})$ the classical flow is the Hartree equation
$$
i\partial_{t}\psi=-\Delta\psi+(V*|\psi|^{2})\psi\,.
$$
We conclude by noticing that for such states ($\varrho_N(\varphi)$ and $\varrho_\varepsilon(\xi)$) the
asymptotic one particle reduced
density matrix $\gamma^{(1)}_{0}(t)$ solves the equation
\begin{equation}
  \label{eq.VNeu}
\left\{
  \begin{array}[c]{ll}
    i\partial_{t}\gamma^{(1)}_{0}=\left[-\Delta+
      (V*n_{\gamma^{(1)}_{0}})\,,\gamma^{(1)}_{0}\right]&\\
   \gamma^{(1)}_{0}(t=0)=|\xi\rangle\langle\xi|\quad\text{for}\;
   \varrho_{\varepsilon}(\xi)\,,
&
\left(\text{resp.}\quad
\gamma^{(1)}_{0}(t=0)=|\varphi\rangle\langle\varphi|\quad\text{for}\;
   \varrho_{N}(\varphi)\,\right),
  \end{array}
\right.
\end{equation}
with $n_{\gamma^{(1)}_{0}}(x)=\gamma^{(1)}_{0}(x,x)$\,.

\subsection{LLL-mean field dynamics for rapidly rotating Bose-Einstein condensates}

The case of bounded interaction terms occurs exactly in the modelling
of rapidly rotating Bose-Einstein condensates in the
Lowest-Landau-Level (LLL) regime.
The (LLL) one particle states can be described (see
\cite{ABN}) within the Bargmann space
$$
\Z=\left\{f\in L^{2}(\cz_{\zeta_{1}}, e^{-\frac{|\zeta_{1}|^{2}}{h}}~L(d\zeta_{1}))\,,\quad \partial_{\bar
    \zeta_{1}}f=0 \right\}
$$
where  $L(d\zeta_{1})$ is the Lebesgue measure on $\cz$,  $h>0$
is a parameter which is small in the rapid rotation regime and where the norm on $\Z$ is given by
$$
|f|_{\Z}^{2}=\int_{\cz}|f(\zeta_{1})|^{2}e^{-\frac{|\zeta_{1}|^{2}}{h}}~\frac{L(d\zeta_{1})}{(\pi
    h)}=\frac{1}{\pi h}|u|_{L^{2}}^{2}\quad,\quad u(\zeta_{1})=f(\zeta_{1})e^{-\frac{|\zeta_{1}|^{2}}{2h}}\,.
$$
The multiparticle bosonic problem has been considered in \cite{LeSe}
and the (LLL)-model has been justified for the stationary states of such a
system not only in the mean field asymptotics. The $k$-particle states are elements of
$$
\bigvee^{k}\Z=\left\{F\in
  L^{2}(\cz_{\zeta}^{k}, e^{-\frac{|\zeta|^{2}}{h}}~L(d\zeta)))\,,\quad \partial_{\bar\zeta}F=0\,,\quad
F(\zeta_{\sigma(1)}\,\ldots, \zeta_{\sigma(k)})=F\,,\quad \forall
\sigma\in \mathfrak{S}_{k}\quad \right\}\,,
$$
with the norm
$$
|F|_{\bigvee^{k}\Z}^{2}=\int_{\cz^{k}}|F(\zeta)|^{2}~\frac{L(d\zeta)}{(\pi
h)^{k}}\,.
$$
With or without the symmetry condition, $\bigotimes^{k}\Z$ and
$\bigvee^{k}\Z$ are closed subspaces of $L^{2}(\cz_{\zeta}^{k},
e^{-\frac{|\zeta|^{2}}{h}}~L(d\zeta))$ and they are the image of the
orthogonal projection (add the symmetry for $\bigvee^{k}\Z$)
$$
(\Pi_{h}^{k}G)(\zeta)=\int_{\cz^{k}}e^{\frac{\zeta.\tau-|\tau|^{2}}{h}}G(\tau)~\frac{L(d\tau)}{(\pi
h)^{k}}\,.
$$
Within the modelling of rapidly rotating Bose-Einstein condensates,
the one particle kinetic energy term is
$A=h\zeta_{1}\partial_{\zeta_{1}}$
and it is associated with
$$
0\leq E_{kin}(f)=\langle f\,, h\zeta_{1}\partial_{\zeta_{1}}f\rangle_{\Z}\,.
$$
The standard  one particle nonlinear energy is given by
$$
\alpha\int_{\cz}|u|^{4}~L(d\zeta_{1})\,,\quad u(\zeta_{1})=f(\zeta_{1})e^{-\frac{|\zeta_{1}|^{2}}{2h}}\,.
$$
where $\alpha>0$ is another parameter provided by the physics
(see\cite{ABN}), but more general energies can be considered
\begin{equation}
  \label{eq.gennonlin}
E_{NL}(f)=\sum_{p=2}^{r}\alpha_{p}\int_{\cz}|u|^{2p}~L(d\zeta_{1})\,,
\quad u(\zeta_{1})=f(\zeta_{1})e^{-\frac{|\zeta_{1}|^{2}}{2h}}\,,\quad \alpha_{p}>0\,.
\end{equation}
The mean field Hamiltonian is thus given by
$$
h(f)=E_{kin}(f)+E_{NL}(f)=\langle f\,,\,  h\zeta_{1}\partial_{\zeta_{1}}f\rangle +
\sum_{p=2}^{r}\alpha_{p}\int_{\cz}|f(\zeta_{1})|^{2p}e^{-\frac{p|\zeta_{1}|^{2}}{h}}~
L(d\zeta_{1})\,.
$$
An important property of these nonlinear energies comes from the
hypercontractivity of the semigroup $(e^{-th\xi\partial_{\xi}})_{t\geq 0}$
proved in \cite{Car}
which can be written as
\begin{equation}
  \label{eq.hypcar}
|U|_{L^{2p}}\leq C_{p,h,d}|U|_{L^{2}}\quad\text{if}\quad
U(\zeta)=F(\zeta)e^{-\frac{|\zeta|^{2}}{2h}}\;,\quad F\in  \mathop{\otimes}^{k} \Z\,,\quad p\in [2,+\infty]\,.
\end{equation}
This implies that the nonlinear energy is a  norm continuous  polynomial
with respect to $f\in \Z$ and therefore the nonlinear mean field
equation
\begin{equation}
  \label{eq.hamBEC}
i\partial_{t}f=h\zeta_{1}\partial_{\zeta_{1}}f+\sum_{p=2}^{r}p\alpha_{p}\Pi_{h}^{1}(|u|^{2(p-1)} u)\Pi_{h}^{1}f
\end{equation}
defines a nonlinear flow on the phase-space $\Z$ according to
Subection~\ref{se.polapprox} (we refer the reader to \cite{Nie} for a
more detailed analysis of the nonlinear dynamics of the LLL-model)\,.
\\
Let us consider the second quantized version $H_{\varepsilon}$ of the
energy $h$ in $\Gamma_s(\Z)$.
The kinetic energy is nothing but $\d\Gamma(A)$:
$$
\d\Gamma(A)\big|_{\bigvee ^{k}\Z}=\varepsilon\sum_{j=1}^{k}h\zeta_{j}
\partial_{\zeta_{j}}= \varepsilon h\zeta.\partial_{\zeta}\,.
$$
and the quantum Hamiltonian $H_{\varepsilon}$ is then
\begin{eqnarray}
  \label{eq.quaBEC}
&&H_{\varepsilon}=\d\Gamma(A)+\sum_{p=2}^{r}\alpha_{p}Q_{p}^{Wick}
\\
\label{eq.QpBEC}
\text{with}
&&
Q_{p}(f)=\int_{\cz}|u(\zeta_{1})|^{2p}~L(d\zeta_{1})=\int_{\cz}|f(z)|^{2p}e^{-\frac{p|\zeta_{1}|^{2}}{h}}~L(d\zeta_{1})=\langle
f^{\otimes p}\,,\tilde{Q}_{p}f^{\otimes p}\rangle\,.
\end{eqnarray}
The operator $\tilde{Q_{p}}$ is easily identified after removing the
center of mass in multiple integrals (see \cite{LeSe} for details) as
$$
\tilde{Q}_{p}F(\zeta)=\Pi_{p}^{h}\left(\left[\prod_{j=1}^{p-1}
\delta(\zeta_{j}')\right]F\right)(\zeta)
= \frac{1}{(\pi h)^{p}}F\left(\frac{\zeta_{1}+\cdots+\zeta_{p}}{p},\ldots,\frac{\zeta_{1}+\cdots+\zeta_{p}}{p}\right)
$$
with
$\zeta_{j}'=\zeta_{j}-\frac{\zeta_{1}+\cdots+\zeta_{p}}{p}$\,. One
easily checks  as well, by using additionally the hypercontractivity
estimate
\eqref{eq.hypcar} with $p=+\infty$,
that $\tilde Q_p\in \L(\bigvee^p\Z)$.

The propagation result of Theorem~\ref{th.main} applies for such a model for all initial
states which fulfill its assumptions (boundedness of all moments and
condition $(PI)$).

\subsection{Fock tensorization}
\label{se.focktensor}
We have already used, and it is the basis of the introduction of
cylindrical observables, the fact that
$\Gamma_{s}(\Z)\sim \Gamma_{s}(\Z_1)\otimes \Gamma_{s}(\Z_{2})$
when $\Z=\Z_{1}\displaystyle\mathop{\oplus}^{\perp}\Z_{2}$\,.
The definition of Wigner measures introduced via cylindrical
observables, yields the next result.
\begin{lem}
  \label{le.focktensor}
Assume $\Z=\Z_{1}\displaystyle\mathop{\oplus}^{\perp}\Z_{2}$ and let
$(\varrho^{1}_{\varepsilon})_{\varepsilon\in (0,\bar\varepsilon)}$,
$(\varrho^{2}_{\varepsilon})_{\varepsilon\in (0,\bar\varepsilon)}$ be
two families of normal states on $\Gamma_{s}(\Z_{1})$ and
$\Gamma_{s}(\Z_{2})$ such that
$\Tr\left[\varrho_{\varepsilon}^{\ell}{\bf N}_{\ell}^{\delta}\right]\leq
C_{\delta}$ holds uniformly for some $\delta>0$ and
$\mathcal{M}(\varrho^{\ell}_{\varepsilon}\,,~\varepsilon\in
(0,\bar\varepsilon))=\left\{\mu^{\ell}\right\}$ for $\ell=1,2$\,. Let
$\varrho_{\varepsilon}$ be  the
state on $\Gamma_{s}(\Z)$ identified with
$\varrho^{1}_{\varepsilon}\otimes
\varrho^{2}_{\varepsilon}$ in the
decomposition $\Gamma_{s}(\Z)\sim \Gamma_{s}(\Z_1)\otimes
\Gamma_{s}(\Z_{2})$. Then the family
$(\varrho_{\varepsilon})_{\varepsilon\in (0,\bar\varepsilon)}$  admits the unique Wigner measure
$\mu=\mu^{1}\times \mu^{2}$ on the phase space $\Z=\Z_{1}\times \Z_{2}$\,.
\end{lem}

Before giving applications and variations on this result it is worth
to notice that the identification of the ``tensor'' state $\varrho_{\varepsilon}$
requires some care. It is not equal in general to
$\varrho^{1}_{\varepsilon}\otimes \varrho^{2}_{\varepsilon}$ since such
a states does not preserve the symmetric Fock space
$\Gamma_{s}(\Z)$\,.\\
Here is a simple example, take $\varphi_{1}\in \Z_{1}$ and
$\varphi_{2}\in \Z_{2}$ with $|\varphi_{\ell}|_{\Z_{\ell}}=1$, $N_{1}\,,
N_{2}\in\nz$, and set $\varrho^{\ell}=|\varphi_{\ell}^{\otimes
  N_{\ell}}\rangle \langle \varphi_{\ell}^{\otimes N_{\ell}}|$ for
$\ell=1,2$\,.
The tensor states $\varrho^{1}\otimes \varrho^{2}$  is the pure state
 $|\varphi_{1}^{\otimes N_{1}}\otimes \varphi_{2}^{\otimes
  N_{2}}\rangle\langle \varphi_{1}^{\otimes N_{1}}\otimes \varphi_{2}^{\otimes
  N_{2}}|$ in $\Gamma_{s}(\Z_{1})\otimes \Gamma_{s}(\Z_{2})$\,.
It suffices to identify the vector $\varphi^{\vee(N_{1},N_{2})}\in
\Gamma_{s}(\Z)$ associated with
$\varphi_{1}^{\otimes N_{1}}\otimes \varphi_{2}^{\otimes N_{2}}$\,.
It is the symmetric vector in
$\bigvee^{N_{1}+N_{2}}\Z$  made with $N_{1}$-times
$\varphi_{1}$ and $N_{2}$-times $\varphi_{2}$ and we can summarize the
situation with
\begin{eqnarray*}
\varphi_{\ell}^{\otimes
    N_{\ell}}&=&\frac{1}{\sqrt{\varepsilon^{N_{\ell}}N_{\ell}!}}
\stackrel{N_{\ell}~\text{times}}{a^{*}(\varphi_{\ell})\ldots
  a^{*}(\varphi_{\ell})}|\Omega_{\ell}\rangle
\quad \text{in}~\Gamma_{s}(\Z_{\ell})\,,\quad \ell=1,2,\\
\varphi^{\vee(N_{1},N_{2})}
&=&
 \sqrt{\frac{(N_{1}+N_{2})!}{\varepsilon^{(N_{1}+N_{2})} N_1! N_2!}}
\mathcal{S}_{N_{1}+N_{2}}(\varphi_{1}^{\otimes N_{1}}\otimes
\varphi_{2}^{\otimes N_{2}})\\
&=&
\frac{1}{\sqrt{ \varepsilon^{N_{1}+N_{2}}N_1!N_2!}}
\stackrel{N_{1}~\text{times}}{a^{*}(\varphi_{1})\ldots
  a^{*}(\varphi_{1})}
\stackrel{N_{2}~\text{times}}{a^{*}(\varphi_{2})\ldots
  a^{*}(\varphi_{2})}|\Omega\rangle \quad\text{in}~\Gamma_{s}(\Z)\,.
\end{eqnarray*}
The tensor decomposition is especially useful when $\Z$ is endowed
with a Hilbert basis $(e_{j})_{j\in \nz^{*}}$\,. An Hilbert basis of
$\Gamma_{s}(\Z)$ is $(e^{\vee
  \alpha})_{\alpha\in\cup_{j=0}^{\infty}(\nz^{*})^{j}}$ given by:
$$
e^{\vee  \alpha}=\sqrt{\frac{|\alpha|!}{\alpha!}}\,\mathcal{S}_{|\alpha|}\left(e^{\otimes\alpha}\right)=
\frac{1}{\sqrt{\varepsilon^{|\alpha|}|\alpha|!}}\,[a^{*}(e)]^{\alpha}|\Omega\rangle
$$
with a natural multi-index notation $\alpha=(\alpha_{1},\ldots,
\alpha_{k})$, $|\alpha|=\alpha_{1}+\cdots+ \alpha_{k}$\,,
$e^{\otimes \alpha}= e_{1}^{\otimes \alpha_{1}}\otimes \cdots\otimes
e_{k}^{\alpha_{k}}$\, and
$$
[a^{*}(e)]^{\alpha}=a^{*}(e_{1})^{\alpha_{1}}\ldots
a^{*}(e_{k})^{\alpha_{k}}\,.
$$
For example, the identification between $\Gamma_{s}(\cz e_{1})\otimes
\Gamma_{s}((\cz e_{1})^{\perp})$ and $\Gamma_{s}(\Z)$ is done via the
mapping defined by
$e^{\bigvee \alpha_{1}}_{1}\otimes e^{\bigvee\alpha'}\to e^{\bigvee
  (\alpha_{1},\alpha')}$, for all $\alpha_{1}\in \nz$ and all
$\alpha'\in
\cup_{k=0}^{\infty}(\nz\setminus\left\{0,1\right\})^{k}$\,.
This can be iterated but remember that the definition of infinite tensor products
requires the additional specification of one vector per component
which is hopefully rather
canonical for Fock spaces endowed with a vacuum vector (see
\cite{Gui})\,.\\
Below is a notation convenient to the definition of tensor states and
which allows some extensions.
Consider the linear isometry $C_{j}$ on $\H=\Gamma_{s}(\Z)$ defined
by its action on the Hilbert basis $(e^{\vee \alpha})_{\alpha\in \cup_{k=0}^{\infty}(\nz^{*})^{k}}$
\begin{equation}
  \label{eq.defCj}
  C_{j}e^{\vee \alpha}= \frac{1}{|a^{*}(e_{j}) e^{\vee \alpha}|}a^{*}(e_{j}) e^{\vee \alpha}=\frac{1}{\sqrt{\varepsilon (\alpha_j+1)}}
  \,a^{*}(e_{j}) e^{\vee \alpha}=e^{\vee (\alpha+1_{j})}\,,
\end{equation}
with $|1_j|=1$ and $(1_j)_j=1$\,.
In the tensor decomposition $\Gamma_{s}(\Z)\sim
\Gamma_{s}(\cz e_{j})\otimes \Gamma_{s}((\cz e_{j})^{\perp})$, this
isometry $C_{j}$ is nothing but the tensor product
$\left[\frac{1}{\sqrt{{\bf N}_{j}}}a^{*}(e_{j})\right]\otimes I$\,.
\begin{definition}
\label{de.Ej}
  Let $\Z$ be endowed with a Hilbert basis $(e_{j})_{j\in\nz^{*}}$,
  for $j\in \nz^{*}$, and take the isometries
$(C_{j})_{j\in\nz^{*}}$ defined in $\H$ by \eqref{eq.defCj}.
For $j\in\nz^{*}$, the operator $E_{j}$ is defined on
  ${\cal L}^{1}(\H)$ by
$$
E_{j}\varrho=C_{j}\varrho
C_{j}^{*}\,,\quad \forall \varrho\in \mathcal{L}^{1}(\H)\,.
$$
For $\lambda=(\lambda_{j})_{j\in\nz^{*}}\in \ell^{1}([0,+\infty))$
such that $\sum_{j=1}^{\infty}\lambda_{j}=1$, the notation
$\lambda.E$ means
$$
\lambda.E=\sum_{j=1}^{\infty}\lambda_{j}E_{j}\,.
$$
\end{definition}
The operators $E_{j}$ and $\lambda.E$ transform normal states on
$\bigvee^{k-1}\Z$ into normal states on $\bigvee^{k}\Z$ and they all
commute. After taking $\varphi_{1}=e_{1}$ and $\varphi_{2}=e_{2}$ the
tensor state on $\Gamma_s(\Z)$ identified with $\varrho^{1}\otimes \varrho^{2}$ and studied above with
$\Z_{1}=\cz e_{1}$ and $\Z_{2}=(\cz e_{1})^{\perp}$ is nothing but
$$
E^{(N_{1},N_{2})}|\Omega\rangle\langle \Omega|=
E_{1}^{N_{1}}E_{2}^{N_{2}}|\Omega\rangle\langle \Omega|
=
E_{2}^{N_{2}}E_{1}^{N_{1}}|\Omega\rangle\langle \Omega|
\,.
$$
Moreover the multinomial formula holds
\begin{equation}
  \label{eq.multinom}
(\lambda.E)^{N}=\sum_{|\alpha|=N}\frac{N!}{\alpha!}\lambda^{\alpha}E^{\alpha}\,.
\end{equation}
We use these notion to formulate the propagation of nontrivial Wigner
measures. The Hamiltonian  is
$$
H_{\varepsilon}=\d\Gamma(A)+\left(\sum_{j=2}^{r}\langle z^{\otimes
  j}\,,\tilde{Q}_{j}z^{\otimes j}\rangle\right)^{Wick}\quad, \quad \varepsilon=\frac{1}{N}
$$
with $(A,\D(A))$ self-adjoint and $\tilde{Q}_{j}=\tilde{Q}_{j}^{*}\in \mathcal{L}(\bigvee^{j}\Z)$\,.
It is associated with the mean field Hamiltonian
$$
h(z,\bar z)=\langle z\,,\,A z\rangle+\sum_{j=2}^{r}Q_{j}(z)
$$
and the flow $({\bf F}_t)_{t\in\rz}$ in the phase space $\Z$\,.
\begin{prop}
  \label{pr.focktensor}
Let $\Z$ be endowed with an orthonormal basis $(e_{j})_{j\in\nz^{*}}$
and let the family $(E_{j})_{j\in\nz^{*}}$ be as in
Definition~\ref{de.Ej}. Once $\varrho_{\varepsilon}(0)$ is fixed
$\varrho_{\varepsilon}(t)$ is defined by $\varrho_{\varepsilon}(t)=e^{-i\frac{t}{\varepsilon}H_{\varepsilon}}\varrho_{\varepsilon}(0)e^{i\frac{t}{\varepsilon}H_{\varepsilon}}$\,.
\\
\noindent\textbf{1)} For $k\in\nz^{*}$ and $(\nu_{1},\ldots, \nu_{k})\in
[0,1]^{k}$ fixed such that $\sum_{\ell=1}^{k}\nu_{\ell}=1$,  assume that
$N_{\ell}$ equals the integer part $[\nu_{\ell}N]$
for $\ell\in \left\{1,\ldots, k\right\}$. Then the family of states
$(\varrho_{\varepsilon}(t))_{\varepsilon=1/N}$ given by
$\varrho_{\varepsilon}(0)=E^{(N_{1},\ldots,
  N_{k})}|\Omega\rangle\langle
\Omega|$
 admits a
unique Wigner measure
$$
\mu_{t}=({\bf F}_{t})_{*}\mu_{0}=({\bf F}_{t})_{*}(\delta_{\sqrt{\nu_{1}}e_{1}}^{S^{1}}\times\cdots\times
\delta_{\sqrt{\nu_{k}}e_{k}}^{S^{1}})\,.
$$
The reduced density matrices $\gamma^{(p)}_{\varepsilon}(t)$ converge
in $\mathcal{L}^{1}(\bigvee^{p}\Z)$ to
\begin{equation}
  \label{eq.convgamma}
\gamma_{0}^{(p)}(t)=\int_{\Z}|z_{t}^{\otimes p}\rangle\langle z_{t}^{\otimes p}|~d\mu_{0}(z)
\end{equation}
by setting $z_{t}={\bf F}_{t}z$\,.\\
\noindent\textbf{2)} Let $\lambda=(\lambda_{j})_{j\in\nz^*}\in
\ell^{1}([0,+\infty))$ be such that
$\sum_{j=1}^{\infty}\lambda_{j}=1$.  Then the family of states
$(\varrho_{\varepsilon}(t))_{\varepsilon=1/N}$ given by
$\varrho_{\varepsilon}=(\lambda.E)^{N}|\Omega\rangle\langle \Omega|$
satisfies the same properties as above with
$$
\mu_{0}=\mathop{\times}_{j=1}^{\infty}\delta_{\sqrt{\lambda_{j}}e_{j}}^{S^{1}}\,.
$$
\end{prop}
\proof
Actually it suffices to identify the measure $\mu_{0}$ and to check the
assumptions of Theorem~\ref{th.main} at time $t=0$\,.\\
\noindent\textbf{1)} It is a simple application of
Lemma~\ref{le.focktensor} with the decomposition
$$
\Gamma_{s}(\Z)\sim\Gamma_{s}(\cz e_{1})\otimes \cdots\otimes
\Gamma_{s}((\cz e_{k-1}))\otimes \Gamma_{s}((\cz
e_{1}\oplus\cdots\oplus \cz e_{k-1})^{\perp})\,.
$$
In this decomposition $E^{(N_{1},\ldots,N_{k})}|\Omega\rangle\langle
\Omega|$ is nothing but a tensor product of Hermite states.
$|e_{\ell}^{\otimes N_{\ell}}\rangle\langle e_{\ell}^{\otimes
  N_{\ell}}|$ and the result is a simple tensorization of the result
for Hermite states  with $\varepsilon=\frac{\nu_{\ell}}{N_{\ell}}$\,.\\
\noindent\textbf{2)} The state $\varrho_{\varepsilon}(0)=(\lambda.E)^{N}|\Omega\rangle\langle
\Omega |$ belongs to $\mathcal{L}^{1}(\bigvee^{N}\Z)$\,. It is
therefore localized in the ball with radius $1$\,. According to
Proposition~\ref{pr.moment}, its Wigner measures are completely
determined if we know the limits of
$$
\Tr\left[\varrho_{\varepsilon}(0)b^{Wick}\right]
$$
for all the $b\in \P_{alg}^{\infty}(\Z)$\,.
Due to Pythagorean summation, the measure
$\mu_{0}=\mathop{\times}_{j=1}^{\infty}\delta_{\sqrt{\lambda_{j}}e_{j}}^{S^{1}}$
is supported in the ball of radius $1$.
The estimates
\begin{eqnarray*}
  &&
\left|\Tr\left[\varrho_{\varepsilon}(0) (b-b')^{Wick}\right]\right|
=
\left|\Tr\left[\varrho_{\varepsilon}(0) \chi({\bf
      N})(b-b')^{Wick}\chi({\bf N})\right]\right|
\leq C_{p,q}|b-b'|_{\P_{p,q}}\,,\\
\text{and}
&&
\left|\int_{\Z} (b(z)-b'(z))~d\mu_{0}(z)\right|=
\left|\int_{\Z} (b(z)-b'(z))\chi^{2}(|z|^{2})~d\mu_{0}(z)\right|=C_{p,q}|b-b'|_{\P_{p,q}}\,,
\end{eqnarray*}
with the first one deduced from the number estimate
\eqref{eq.2bis} in Proposition~\ref{wick-estimate},
hold for all $b,b'\in \P_{p,q}^{\infty}(\Z)$, $p,q\in\nz$ as soon as
$\chi\in \mathcal{C}^{\infty}_{0}([0,+\infty))$ is chosen
 such that $\chi\equiv 1$ on $[0,1]$\,.
Hence it suffices to prove $\lim_{\varepsilon\to
  0}\Tr\left[\varrho_{\varepsilon}(0)b^{Wick}\right]=
\int_{\Z}b(z)~d\mu_{0}(z)$ for a total set of
$\P_{alg}^{\infty}(\Z)$\,.
With the compact kernel condition, any $\tilde{b}\in
\mathcal{L}^{\infty}(\bigvee^{p}\Z, \bigvee^{q}\Z)$ can be approximated
by a linear combination of rank one operators of the form $|e^{\vee \gamma}\rangle\langle
e^{\vee \beta}|=\sqrt{\frac{\beta !\gamma!}{|\beta|! |\gamma|!}}{\cal S}_{|\gamma|}|e^{\otimes \gamma}\rangle\langle
e^{\otimes \beta}|{\cal S}_{|\beta|}$, $|\beta|=p$, $|\gamma|=q$\,. With
\begin{eqnarray*}
&&
( \langle z^{\otimes
  q}\,,\, e^{\otimes \gamma}\rangle\langle e^{\otimes \beta}\,,\, z^{\otimes
  p}\rangle)^{Wick}=
[a^{*}(e)]^{\gamma}\,[a(e)]^{\beta}
\\
\text{and}
&&
\varrho_{\varepsilon}(0)=\sum_{|\alpha|=N}\lambda^\alpha \,\frac{N!}{\alpha!}
|e^{\vee \alpha}\rangle\langle e^{\vee \alpha}|\,,
\end{eqnarray*}
we can compute directly
$$
\Tr\left[\varrho_{\varepsilon}(0)(\langle z^{\otimes q}\,,\,e^{\otimes \gamma}\rangle\langle
e^{\otimes\beta}\,,\, z^{\otimes p}\rangle)^{Wick}\right]
=
\sum_{|\alpha|=N}\frac{N!}{\alpha
  !}\lambda^{\alpha}
\langle a(e)^{\gamma}e^{\vee \alpha}\,,a(e)^{\beta}e^{\vee \alpha}\rangle\,.
$$
Actually
$$
a(e)^{\beta}e^{\vee \alpha}=
\left\{
  \begin{array}[c]{ll}
    \sqrt{\varepsilon^{p}\frac{\alpha!}{\alpha'!}}e^{\vee \alpha'}
&\text{if}\quad \alpha=\alpha'+\beta\,,\\
0 &\text{else}\,,
  \end{array}
\right.
$$
with a similar identity for $\gamma$ yields
\begin{eqnarray*}
\Tr\left[\varrho_{\varepsilon}(0)(\langle z^{\otimes q}\,,\,e^{\otimes \gamma}\rangle\langle
e^{\otimes \beta}\,,\, z^{\otimes p}\rangle)^{Wick}\right]
&=&
\delta_{\beta,\gamma}\varepsilon^{^{p}}
\frac{N!}{(N-p)!}
\left(
\sum_{|\alpha'|=N-p}\frac{(N-p)!}{\alpha'
  !}\lambda^{\alpha'}\right) \lambda^{\beta}
\\
&=&
\delta_{\beta,\gamma}\varepsilon^{p}N(N-1)\ldots(N-p+1)\lambda^{\beta}\,.
\end{eqnarray*}
With $\varepsilon=1/N$ and $(p,q)$ fixed, we obtain
$$
\Tr\left[\varrho_{\varepsilon}(0)(\langle z^{\otimes q}\,,\,e^{\otimes \gamma}\rangle\langle
e^{\otimes \beta}\,,\, z^{\otimes p}\rangle)^{Wick}\right]
=\delta_{\beta,\gamma}\lambda^{\beta}=\int_{\Z}
\langle z^{\otimes q}\,,\,e^{\otimes \gamma}\rangle\langle
e^{\otimes \beta}\,,\, z^{\otimes p}\rangle~d\mu_{0}(z)\,.
$$
\fin

We conclude with two remarks:
\begin{itemize}
\item The tensorized Hermite state $E^{(N_{1},\ldots
    N_{\ell}\ldots)}|\Omega\rangle \langle\Omega|$ with
   $N_{\ell}=[\lambda_{\ell}N]$ and  $\sum_{j=1}^{\infty}\lambda_{j}=1$
  can
  be studied and behaves asymptotically like
  $(\lambda.E)^{N}|\Omega\rangle \langle\Omega|$\,.
\item When those tensor states are not Hermite states, the reduced
  density matrices satisfy no closed equation and all the hierarchy
  has to be considered. In the example leading to \eqref{eq.VNeu} for
  Hermite states the general equation for $\gamma^{(1)}_{0}(t)$ writes
$$
i\partial_{t}\gamma_{0}^{(1)}(x,y)=[-\Delta, \gamma_{0}^{(1)}](x,y)
+ \int_{\rz^{d}}V(x-x')\gamma_{0}^{(2)}(x',x,x',y)-\gamma_{0}^{(2)}(x',x,x',y)V(y-x')~dx'\,,
$$
and the equation for $\gamma^{(2)}_{0}$ involves $\gamma^{(3)}_{0}$
and so on\ldots The propagation of Wigner measures gathers all the
asymptotic information in this case. Geometrically it is interesting
to notice that if the  initial Wigner
measure is
$\delta_{\sqrt{\lambda_{1}}e_{1}}^{S^{1}}\times\delta_{\sqrt{\lambda_{2}}e_{1}}^{S^{1}}$,
with $\lambda_{1}+\lambda_{2}=1$, it is supported by a 2-dimensional
torus. After the action of the continuous flow, the support of
$\mu_{t}$ remains topologically a 2-dimensional torus but in general
deformed in the infinite dimensional phase space with no exact
finite dimensional reduction.
\end{itemize}

\subsection{Condition $(PI)$ for Gibbs states}
\label{se.PIGibbs}

For $\sigma(\varepsilon)\in \mathcal{L}^{1}(\Z)$, which is a non negative
strict contraction:
$$
\sigma(\varepsilon)=\sum_{i=1}^{\infty}\sigma_{i}(\varepsilon)|e_{i}(\varepsilon)\rangle \langle
e_{i}(\varepsilon)|\quad,\quad  0\leq \sigma_{i}(\varepsilon)<
1\,,\quad \sum_{i=1}^{\infty}\sigma_{i}(\varepsilon)
<+\infty\,,
$$
where $(e_{i}(\varepsilon))_{i\in\nz^{*}}$ is a Hilbert basis of $\Z$, the operator
$\Gamma(\sigma(\varepsilon))$ belongs to $\mathcal{L}^{1}(\H)$\,. It equals
$\Gamma(\sigma(\varepsilon))=\sum_{n=0}^{\infty}\mathcal{S}_{n}(\sigma(\varepsilon))^{\otimes
  n}\mathcal{S}_{n}$ and the tensor decomposition gives
$$
\Tr\left[\Gamma(\sigma(\varepsilon))\right]=\prod_{i=1}^{\infty}\frac{1}{1-\sigma_{i}(\varepsilon)}\in \rz_{+}\,.
$$
Hence we can consider the quasi-free state
$$
\varrho_{\varepsilon}=
\frac{1}{\Tr\left[\Gamma(\sigma(\varepsilon))\right]}
\Gamma(\sigma(\varepsilon))\,.
$$
It is more convenient to write
$$
\sigma_{i}(\varepsilon)=\frac{\nu_{i}(\varepsilon)}{\nu_{i}(\varepsilon)+\varepsilon}\quad
\text{with}\quad
\nu_{i}(\varepsilon)\in [0,+\infty)\,,
$$
and the condition $\sum_{i=1}^{\infty}\sigma_{i}(\varepsilon)<
+\infty$ is equivalent to $\sum_{j=1}\nu_{i}(\varepsilon)< +\infty$\,.
\begin{lem}
  \label{le.gamsig}
For $\sigma(\varepsilon)=\sum_{i=1}^{\infty}
\frac{\nu_{i}(\varepsilon)}{\nu_{i}(\varepsilon)+\varepsilon}|e_{i}(\varepsilon)\rangle\langle
e_{i}(\varepsilon)|\in \mathcal{L}^{1}(\Z)$, the quasi-free state
$\varrho_{\varepsilon}=\frac{1}{\Tr\left[\Gamma(\sigma(\varepsilon))\right]}
\Gamma(\sigma(\varepsilon))$ satisfies
$$
\forall k\in\nz\,,
\sup_{\varepsilon\in
  (0,\bar\varepsilon)}\Tr\left[\varrho_{\varepsilon}{\bf N}^{k}\right]
< +\infty
$$
if and only if there exists $C>0$ such that
$\sum_{i=1}^{\infty}\nu_{i}(\varepsilon)\leq C$\,.
In such a case,  the quantity $\Tr\left[\varrho_{\varepsilon}{\bf
    N}^{k}\right]$, $k\in\nz$, is equivalent to
$$
k!\sum_{|\alpha|=k}\nu(\varepsilon)^{\alpha}
$$
as $\varepsilon\to 0$,
with the usual multi-index convention,
$\nu(\varepsilon)^{\alpha}=\prod_{k=1}^{\infty}\nu_{k}(\varepsilon)^{\alpha_{k}}$\,.
\end{lem}
\proof
Consider for $x\in [-c,c]$, $c>0$,  the quantity
$$
\Tr\left[\varrho_{\varepsilon}(1+\varepsilon x)^{\frac{{\bf
        N}}{\varepsilon}}\right]
=
\frac{
\prod_{i=1}^{\infty}\frac{1}{1-\frac{\nu_{i}(\varepsilon)}{\nu_{i}+\varepsilon}
(1+\varepsilon x)}}{
\prod_{i=1}^{\infty}\frac{\nu_{i}(\varepsilon)+\varepsilon}{\varepsilon}}
=
\prod_{i=1}^{\infty}\frac{1}{1-\nu_{i}(\varepsilon)x}
$$
When $\Tr\left[\varrho_{\varepsilon}\N^{k}\right]$ is uniformly bounded
w.r.t $\varepsilon\in (0,\bar\varepsilon)$, for all
$k\in\nz$ it is a $\mathcal{C}^{\infty}$ function around $x=0$ with
$$
\partial_{x}^{k}\Tr\left[\varrho_{\varepsilon}(1+\varepsilon x)^{\frac{{\bf
        N}}{\varepsilon}}\right]\big|_{x=0}
=\Tr\left[\varrho_{\varepsilon}{\bf N}({\bf N}-\varepsilon)\ldots({\bf
  N}-(k-1)\varepsilon)\right]\sim
\Tr\left[\varrho_{\varepsilon}{\bf N}^{k}\right]\quad\text{as}\,
\varepsilon\to 0\,.
$$
But the first derivative is nothing but
$$
\partial_{x}\Tr\left[\varrho_{\varepsilon}(1+\varepsilon x)^{\frac{{\bf
        N}}{\varepsilon}}\right]\big|_{x=0}=
\sum_{i=1}^{\infty}\nu_{i}(\varepsilon)\,,
$$
which says that the uniform bound
$\sum_{i=1}^{\infty}\nu_{i}(\varepsilon)\leq C$  is a necessary
condition.\\
Reciprocally when $\sum_{i=1}^{\infty}\nu_{i}(\varepsilon)\leq C$,
then the function $\prod_{i=1}^{\infty}(1-\nu_{j}(\varepsilon)x)^{-1}$ is
analytic with respect to $x$ in a disc of radius $R_{C}$ and equals
$$
\prod_{i=1}^{\infty}(1-\nu_{j}(\varepsilon)x)^{-1}=\prod_{i=1}^{\infty}(\sum_{j=0}^{\infty}\nu_{i}(\varepsilon)^{j}x^{j})
=\sum_{k=0}^{\infty}x^{k}\left[\sum_{|\alpha|=k}\nu(\varepsilon)^{\alpha}\right]\,,
$$
which yields the result.
\fin

\noindent A Gibbs state is a quasi-free state with
$\sigma(\varepsilon)=e^{-\varepsilon L(\varepsilon)}$ where
$L(\varepsilon)$ is a strictly positive operator assumed here with a
discrete spectrum:
\begin{equation}
  \label{eq.defL}
L(\varepsilon)=\sum_{i=1}^{\infty}\ell_{i}(\varepsilon)|e_{i}\rangle
\langle e_{i}|\,,\quad \ell_{i}(\varepsilon)\leq \ell_{i+1}(\varepsilon)\,,
\end{equation}
where the  basis $(e_{j})_{j\in\nz^{*}}$ is
assumed independent of $\varepsilon\in (0,\bar\varepsilon)$  for the
sake of simplicity\,. There is a simple traduction of the assumptions
of Theorem~\ref{th.main}, the non obvious one being the
condition~$(PI)$ hidden in the assumption~\eqref{eq.hypconv}.
\begin{prop}
  The Gibbs state
  $\varrho_{\varepsilon}=\frac{1}{\Tr\left[\Gamma(e^{-\varepsilon
        L(\varepsilon)})\right]}\Gamma(e^{-\varepsilon
    L(\varepsilon)})$ with $L(\varepsilon)$ given in \eqref{eq.defL}
  satisfies the assumptions of Theorem~\ref{th.main} if and only if
  \begin{itemize}
  \item For all $i\in\nz^{*}$ the limit $\lim_{\varepsilon\to
      0}\ell_{i}(\varepsilon)=\ell_{i}(0)$ exists in $(0,+\infty]$\,.
\item If $J\in \nz^{*}\cup\left\{\infty\right\}$ denotes the largest
  element in $\nz^{*}\cup\left\{\infty\right\}$ such that
  $\ell_{i}(0)<+\infty$ for all $i\leq J$, the two conditions are
  verified
  \begin{eqnarray}
\label{eq.elliJ}
&&    \sum_{i=1}^{J}\frac{1}{\ell_{i}(0)}<+\infty\,,\\
\label{eq.Jelli}
\text{and}&&
\lim_{\varepsilon\to 0}\sum_{i> J}\frac{\varepsilon
  e^{-\varepsilon\ell_{i}(\varepsilon)}}{(1-e^{-\varepsilon \ell_{i}(\varepsilon)})}=0\,.
  \end{eqnarray}
  \end{itemize}
\end{prop}
\proof
First of all, writing $\sigma(\varepsilon)=e^{-\varepsilon
  L(\varepsilon)}$ allows to apply Lemma~\ref{le.gamsig} with
$\nu_{i}(\varepsilon)=\frac{\varepsilon
  e^{-\varepsilon\ell_{i}(\varepsilon)}}{1-e^{-\varepsilon
    \ell_{i}(\varepsilon)}}$\,. From $e^{-\varepsilon
  \ell_{i}(\varepsilon)}\geq 1-\varepsilon \ell_{i}(\varepsilon)$ we
deduce
$$
\nu_{i}(\varepsilon) \geq \frac{e^{-\varepsilon \ell_{i}(\varepsilon)}}{\ell_{i}(\varepsilon)}\,.
$$
Hence the uniform boundedness of $\Tr\left[\varrho_{\varepsilon}{\bf
    N}^{k}\right]$ for $k\in\nz$, which is equivalent to
$\sum_{i=1}^{\infty}\nu_{i}(\varepsilon)\leq C$ implies
\begin{equation}
  \label{eq.infelli}
\inf_{j\in\nz^{*}, \varepsilon\in
  (0,\bar\varepsilon)}\ell_{j}(\varepsilon)=\kappa >0\,.
\end{equation}
We now use the assumption that the family
$(\varrho_{\varepsilon})_{\varepsilon\in (0,\bar\varepsilon)}$ admits
a unique Wigner measure $\mu_{0}$\,. As a quasi-free state,
$\varrho_{\varepsilon}$ is given by its characteristic function (see
for example \cite{BrRo} and \cite{AmNi1} for the
$\varepsilon$-dependent version)
$$
\Tr\left[\varrho_{\varepsilon}W(f)\right]=e^{-\frac{\varepsilon}{4}\langle
f\,,\, \frac{1+e^{-\varepsilon L(\varepsilon)}}{1-e^{-\varepsilon L(\varepsilon)}}f\rangle}\,.
$$
But the Wigner measure is characterized by its characteristic function
$$
G(\xi)= \int_{\Z}e^{-2i\pi S(z,\xi)}~d\mu_{0}(z)=\lim_{\varepsilon\to  0}\Tr\left[\varrho_{\varepsilon}W(\sqrt{2}\pi\xi)\right]\,.
$$
By taking $\xi=e_{i}$, $i\in\nz^{*}$, this implies that the limit
$$
\lim_{\varepsilon\to 0}e^{-\frac{\varepsilon
    \pi^{2}}{2}\frac{1+e^{-\varepsilon
      \ell_{i}(\varepsilon)}}{1-e^{-\varepsilon \ell_{i}(\varepsilon)}}}
$$
exists in $\rz$\,. With the constraint \eqref{eq.infelli} there are
two possibilities: either $\lim_{\varepsilon\to
  0}\ell_{i}(\varepsilon)=\ell_{i}(0)\in [\kappa,+\infty)$  and
$G(e_{i})=e^{-\frac{\pi^{2}}{\ell_{i}(0)}}$ or
$\lim_{\varepsilon\to 0}\ell_{i}(\varepsilon)=+\infty$ and
$G(e_{i})=1$\,. After recalling that the $\ell_{i}(\varepsilon)$ are
ordered and by introducing the index $J$ like in our statement, we get
for $\xi=\sum_{i=1}^{\infty}\xi_{i}e_{i}\in\Z$
$$
G(\xi)= e^{-\pi^{2}\sum_{i=1}^{J}\frac{|\xi|^{2}}{\ell_{i}(0)}}\,.
$$
The measure $\mu_{0}$ has to be the gaussian measure
$$
\mu_{0}=\mathop{\times}_{i=1}^{J}\left[\frac{\ell_{i}(0)}{\pi}e^{-\ell_{i}(0)|z_{i}|^{2}}~L(dz_{i})\right]\,,
\quad
z=\sum_{i=1}^{\infty}z_{i}e_{i}\,.
$$
Our assumptions imply that the integral $\int_{\Z}|z|^{2}~d\mu_{0}(z)$ equals
$$
\sum_{i=1}^{J}\frac{1}{\ell_{i}(0)}=
\int_{\Z}|z|^{2}~d\mu_{0}(z)
=
\lim_{\varepsilon\to 0}\Tr\left[\varrho_{\varepsilon}{\bf N}\right]\,.
$$
After Lemma~\ref{le.gamsig} we know that
$$
\sum_{i=1}^{J}\frac{1}{\ell_{i}(0)}=\lim_{\varepsilon\to
  0}\sum_{i=1}^{\infty}\nu_{i}(\varepsilon)=
\lim_{\varepsilon\to 0}\sum_{i=1}^{\infty}\frac{\varepsilon
  e^{-\varepsilon \ell_{i}(\varepsilon)}}{1-\varepsilon \ell_{i}(\varepsilon)}\,,
$$
which enforces the two conditions \eqref{eq.elliJ} and
\eqref{eq.Jelli}\,.\\
Conversely assume that all the conditions are satisfied. Reconsidering
the final argument in the proof of Lemma~\ref{le.gamsig} says that the function
$$
\prod_{i=J+1}^{\infty}(1-\nu_{i}(\varepsilon)x)^{-1}
$$
converges to $1$ in a given neighborhood of $x=0$\,. Hence
$$
\lim_{\varepsilon\to 0}\Tr\left[\varrho_{\varepsilon}{\bf
    N}^{k}\right]
=\lim_{\varepsilon\to 0}k!\sum_{
  \begin{array}[c]{c}
\scriptstyle|\alpha|=k\,,\\
\scriptstyle \alpha_{i}=0
~\text{for}~i> J
\end{array}
}\left(\frac{\varepsilon
  e^{-\varepsilon \ell_{i}(\varepsilon)}}{1-e^{-\varepsilon
    \ell_{i}(\varepsilon)}}\right)^{\alpha}=
k!\sum_{
  \begin{array}[c]{c}
\scriptstyle|\alpha|=k\,,\\
\scriptstyle \alpha_{i}=0
~\text{for}~i> J
\end{array}}
\ell(0)^{-\alpha}\,,
$$
which is easily checked to be equal to $\int_{\Z}|z|^{2k}~d\mu_{0}(z)$\,.
\fin

In the Bose-Einstein condensation of the free Bose gas in dimension
$3$, considered in
\cite{AmNi1}, the first eigenvalue is tuned so that $\ell_{1}(0)\in
(0,+\infty)$ and all the other eigenvalues are such that
$\ell_{i}(0)=+\infty$\,. The condition which fails and gives rise to a
physical example of dimensional defect of compactness is \eqref{eq.Jelli}\,.

\subsection{The Hartree-von Neumann limit}
Let $\varrho_{0}$
 be a non-negative trace class operator on $L^2(\rz^d)$ satisfying ${\rm Tr}[\varrho_{0}]=1$ and let
$$
\varrho^{\otimes N}=\varrho \otimes \cdots\otimes \varrho.
$$
Consider the time-dependent von Neumann equation for a system of  $N$ particles
\begin{equation}
\label{vnNeu}
\left\{ \begin{array}{ccl}
i \partial_t \varrho_N(t)&=& [\mathbb{H}_N, \varrho_N(t)] \\ \nm\ds
\varrho_N(0)&=& \varrho_{0}^{\otimes N},
\end{array} \right.
\end{equation}
with  $\varrho_N(t)$ is a trace class operator on
$L^{2}(\rz^{d})^{\otimes N}\sim L^{2}(\rz^{dN})$.
Here $\mathbb{H}_N$ is the Hamiltonian of the $N$ particles system
$$
{\mathbb H}_N=\sum_{i=1}^N 1\otimes\cdots \otimes A\otimes \cdots \otimes 1+ \frac{1}{N}\sum_{i<j} V(x_i-x_j) \,,
$$
with $A$ is a self-adjoint operator and $V\in L^\infty(\rz^d)$ real-valued satisfying $V(x)=V(-x)$.
As will appear in the proof, more general interactions could be
considered in the spirit of Theorem~\ref{th.main}, but we prefer to
stick to the usual presentation for an example.\\
The next result concerns the
 limit of the von Neumann dynamics (\ref{vnNeu}) in the mean field
 regime $N\to \infty$ already studied
in \cite{BeMa}\cite{AnSi}. We shall see that although the particles
are not assumed to be bosons, our bosonic mean field result apply to
this case due to the symmetry of the tensorized initial state
$\varrho_{0}^{\otimes N}$\,.
\begin{prop}
\label{pr.vN}
Let  $(\varrho_{N}(t))$ denote the solution to \eqref{vnNeu}, and
consider the trace class operator $\sigma_{N}^{(k)}(t)\in
\mathcal{L}^{1}(L^{2}(\rz^{kd}))$ defined by
relation
$$
\forall B\in {\cal L}(L^{2}(\rz^{kd}))\,,\quad
\Tr\left[\sigma_{N}^{(k)}(t)B\right]=\Tr\left[\varrho_{N}(t)(B\otimes
  I_{L^{2}(\rz^{d(N-k)})})\right]\,.
$$
Then the convergence
\begin{equation}
  \label{limvn}
\lim_{N\to \infty}\sigma^{(k)}_{N}= \varrho(t)^{\otimes k}
\end{equation}
holds in $\mathcal{L}^{1}(L^{2}(\rz^{dk}))$ for all $t\in \rz$ and
when $\varrho(t)$ solves the Hartree-von Neumann equation
\begin{equation}
\label{Hartree-von-Neumann}
\left\{ \begin{array}{ccl} i \partial_t \varrho(t)&=&[A +(V*n_{\varrho(t)}), \varrho(t) ] \\
\varrho(0)&=&\varrho_{0}\,,
\end{array} \right.
\end{equation}
with $n_\varrho(x, t):= \varrho (x; x, t)$\,.
\end{prop}
\proof
The proof will be done in three steps: Bosonisation, Liouvillian and
mean field limit.\\
\noindent\textbf{Bosonization:}
The phase space that we will consider is not
the one particle space $L^{2}(\rz^{d})$ but
$$
\Z=\mathcal{L}^{2}(L^{2}(\rz^{d}))\,,
$$
 the space of Hilbert-Schmidt
operators on $L^{2}(\rz^{d})$\,. It is endowed with the inner product
$$
\langle \omega_1,\omega_2\rangle_{\Z}=\Tr_{L^{2}}[\omega_1^* \omega_2]\,
$$
where $\Tr_{L^{2}}[.]$ here denotes the trace on $L^{2}(\rz^d)$ and
$\omega^*_1$ is the adjoint of $\omega_1$\,.\\
The cyclicity of the trace leads to
\begin{equation}
  \label{eq.trscal}
\Tr_{(L^{2})^{\otimes N}}\left[\varrho_{N}(t)(B\otimes
  I_{L^{2}(\rz^{d(N-k)})})\right]
= \langle \Psi_{N}(t)\,,\, (B\otimes
I_{L^{2}(\rz^{d(N-k)})})\Psi_{N}(t)\rangle_{\Z^{\otimes N}}
\end{equation}
with $\Psi_{N}(t)=e^{-it\mathbb{H}_{N}}\sqrt{\varrho_{0}}^{\otimes
  N}e^{it\mathbb{H}_{N}}$\,. \\
The important point is that at time $t=0$, $\Psi_{N}(0)=\sqrt{\varrho_{0}}^{\otimes
  N}$ , is a Hermite state in
$\bigvee^{N}\Z$ and that the evolution preserves this symmetry so that
$$
\forall t\in \rz\,,\quad \Psi_{N}(t)\in \bigvee\,^{N}\Z\,,\quad
\Psi_{N}(0)=\sqrt{\varrho_{0}}^{\otimes N}\,.
$$
With any  bounded operator $B: L^{2}(\rz^{dk})\to L^2(\rz^{dk})$, the
action by left (resp. right) multiplication is defined by
\begin{eqnarray*}
L_B~(\text{resp.}~R_{B}):\bigvee\,^k\Z &\to& \bigvee\,^k\Z\\
\omega^{\otimes k} &\mapsto& \mathcal{S}_{k}(B\omega^{\otimes k})\,,\quad
(\text{resp.}~\mathcal{S}_{k}(\omega^{\otimes k}B))\,,
\end{eqnarray*}
where $\mathcal{S}_{k}$ is the orthogonal projection from $\otimes^{k}\Z$
onto $\bigvee^{k}\Z$\,.
Since  $(\omega^{\otimes k})_{\omega\in\Z}$ is a total family in
$\bigvee^k\Z$ this defines a bounded operator $L_{B}\in
\mathcal{L}(\bigvee^{k}\Z)$ (resp. $R_{B}\in
\mathcal{L}(\bigvee^{k}\Z)$) such that $L_{B}^{*}=L_{B^{*}}$
(resp. $R_{B}^{*}=R_{B^{*}}$)\,. When $B(x_{1},\ldots, x_{k},
y_{1},\ldots, y_{k})$ is the Schwartz kernel of $B\in
\mathcal{L}(L^{2}(\rz^{dk}))$\,, $L_{B}$ (resp. $R_{B}$) is the left
(resp. right)
multiplication by the operator with kernel
$$
\frac{1}{k!}\sum_{\sigma\in\mathfrak{S}_{k}}B(x_{\sigma(1)},\ldots, x_{\sigma(k)},
  y_{\sigma(1)},\ldots, y_{\sigma(k)})\,.
$$
Hence the trace \eqref{eq.trscal} equals
$$
\Tr_{(L^{2})^{\otimes N}}\left[\varrho_{N}(t)(B\otimes
  I_{L^{2}(\rz^{d(N-k)})})\right]
=
\langle \Psi_{N}(t)\,, L_{[B\otimes I^{\otimes (N-k)}]} \Psi_{N}(t)\rangle_{\bigvee^{N}\Z}\,.
$$
With an operator $B\in \mathcal{L}(L^{2}(\rz^{dk}))$, we can now
associate a symbol
$$
b_B(\omega)=\langle \omega^{\otimes k},L_B \omega^{\otimes
  k}\rangle_{\bigvee^k\Z}=\Tr_{(L^{2})^{\otimes
    k}}\left[(\omega^{*})^{\otimes k}B\omega^{\otimes k}\right]\in \P_{k,k}(\Z).
$$
Since $L_{[B\otimes I^{\otimes (N-k)}]}$ is nothing but $L_{B}\bigvee
I_{\bigvee^{N-k}\Z}$ we get
$$
\Tr_{(L^{2})^{\otimes N}}\left[\varrho_{N}(t)(B\otimes
  I_{L^{2}(\rz^{d(N-k)})})\right]
=\frac{(N-k)!}{N!\varepsilon^{k}}
\langle \Psi_{N}(t)\,, b_{B}^{Wick}
\Psi_{N}(t)\rangle_{\bigvee^{N}\Z}\,,\quad \varepsilon=\frac{1}{N}\,.
$$
\noindent\textbf{Liouvillian:} Let us now determine the appropriate Hamiltonian $H_\varepsilon$ of this problem which is actually a  Liouvillian.
The map
$$
\rz\ni t\mapsto e^{-i t A} \omega e^{it A}
$$
defines a continuous unitary group on $\Z$ with a self-adjoint generator
\begin{eqnarray*}
\mathfrak{L}_A:\Z&\to&\Z\\
\omega&\mapsto&[A,\omega]\,.
\end{eqnarray*}
The interaction is a bounded  self-adjoint operator $\tilde
Q:\bigvee\,^2\Z\to\bigvee\,^2\Z$
given by $\tilde{Q}=\frac{1}{2}(L_{V}-R_{V})\in
\mathcal{L}(\bigvee^{2}\Z)$ and we associate the symbol
$Q(\omega)=\langle \omega^{\otimes 2}\,,\, \tilde{Q}\omega^{\otimes 2}\rangle$\,.
For any $\omega\in\Z$ the kernel of $\tilde Q\omega^{\otimes 2}\in\bigvee^{2}\Z$ is given by
\begin{eqnarray*}
(\tilde Q\omega^{\otimes 2})(x_1,y_1;x_2,y_2)= \frac{1}{2} V(x_1-x_2) \omega(x_1,y_1) \omega(x_2,y_2)-
\frac{1}{2} V(y_1-y_2) \omega(x_1,y_1) \omega(x_2,y_2)\,.
\end{eqnarray*}
After introducing the Hamiltonian
$$
H_\varepsilon=\d\Gamma(\mathfrak{L}_A)+Q^{Wick}\,,
$$
acting as a self-adjoint operator on $\Gamma_s(\Z)$, we get for
 $\Theta\in\bigvee\,^N\Z\cap \D(\d\Gamma(\mathfrak{L}_{A}))$\,,
$$
\varepsilon^{-1}H_{\varepsilon}\,\Theta=[{\mathbb H}_N,\Theta]\,\quad \mbox{ with } \quad \varepsilon=1/N.
$$
 This implies
$$
\Psi_{N}(t)=e^{-i t \mathbb{H}_N}(\sqrt{\varrho_{0}})^{\otimes N}e^{i t
  \mathbb{H}_N}=
e^{-i\frac{t}{\varepsilon} H_\varepsilon}(\sqrt{\varrho_{0}})^{\otimes N}
\in\bigvee\,^N\Z\,.
$$
\noindent\textbf{Mean field limit:} The initial data
$\varrho_{\varepsilon}(0)=|\sqrt{\varrho_{0}}^{\otimes N}\rangle
\langle \sqrt{\varrho_{0}}^{\otimes N}|$ is a Hermite state which fulfills
the assumptions of Theorem~\ref{th.main} with
$$
\mu_{0}= \delta_{\sqrt{\varrho_{0}}}^{S^{1}}\;.
$$
The classical energy associated with the Hamiltonian $H_{\varepsilon}$
is
$$
h(\omega)= \langle \omega\,,\, \mathfrak{L}_{A}\omega\rangle_{\Z} +
\frac{1}{2}\langle \omega^{\otimes 2}\,,\,
(L_{V}-R_{V})\omega^{\otimes 2}\rangle_{\Z}
$$
and the mean field flow ${\bf F}_{t}$ is nothing but the one given by
$$
i\partial_{t}\omega=\partial_{\bar
  \omega}h(\omega)=\left[A,\omega\right]
+ (V*n^1_\omega) \; \omega-\omega \; (V*n^2_\omega)\,,
$$
where $V*n^i_\omega$ are multiplication operators  and
$n^1_\omega(x)=\int_{\rz^d} |\omega(x,y)|^2 dy$,
$n^1_\omega(y)=\int_{\rz^d} |\omega(x,y)|^2 dx$  when
$\omega(x,y)$ denotes the kernel of $\omega$\,.
Beside the invariance  $|{\bf F}_t (\omega)|_\Z=|\omega|_\Z$ and
$\quad {\bf F}_t(e^{-i\theta} \omega)=e^{-i\theta} {\bf
  F}_t(\omega)$\,, the flow ${\bf F}_{t}$ also satisfies
\begin{equation}
\label{invariance}
\quad
{\bf F}_t(\omega^{*})={\bf F}_t(\omega)^*\,.
\end{equation}
Thus previous equation becomes equivalent to the Hartree-von Neumann
equation \eqref{Hartree-von-Neumann} with $\varrho(t)=\omega(t)^{2}$
when $\omega(0)=\sqrt{\varrho_{0}}$\,.
The Theorem~\ref{th.main} says
$$
\forall b\in \P_{k,k}(\Z)\,,\quad \lim_{N\to
  \infty}\Tr_{\bigvee^{N}\Z}\left[|\Psi_{N}(t)\rangle\langle
  \Psi_{N}(t)| b^{Wick}\right]=\int_{\Z}b(\omega_{t})~\delta_{\sqrt{\varrho_{0}}}^{S^{1}}= b(\sqrt{\varrho(t)})\,.
$$
In particular when $B\in \mathcal{L}(L^{2}(\rz^{dk}))$, this implies
$$
\lim_{N\to\infty}\Tr\left[\varrho_{N}(t)(B\otimes
  I_{L^{2}(\rz^{d(N-k)})})\right]
=\Tr_{L^{2}(\rz^{dk})}\left[\varrho(t)^{\otimes
  k}B\right]\,.
$$
This proves the weak convergence in \eqref{limvn}, but since it is
concerned with non negative trace class operator and
$\Tr\left[\sigma_{N}^{(k)}(t)\right]=1=\Tr\left[\varrho(t)^{\otimes
    k}\right]$ the convergence holds in the $\mathcal{L}^{1}$-norm.
\fin\\
We end with three remarks:
\begin{itemize}
\item When $\varrho$ is a pure state, the result of
  Proposition~\ref{pr.vN} is the same as \eqref{eq.VNeu}.
\item When $\varrho$ is not a pure state the
  Subsection~\ref{se.focktensor} has already shown that one has to be
  very careful with tensor products. Actually $\varrho^{\otimes N}\in
  \mathcal{L}^{1}(\mathop{\otimes}^{N}\Z)$ commutes with the
  symmetrization projection $\mathcal{S}_{N}$ (or the
  antisymmetrization $\mathcal{A}_{N}$ for fermions) but the
  corresponding states in $\L^1(\bigvee^N\Z)$ (resp.~$\L^1(\bigwedge^N\Z)$) are
$$
\mathcal{S}_{N}\varrho^{\otimes N}\mathcal{S}_{N}\quad(\text{resp.}~\mathcal{A}_{N}\varrho^{\otimes N}\mathcal{A}_{N})\,.
$$
But as shows the formula
$\Tr\left[\Gamma_{s}(\varrho)\right]=\prod_{\lambda\in
  \sigma(\varrho)}\frac{1}{1-\lambda}$
(resp. $\Tr\left[\Gamma_{a}(\varrho)\right]=\prod_{\lambda\in
  \sigma(\varrho)}\frac{1}{1+\lambda}$), the trace of
$\mathcal{S}_{N}\varrho^{\otimes N}\mathcal{S}_{N}$
(resp. $\mathcal{A}_{N}
\varrho^{\otimes N}\mathcal{A}_{N}$) converges to $0$ as $N\to
\infty$\,. We leave  for subsequent works, the question whether normalizing these states
would lead to the same asymptotics as in Proposition~\ref{pr.vN}.
\item We recall that a tensorization based on the tensor decomposition
  of Fock spaces in Subsection~\ref{se.focktensor} led to the
  evolution of Wigner measures which cannot be translated in terms of
  Hartree-von Neumann equations.
\end{itemize}

\bigskip
\noindent\textbf{Acknowledgements:} This work was finished while the
second author had a CNRS-sabbatical semester in Ecole Polytechnique.

\bigskip
\bibliographystyle{amsalpha}

\end{document}